\documentclass[a4paper,11pt]{article}
\pdfoutput=1 

\usepackage{jcappub} 

\usepackage[T1]{fontenc} 

\title{Measurement of the B-band Galaxy Luminosity Function with Approximate Bayesian Computation}

\author[a]{Luca Tortorelli,}
\author[a]{Martina Fagioli,}
\author[a]{J\"org Herbel,}
\author[a]{Adam Amara,}
\author[a]{Tomasz Kacprzak}
\author[a]{and Alexandre Refregier}

\affiliation[a]{Institute for Particle Physics and Astrophysics, ETH Z\"urich, Wolfgang-Pauli-Str. 27, 8093 Z\"urich, Switzerland}

\emailAdd{torluca@phys.ethz.ch}

\abstract{The galaxy Luminosity Function (LF) is a key observable for galaxy formation, evolution studies and for cosmology. In this work, we propose a novel technique to forward model wide-field broad-band galaxy surveys using the fast image simulator UFig and measure the LF of galaxies in the B-band. We use Approximate Bayesian Computation (ABC) to constrain the galaxy population model parameters of the simulations and match data from the Canada-France-Hawaii Telescope Legacy Survey (CFHTLS). We define a number of distance metrics between the simulated and the survey data. By exploring the parameter space of the galaxy population model through ABC to find the set of parameters that minimize these distance metrics, we obtain constraints on the LFs of blue and red galaxies as a function of redshift. We find that $\mathrm{M^*}$ fades by $\Delta \mathrm{M}^*_{\mathrm{0.1-1.0,b}} = 0.68 \pm 0.52$ and $\Delta \mathrm{M}^*_{\mathrm{0.1-1.0,r}} = 0.54 \pm 0.48$ magnitudes between redshift $\mathrm{z = 1}$ and $\mathrm{z = 0.1}$ for blue and red galaxies, respectively. We also find that $\phi^*$ for blue galaxies stays roughly constant between redshift $\mathrm{z = 0.1}$ and $\mathrm{z=1}$, while for red galaxies it decreases by $\sim 35\%$. We compare our results to other measurements, finding good agreement at all redshifts, for both blue and red galaxies. To further test our results, we compare the redshift distributions for survey and simulated data. We use the spectroscopic redshift distribution from the VIMOS Public Extragalactic Redshift Survey (VIPERS) and we apply the same selection in colours and magnitudes on our simulations. We find a good agreement between the survey and the simulated redshift distributions. We provide best-fit values and uncertainties for the parameters of the LF. This work offers excellent prospects for measuring other galaxy population properties as a function of redshift using ABC.}

\begin{document}
\maketitle
\flushbottom

\section{Introduction}
\label{section:introduction}

The study of the origins and the growth of structures that form galaxies is one of the main topics of modern cosmological research. The structure formation paradigm states that the seeds of galaxies reside in the primordial density perturbations \cite{blumenthal84,davis85}, while their subsequent formation happens via gravitational collapse following dark matter clustering \cite{white78}. Galaxies then assemble their mass with cosmic time following this hierarchical pattern. We cannot follow the evolution of each individual seed with cosmic time, but only see snapshots of the galaxy population at different redshifts. A way to overcome this limitation is by looking at tomographic precision counts of galaxies over cosmological time, i.e. the galaxy Luminosity function (LF). The LF is defined as the number of galaxies in a given comoving volume with a given luminosity. The galaxy LF gives a direct estimate of the total amount of light present in galaxies and its evolution provides information about how galaxies build-up their stellar mass with cosmic time through either star-formation or hierarchical assembly. Thus, the LF is an important observational ingredient both to understand and test models of galaxy formation and evolution \cite{benson03} and for cosmology \cite{Wechsler2018}. They provide information about the physical processes that convert mass into light, about the mechanisms that change galaxy morphology and about the estimate of the luminosity and baryonic density of the Universe.

Since the first galaxy redshift surveys, the probed volume and the number of galaxies with spectroscopic redshift has largely grown. Thanks to large homogeneous imaging and spectroscopic surveys of the low-redshift Universe (e.g., 2dFGRS \cite{norberg02} and SDSS \cite{blanton03}), the local ($\mathrm{z < 0.2}$) galaxy optical LF is well constrained near L$^*$\footnote{L$^*$ is a characteristic galaxy luminosity where the power-law form of the function cuts off. Its typical value from local B-band measurement is $\mathrm{L^*_B = (1.2 \pm 0.1) h^{-2} \times 10^{10} L_{\odot}}$.}, providing a fundamental measurement of the contents of the local Universe. Most of the studies were either performed in spectroscopic low-redshift pencil beam surveys or in shallow large-area low-redshift photometric surveys. Recently, thanks to more powerful telescopes and instrumentation, LF studies started to be carried out in deep spectroscopic high-redshift surveys and in high-depth large-area photometric galaxy surveys (e.g., \cite{beare15,capozzi17}). This allowed millions of galaxies to be targeted, probing the LF beyond $\mathrm{z=0.5}$. The power of wide-field photometric surveys relies on the ability to cover large areas on the sky, therefore greatly reducing the impact of cosmic variance. However, they lack precise redshift estimation of galaxies which is essential for a tomographic analysis. Furthermore, photometric redshifts can suffer from catastrophic errors \cite{Hearin2010}, i.e. galaxy redshift completely different from the one determined through spectroscopic features estimation. This may result in systematic errors that are hard to quantify and they impact the LF estimation. On the other hand, spectroscopic surveys provide precise redshift estimation, but they often probe volumes that are too small to be representative of the entire galaxy population, i.e. it can lead to biases due to cosmic variance. Spectroscopic surveys also require larger telescopes and longer integration times than comparably deep photometric redshift surveys. Considering these limitations, measuring the LF evolution to high-redshifts still represents a significant challenge.

An additional challenge comes from the different sample definition that different studies adopt when building the LF. One would like to derive independent LFs for different groups of galaxies with different physical properties, e.g., red and blue galaxies. With the advent of very large and deep galaxy surveys, this has become possible. However, the samples are not always consistent between different works. Galaxies are generally separated into red and blue according to rest-frame colours, morphologies, spectral types, but it is well known that the overlap between these classifications changes with the sample selection \cite{tortorelli18a}. Furthermore, it is difficult to measure the galaxy number density accurately, especially for red galaxies, since their strong spatial clustering (e.g., \cite{brown03}) results in significant cosmic variance and therefore one requires large samples over a large number of statistically uncorrelated regions.

The most popular parametrization of the galaxy LF was proposed by Schechter in 1976 \cite{schechter76}:
\begin{equation}
\Phi (\mathrm{L,z}) \mathrm{dL} = \mathrm{ \phi^*(z) \left ( \frac{L}{L^*(z)} \right)^{\alpha(z)} \exp{\left[ -L/L^*(z) \right]} \frac{dL}{L^*(z)} }
\end{equation}
where L is the galaxy luminosity, $\phi^*$ sets the normalization of the Schechter function, L$^*$ defines the `knee' of the LF where it transitions from a power law to a decaying exponential behaviour and $\alpha$ is the slope of the power law at the faint end. In general, $\phi^*$ and L$^*$ are treated as free parameters, while $\alpha$ is kept fixed at low-redshift values, since it becomes increasingly difficult to determine at high-redshift, where intrinsically faint galaxies are harder to detect. This functional form is usually conveniently expressed in terms of magnitudes, rather than luminosities:
\begin{equation}
\Phi (\mathrm{M,z}) \mathrm{dM} = 0.4 \ln{(10)} \phi^*(z) 10^{0.4 (\mathrm{M^*(z) - M}) (\alpha(z) + 1)} \exp{{\left[ -10^{0.4 (\mathrm{M(z)^* - M})} \right]}} \mathrm{dM}
\label{equation:lf}
\end{equation}
where M is the absolute magnitude of a galaxy. We adopt the latter parametrization throughout our work.

A large number of LF studies for blue and red galaxies have been carried out (e.g., \cite{norberg02,blanton03,giallongo05,ilbert06,zucca06,brown07,faber07,zucca09,ramos11,loveday12,cool12,fritz14,beare15,Weigel2016,capozzi17,Kawinwanichakij2020}). They find that the global rest-frame B-band LF evolves at the bright end up to $z \sim 3$, with $\mathrm{M^*}$ becoming fainter with time. However, there is still no consensus on how much fainter it has become and whether $\mathrm{M^*}$ for red galaxies fades faster than that for blue galaxies or viceversa. The cited studies show that $\mathrm{M^*}$ brightens for all galaxy types of about $\mathrm{2\ mag}$ in the B-band from $z \sim 0$ to $z \sim 2$. $\phi^*$ for blue galaxies stays roughly constant from $z \sim 1$ to today. For red galaxies it increases with decreasing redshift of roughly $50\%$, although different studies give widely differing estimates for this factor, depending, for instance, on whether red galaxies LFs are described by a single or a double Schechter function. The studies find that the number of blue galaxies in a given comoving volume is greater than that of red galaxies at all redshifts. The number density of red galaxies decreases with increasing redshift and their B-band LF shows a rather flat slope and evolves only mildly. Most of the uncertainties in literature about the estimate of $\phi^*$ and $\mathrm{M^*}$ is due to the highly degenerate nature of the Schechter parameters with respect to the adopted value of the faint-end slope $\alpha$. The latter steepens going from blue to red galaxies and the average value found at low-redshift is $\alpha=-1.3$ for blue and $\alpha=-0.5$ for red galaxies. 

These studies show a notable dispersion in their measurements. They use different number of galaxies and different sample selections, leading to different estimates of the shape of LFs and degree of fading over time. Additionally, there is the difficulty in assembling large samples of accurately classified galaxies with spectroscopic redshifts or the systematic uncertainties affecting the photo-z estimation. The development of new methods to measure the optical LF evolution with redshift may therefore help to overcome part of these limitations.

To address these problems, we propose a new method to measure the galaxy LF as a function of redshift with Approximate Bayesian computation (hereafter, ABC, \cite{akeret15}) by forward modeling wide-field photometric and spectroscopic galaxy surveys. Our method tries to overcome part of the limitations highlighted in the previous paragraph. For instance, we do not have to estimate photometric redshifts for each object, since redshifts for simulated galaxies are given as input property. Another effect that we take into account is the separation in blue and red galaxies. Our model has two different redshift-dependent LFs, one for blue and one for red objects (see section \ref{section:method}), from which we sample absolute magnitudes and redshifts of galaxies. This means that we do not use a pre-defined selection of galaxies to estimate the LF. Furthermore, we assume a parametric model for the galaxy population and we create a large number of simulated images for a given parameter set. This allows us to measure the LF for samples of galaxies that contains up to $\sim 3 \times 10^5$ objects for a single parameter set.

Likelihood-free inference methods, such as ABC, have recently started to provide competitive results in measuring cosmological parameters (e.g., \cite{Alsing2018a,Alsing2018b,Alsing2019a,Alsing2019b}). Attempts have already been made to use ABC to measure size and photometric evolution of galaxies from image simulations \cite{Carassou2017} and star-formation histories of galaxies \cite{Aufort2020}, showing very promising results for this method. They were however limited in the computational time required to render realistic images \cite{Plazas2020} and in the use of trivial distances, such as `L1' and `L2' distances. These limitations are overcomed in our work by using a fast and realistic image simulator, a new algorithmic approach and a large physically motivated set of distances.

The forward modeling approach relies on producing realistic simulations \cite{refregier14,bruderer18} and on directly comparing measured properties of the survey data with those measured on the simulations, e.g., magnitudes and sizes of galaxies. The parameters of the simulations are then changed until certain diagnostics of the measured galaxy properties from the survey data agree with those from the simulations. In order for this approach to be successful, one needs to perform the exact same data analysis steps on survey data and on simulations and be able to produce realistic and fast image simulations. The Ultra Fast Image Generator (UFig, \cite{berge13,bruderer16}) was developed for this purpose. The core of UFig is a simple yet realistic galaxy population model developed in \cite{herbel17}, which has already proven to give a good description of the overall galaxy population properties as shown by forward modeling the Dark Energy Survey (DES) \cite{bruderer18,Kacprzak2019}, the Physics of the Accelerating Universe (PAUS) narrow-band galaxy survey \cite{tortorelli18b} and SDSS/CMASS spectra of red and bluer galaxies \cite{fagioli18,Fagioli2020}.

In this paper we constrain the galaxy population model parameters in UFig by means of wide-field photometric legacy data, such as those from the Canada-France-Hawaii Telescope Legacy Survey (CFHTLS, \cite{cuillandre12}). We provide a measurement of the B-band galaxy LF and we compare it with other measurements in the literature. Given that there is no clear empirical likelihood that can be calculated for our images, in order to constrain the galaxy population model in a probabilistic Bayesian approach, we need to use a likelihood-free inference method, namely ABC. We approximate the Bayesian posterior by iteratively restricting the prior space on the basis of a distance metric between a real and a simulated dataset to obtain a set of posterior samples. In our work, we explore the use of different distance metrics to constrain the parameters of interest.

To test the applicability of the framework for cosmology studies, we also compare the set of redshift distributions estimated from our posterior samples with the n(z) from the VIPERS survey \cite{Garilli2014,Scodeggio2018}. Since the imaging survey used for the sample selection is CFHTLS itself, we apply the same magnitude and colour cuts to our simulated images (see section \ref{subsection:red_distr}) to obtain the sample of objects that we use to build our redshift distributions.

The paper is structured as follows. Section \ref{section:data} gives a description of the data that we use in this work, while the image simulator and the galaxy population model are described in section \ref{section:galpopmodel}. Section \ref{section:method} describes the method we develop in our work. Section \ref{section:results} summarizes the results of our paper. We draw our conclusions and provide future directions in section \ref{section:conclusions}. Throughout this work, we use a standard $\mathrm{\Lambda CDM}$ cosmology with $\Omega_{\mathrm{m}} = 0.3$, $\Omega_{\Lambda} = 0.7$ and $\mathrm{H_0 = 70\ km\ s^{-1}\ Mpc^{-1} }$. Throughout the paper, unless stated otherwise, the uncertainties on the galaxy population model parameters and photometric measurements refer to the 68\% confidence level (hereafter, $1 \sigma$).

\section{Data}
\label{section:data}

This section describes the data we use to constrain our galaxy population model and to test our results for cosmology applications. We use data from the Canada-France-Hawaii Telescope Legacy Survey (CFHTLS) and from the VIMOS Public Extragalactic Redshift Survey (VIPERS). The main reasons for this choice are their public availability, the very good photometric quality, the wide area covered on the sky and the simple and reproducible target selection for spectroscopy.

\subsection{Canada-France-Hawaii Telescope Legacy Survey (CFHTLS)}

The Canada-France-Hawaii Telescope Legacy Survey (CFHTLS) is a photometric galaxy survey conducted with the MegaCam \cite{boulade00} camera mounted on the CFHT telescope on top of Mauna Kea (Hawaii, USA). CFHTLS is a deep sub-arcsecond seeing wide-field ($157$ square degrees) optical survey ($\mathrm{u^{*}}$, $\mathrm{g'}$, $\mathrm{r'}$, $\mathrm{i'}$, $\mathrm{z'}$ bands), comprising two components: CFHTLS `Wide' consists of $155$ square degrees in four independent contiguous patches (called `W1', `W2', `W3', `W4') with a $80\%$ completeness limit in AB of $\mathrm{u^{*}}=25.2$, $\mathrm{g'}=25.5$, $\mathrm{r'}=25.0$, $\mathrm{i'}=24.8$, $\mathrm{z'}=23.9$ for point sources, while CFHTLS `Deep' consists of $4$ independent $1$ square degree ultra deep pointings (called `D1', `D2', `D3', `D4') reaching a $80\%$ completeness limit in AB of $\mathrm{u^{*}}=26.3$, $\mathrm{g'}=26.0$, $\mathrm{r'}=25.6$, $\mathrm{i'}=25.4$, $\mathrm{z'}=25.0$ for point sources. We use data from the 7th and final data release of CFHTLS produced by Terapix \cite{cuillandre12} that provides a data collection which is photometrically calibrated at better than one percent over the total $155$ square degrees of the survey in all five photometric bands. The camera covers one square degree with a pixel scale of $0.186$ arcsec to properly sample the median seeing of CFHT ($\sim 0.7$ arcsec). Three main science cases drove the design of the CFHTLS: the search for Type Ia Supernovae at high-redshift, the cosmic shear analysis and the census of Kuiper belt objects. These scientific themes shaped the observing strategy of the two distinct components of the CFHTLS: the `Deep' fields to detect Type Ia Supernovae at high-redshifts and the `Wide' fields to cover a large areas of the sky at intermediate depth to look for Kuiper belt objects and to perform a cosmic shear analysis. We perform our analysis using only the `Wide' fields for two main reasons. The `Wide' images simulation time is faster than that required to simulate the `Deep' images, given the different depth of the images. The second reason is that the area covered on the sky by the CFHTLS `Wide' images is larger than that of the `Deep' images.

\subsection{The VIMOS Public Extragalactic Redshift Survey (VIPERS)}

The VIMOS Public Extragalactic Redshift Survey (VIPERS) \cite{Garilli2014,Scodeggio2018} is an ESO Large Programme conducted with the VIMOS spectrograph \cite{lefevre00} on the Very Large Telescope (VLT) at Cerro Paranal (Chile). The survey is aimed at studying the clustering of galaxies, measuring redshift-space distortions, characterizing the galaxy population and the density field at $\left \langle \mathrm{z} \right \rangle \simeq 0.8$. Furthermore, the survey has been designed to create a spectroscopic sample of nearly $10^5$ galaxies with $\mathrm{i_{AB}}< 22.5$ at $\mathrm{0.5 < z < 1.2}$. The fields covered by VIPERS are inside the `W1' and `W4' fields of CFHTLS. The surveyed area is about $16$ square degrees in `W1' and $8$ square degrees in `W4'.

The target selection in VIPERS is based on a magnitude-limited selection. Since VIPERS was designed to study clustering and redshift-space distortions at $\mathrm{z \simeq 0.5 - 1.2}$, the desired redshift range for objects classified as galaxies is selected requiring that the magnitudes and colours measured with CFHTLS satisfy:
\begin{equation}
\mathrm{ \left ( r' - i' \right) > 0.5 \left( u^{*} - g' \right) \quad \vee \quad  \left ( r' - i' \right) > 0.7}
\end{equation}
The upper limit in redshift is ensured by the fact that a magnitude-limited sample with $\mathrm{i'_{AB}}< 22.5$ would cover a redshift range up to $\mathrm{z \sim 1.2}$. Stars and galaxies are separated using measured sizes and spectral energy distributions from the CFHTLS 5-band photometry. We use catalogues coming from the second data release of VIPERS\footnote{\url{http://vipers.inaf.it/rel-pdr2.html}}. This consists of spectroscopic and photometric properties for roughly $9 \times 10^4$ galaxies.

\section{UFig and the Galaxy Population model}
\label{section:galpopmodel}

This section provides a short description of the basic principle of the image simulator and of the galaxy population model developed in \cite{herbel17}. The Ultra Fast Image generator (UFig, \cite{berge13,bruderer16}) is a fast code that simulates astronomical images in different optical filter bands. UFig first generates catalogues of galaxies and then renders their pixelated light profiles, including observational and instrumental effects, such as noise, PSF and pixel saturation. UFig was developed for forward modeling purposes in the framework of the Monte Carlo Control Loops (MCCL, \cite{refregier14,Kacprzak2019}) pipeline. To be able to run the MCCL pipeline, one needs to produce thousands of simulated images and therefore, the speed in rendering images is an essential feature. Its run-time is comparable to the timescale of running a commonly used analysis software such as \textsc{Source Extractor} (SE, \cite{bertin96}), e.g., less than one minute for a $0.25$ square degrees optical image. The speed of UFig lies both in the highly optimized code and in the use of simplifying models that can be made more complex if required.

\begin{figure}[t!]
\centering
\includegraphics[width=17cm]{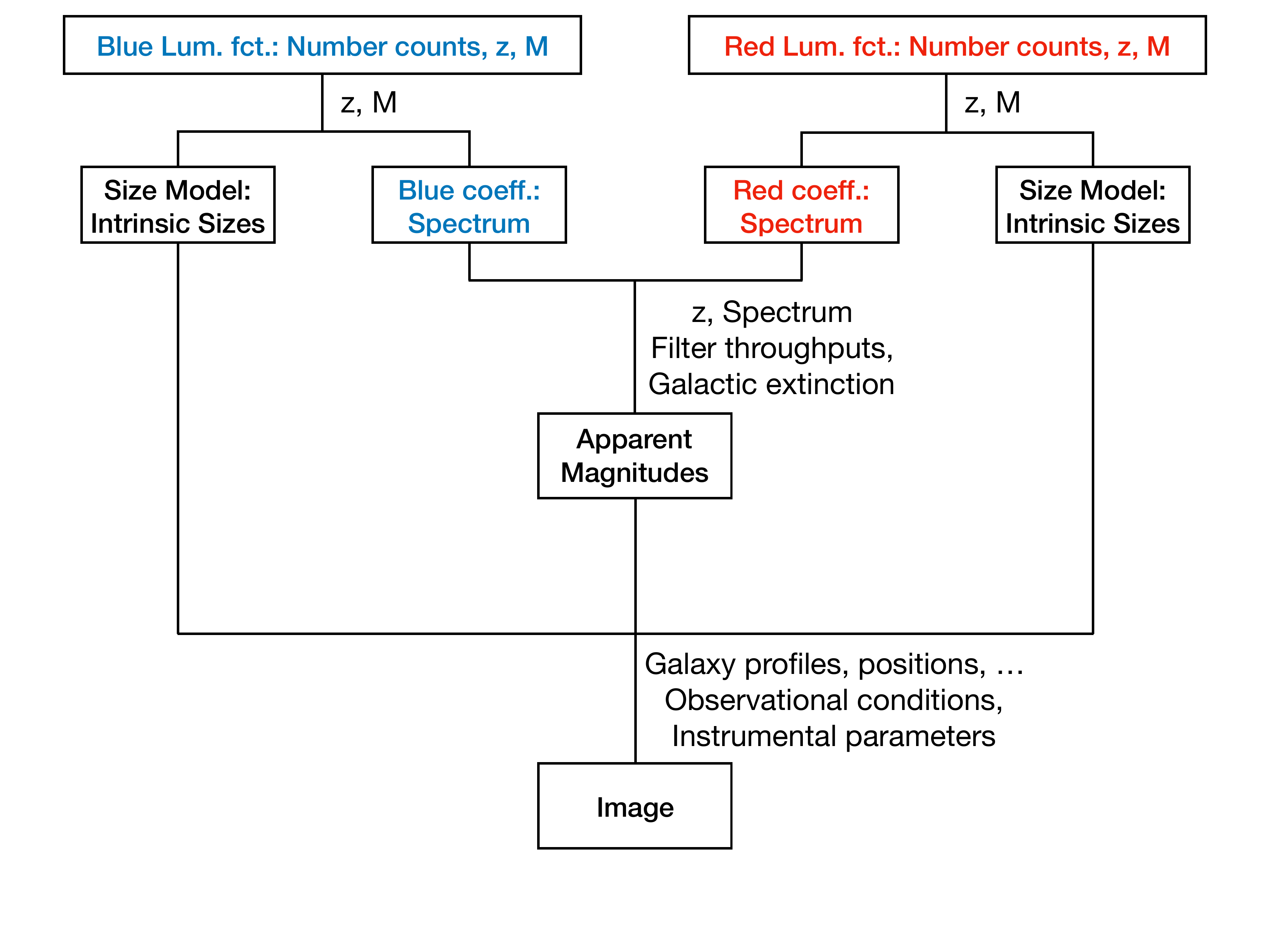}
\caption{Flowchart describing the galaxy population model developed in \cite{herbel17}. Coloured boxes refer to red and blue galaxies specific properties, while black boxes refer to parts of the model that are in common for blue and red galaxies.}
\label{fig:tortorelli_fig1}
\end{figure}

A flowchart of the galaxy population model at the core of UFig is described in figure \ref{fig:tortorelli_fig1} adapted from \cite{herbel17}. We give a short description of the model, while a more detailed one can be found in section 3 of \cite{herbel17}. Blue and red galaxies (see Appendix \ref{appendix:blue_red_uvj}) are sampled from two different redshift-dependent B-band LFs. Their functional form is given by the Schechter function in equation \ref{equation:lf}, where $\mathrm{M^*}$ and $\phi^*$ are free parameters, different for the blue and the red populations. We decide to keep the values of the faint-end slopes to $\alpha=-1.3$ and $\alpha=-0.5$ for blue and red galaxies, respectively, fixed at their low-redshift values quoted in \cite{beare15}. The evolution with redshift of $\mathrm{M^*}$ and $\phi^*$ is parametrized as
\begin{equation}
\begin{split}
\mathrm{M^*(z)} &= \mathrm{M^*_{slope}\ z + M^*_{intcpt} } \\
\mathrm{\ln{\phi^*(z)}} &= \mathrm{\ln{\phi^*_{amp}} + \phi^*_{exp}\ z} \\
\end{split}
\end{equation}
where $\mathrm{M^*_{slope}}$ and $\mathrm{M^*_{intcpt}}$ are the slope and the intercept of the linear evolution of $\mathrm{M^*}$ with redshift, while $\mathrm{\phi^*_{amp}}$ and $\mathrm{\phi^*_{exp}}$ are the amplitude and the exponential decay rate of $\phi^*$ with redshift that we linearize in log-space. From these two LFs, UFig computes the number of galaxies per unit volume per unit magnitude and assign redshift and absolute magnitudes to those galaxies.

The physical sizes of galaxies are drawn from a log-normal distribution \cite{shen03} with mean $\mu_{\mathrm{phys}}$ and standard deviation $\sigma_{\mathrm{phys}}$. The mean itself is linear in the absolute magnitude $\mathrm{M}$ drawn from the LF,
\begin{equation}
\mu_{\mathrm{phys}} \left( \mathrm{M} \right) = \mathrm{r_{50,slope}^{phys}}\ \mathrm{M} + \mathrm{r_{50,intcpt}^{phys}}
\end{equation}
while $\sigma_{\mathrm{phys}}$ is fixed for all galaxies. The physical size is then transformed into an angular size according to the redshift of the object and the chosen cosmology.

Galaxies are rendered on the image in random positions on the sky. To do that, we create an \textsc{HEALPix} pixelization of the image. Then, we uniformly sample \textsc{HEALPix} pixels and we convert them into projected x and y coordinates. The absence of an accurate angular correlation function might affect the completeness of faint sources, especially in highly clustered areas \cite{zhang2015}. However, the development of a more accurate clustering prescription is left to future work. Galaxy photons are distributed on the image according to a S\'ersic light profile:
\begin{equation}
\mathrm{I(r) = I(r_{50}) \exp{\left( -k \left[ \left( \frac{r}{r_{50}} \right)^{1/n} -1 \right] \right)}}
\end{equation}
where $\mathrm{r_{50}}$ is the radius enclosing $50\%$ of the light from the object, $\mathrm{n}$ is the S\'ersic index and $\mathrm{k}$ satisfies the equation $2 \gamma (2n,k) = \Gamma(2n)$, with $\gamma$ and $\Gamma$ being the lower incomplete gamma-function and the gamma-function, respectively. Following the prescription in \cite{berge13}, galaxies having magnitude less than $20$ have S\'ersic indices drawn from a probability density function (p.d.f.) given by $\mathrm{f(n) = \exp{\left( \mathcal{N}(0.3,0.5) + \mathcal{N}(1.6,0.4) \right) + 0.24} }$, where $\mathcal{N}$ is a normal distribution. For fainter galaxies, the p.d.f. is $\mathrm{f(n) = \exp{\left( \mathcal{N}(0.2,1) \right)}+0.2 }$.

We include the effect of the PSF on the image by modeling it as a circular Moffat profile \cite{Moffat1969}:
\begin{equation}
\mathrm{I(r) = I_0 \left[ 1 + \left( \frac{r}{\gamma} \right)^2 \right]^{-\beta}}
\end{equation}
where $\mathrm{I_0}$ is the flux value in the center, while $\gamma$ and $\beta$ are scale parameters that depend on the observational conditions. The PSF affects both stars and galaxies. Stars in particular are simulated on the image using the Besancon model \cite{robin03} based on stellar population synthesis. The PSF parameters measurement on CFHTLS images is described in section \ref{subsection:photometry}.

UFig also assigns a spectrum to each galaxy as a linear combination of five basis spectra taken from the Kcorrect templates \cite{blanton07}
\begin{equation}
\mathrm{f_e \left( \lambda \right) = \sum_i c_i\ f_{e,i}\left( \lambda \right)}
\end{equation}
where the subscript `e' refers to rest-frame. This is empirically motivated by SDSS data \cite{herbel17}. The $\mathrm{c_i}$ coefficients are jointly drawn from a Dirichlet distribution of order $5$, which is parametrized by five parameters $\mathrm{a}_{\mathrm{i}}$ (hereafter, spectral coefficients). They, in turn, evolve with redshift according to 
\begin{equation}
\mathrm{ a_i \left( z \right) = \left( a_{i,0} \right)^{1 - z / z_1} \times  \left( a_{i,1} \right)^{z/z_1}}
\end{equation}
where $\mathrm{a}_{\mathrm{i},0}$ describes the galaxy population at $\mathrm{z = 0}$, while $\mathrm{a}_{\mathrm{i},1}$ at redshift $\mathrm{z = z_1 > 0}$. We use $\mathrm{z_1 = 1}$ in our work and different distributions for blue and red galaxies. The reason we choose the Dirichlet distribution is that after drawing the $\mathrm{c_i}$ and calculating $\mathrm{f_e \left(\lambda \right)}$, the spectrum is rescaled to match the absolute magnitude of the object. The rest-frame spectra are then redshifted to the specific galaxy redshift z. We apply a position and wavelength-dependent Milky-Way extinction factor and then we integrate the spectrum in the particular probed waveband to obtain the apparent magnitude of each galaxy.

These galaxy and star properties are collected in a catalogue and used by UFig to render the objects on a pixelated grid. Then, UFigs adds the instrumental effects of the survey we are modeling. We add Poisson noise for galaxies and stars, we add a Gaussian noise to mimic the noise arising from sky brightness, readout and errors during data processing and we correlate the noise using a Lanczos resampling \cite{duchon79} of order 3. Means and standard deviations for the Gaussian noise are measured on survey images, while the PSF parameters are measured on survey images using an external reference catalogue of stars (see section \ref{subsection:photometry} for both measurements).

\section{Method}
\label{section:method}

This section summarizes the method we develop to measure the LF using ABC. We describe the steps we perform to measure galaxy properties and to simulate CFHTLS images. Then, we describe the ABC inference we perform along with the defined prior space and the different distance metrics we define for our problem. In Appendix \ref{appendix:abc_inferece_toy_problem} we test an 8-dimensional multivariate Gaussian case. In Appendix \ref{appendix:simonsim_run}, we define a target known set of parameters for the galaxy population model and we test whether we are able to recover the true input value from our ABC inference scheme.

\subsection{Photometry with \textsc{Source Extractor}}
\label{subsection:photometry}

In order for the forward modeling approach to be successful, the same analysis steps need to be performed on survey data and simulated data. Therefore, we need a fast and robust software to perform photometry that has already been successfully used in a previous work \cite{tortorelli18b}, namely \textsc{Source Extractor} (SE, \cite{bertin96}).

The SE configuration file we use in our work is described in Appendix \ref{appendix:sexconfig}. It requires as input gain, saturation, pixel scale and magnitude zero-point of each image that we read from the images header. It also requires an estimate of the full width half maximum (hereafter, FWHM) of the point spread function (hereafter, PSF) for each image. We estimate the FWHM of the PSF using `GAIA DR1' \cite{gaia,gaiadr1} stars. We select stars in the image which have g-band magnitude in the range $\mathrm{19 < m < 20}$. We choose this magnitude range in order to have high signal-to-noise stars that are not saturated. We cut $30 \times 30$ pixel stamps around the stars. We fit their light profiles using a 2D circular Moffat profile to measure the $\gamma$ and the $\beta$ parameters of the Moffat profile (see section \ref{section:galpopmodel}). The FWHM for each star is given by $\mathrm{FWHM = 2 \gamma \sqrt{2^{1/\beta}-1} }$. The final estimate of the $\mathrm{FWHM}$ and of the $\beta$ for the full image is the mean value from all the stars on the image.

Very bright stars constitute a nuisance in wide-field surveys and CFHTLS makes no exception. The contamination due to stars halos and reflections may bias the photometric measurements, especially those of faint sources. For this reason, different deep extragalactic studies \cite{Erben2009,Gwyn2012} use masks to avoid stars contamination, which may cover up to 40\% of the total imaged area \cite{Heymans2012}. To avoid biasing our photometric measurements, we also use masks to take into account stars contamination. Specifically, we use masks provided by the CFHTLS collaboration. However, they do not cover perfectly the halo of bright stars, defined as those brighter than $\mathrm{m < 10}$ in the `GAIA' g-band. Therefore, we extend them by masking the star halos with a circle of 4 arcminutes radius. This radius is estimated by measuring the area of the star halos in the CFHTLS images. These enhanced masks are then provided to SE to be used as weight images. This allows the software to ignore all the sources that fall inside stars halos or close to bleed trails artifacts during the photometric measurement.

We use forced photometry to measure galaxy properties with the available multi-wavelength data. We run SE in the dual-image mode: we provide SE with a detection and a measurement image. We create the detection image following the prescription in \cite{coe06} and references therein. We stack the CFHTLS images in the $5$ different wavebands by normalizing them according to the rms of the background noise in each waveband. The detection image has the average PSF of all the stacked images. SE estimates the isophotal apertures and sizes from this detection image. To estimate the background noise, we mask each image with the SE output segmentation map (see \cite{bertin96} for a description), we apply a sigma clipping on this masked image and we define as mean and standard deviation of the background noise the ones measured on the sigma clipped masked image.

We then apply selection cuts on the output catalogues of galaxy properties. In particular, to avoid spurious detections and contaminations, we select objects having the SE parameter `$\mathrm{FLAGS < 4}$'. To separate galaxies from stars, we select objects having the parameter `$\mathrm{CLASS\_STAR < 0.9}$' in every waveband. In order to keep galaxies with reliable photometry, we select objects having `$\mathrm{MAG\_ISO < 99}$', `$\mathrm{MAG\_AUTO < 99}$' and `$\mathrm{FLUX\_RADIUS < 30}$' in every waveband. For simulated images (see \ref{subsection:imagesims}), we also matched the SE output catalogue with the UFig input one to provide each simulated object with its assigned input redshift. We also apply the PSF and aperture correction to our magnitude measurements following \cite{coe06} (hereafter, `$\mathrm{MAG\_ISO\_CORR}$'). This final magnitudes are then used to compare observations and simulations.

\subsection{CFHTLS image simulations}
\label{subsection:imagesims}

\begin{figure}[t!]
\centering
\includegraphics[width=7.65cm]{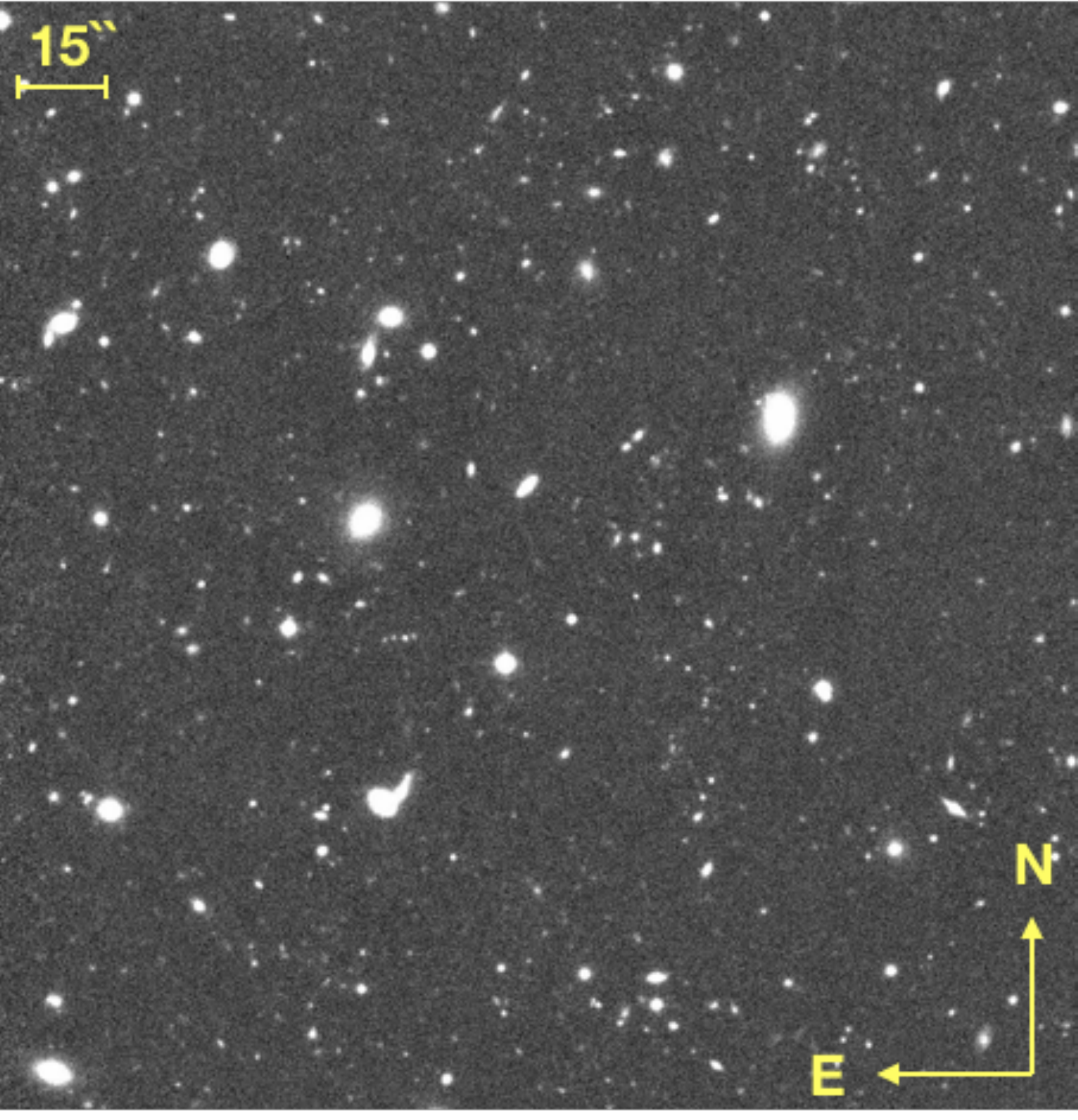}
\includegraphics[width=7.65cm]{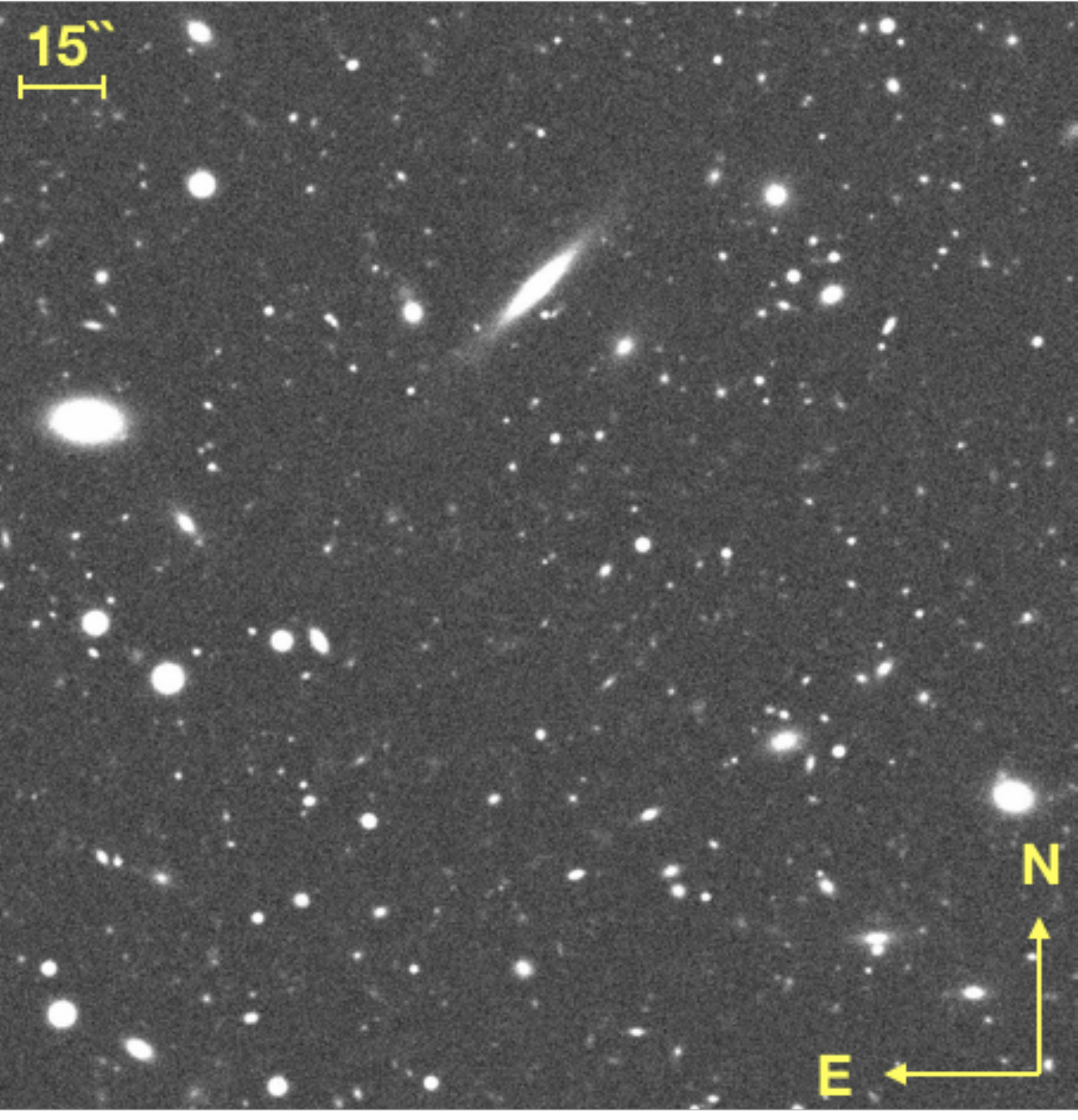}
\caption{Comparison between a CFHTLS `Wide' observed (left panel) and simulated (right panel) image. Both images represent the same patch on the sky in the $\mathrm{i'}$-band. Bottom right arrows in each image represent the north and east celestial directions. The upper left thicks represent the angular size on the image corresponding to 15 arcsec.}
\label{fig:tortorelli_fig2}
\end{figure}

Besides the galaxy population parameters described in \ref{section:galpopmodel}, the simulated images generated by UFig depend also on instrumental and observational parameters. These are gain, saturation, exposure time, magnitude zero-point, PSF FWHM and $\beta$ of the Moffat profile, background noise amplitude and standard deviation. These parameters are estimated for CFHTLS images as described in \ref{subsection:photometry}. To ensure a more homogeneous coverage and avoid clustering impact, we cut each $\sim1$ square degree CFHTLS `Wide' image (and corresponding masks) in $16$ patches of $4000 \times 4000$ pixels, each corresponding to $\sim 13\ \mathrm{arcmin} \times 13\ \mathrm{arcmin}$. We use these simulated $4000 \times 4000$ pixels patches for our ABC inference. Hereafter, we refer to them as simulated images.

Gain, saturation, exposure time and magnitude zero-point for simulated images are taken from the survey image headers. The background noise amplitude and standard deviation are measured on survey data as described in \ref{section:galpopmodel}. We assume the simulated image noise to be the drawn from the same position-independent gaussian distribution. This is motivated by checking that the survey $4000 \times 4000 $ pixels patches have constant noise distribution throughout the image. The same applies to the PSF modeling. We find that the PSF is uniform on the survey patches, with a variation of less than $\sim 2 \%$. We model it in simulations as a 2D circular Moffat profile with the same $\mathrm{FWHM}$ and $\beta$ throughout the patch. Each simulated image has also a different noise seed, such that simulations of the same patch of the sky with different LF parameters have different galaxy positions and properties. We simulate images in all $5$ CFHTLS wavebands.

Following the forward-modeling approach, the same SE configuration is used for the survey and simulated images, together with the same detection image prescription and the same cuts. Given the simplistic treatment of the stars in our simulations compared to the observed CFHTLS stars, we do not create new masks based on simulated stars position. We simulate only stars with magnitude $\mathrm{m > 15}$ to avoid saturation effects and we assume that simulated saturated stars and their artifacts lie in the exact same position as the real ones. This allows us to use the exact same masks we use for survey data.

UFig also provides the input catalogue with the intrinsic properties of galaxies as drawn from the LF and the size distribution. We match this catalogue to the SE output such that, for each detected galaxy, we know its redshift. This is a crucial aspect since it allows us to perform a tomographic LF measurement and to avoid photometric object-by-object redshift estimation.

\begin{figure}[t!]
\centering
\includegraphics[width=14cm]{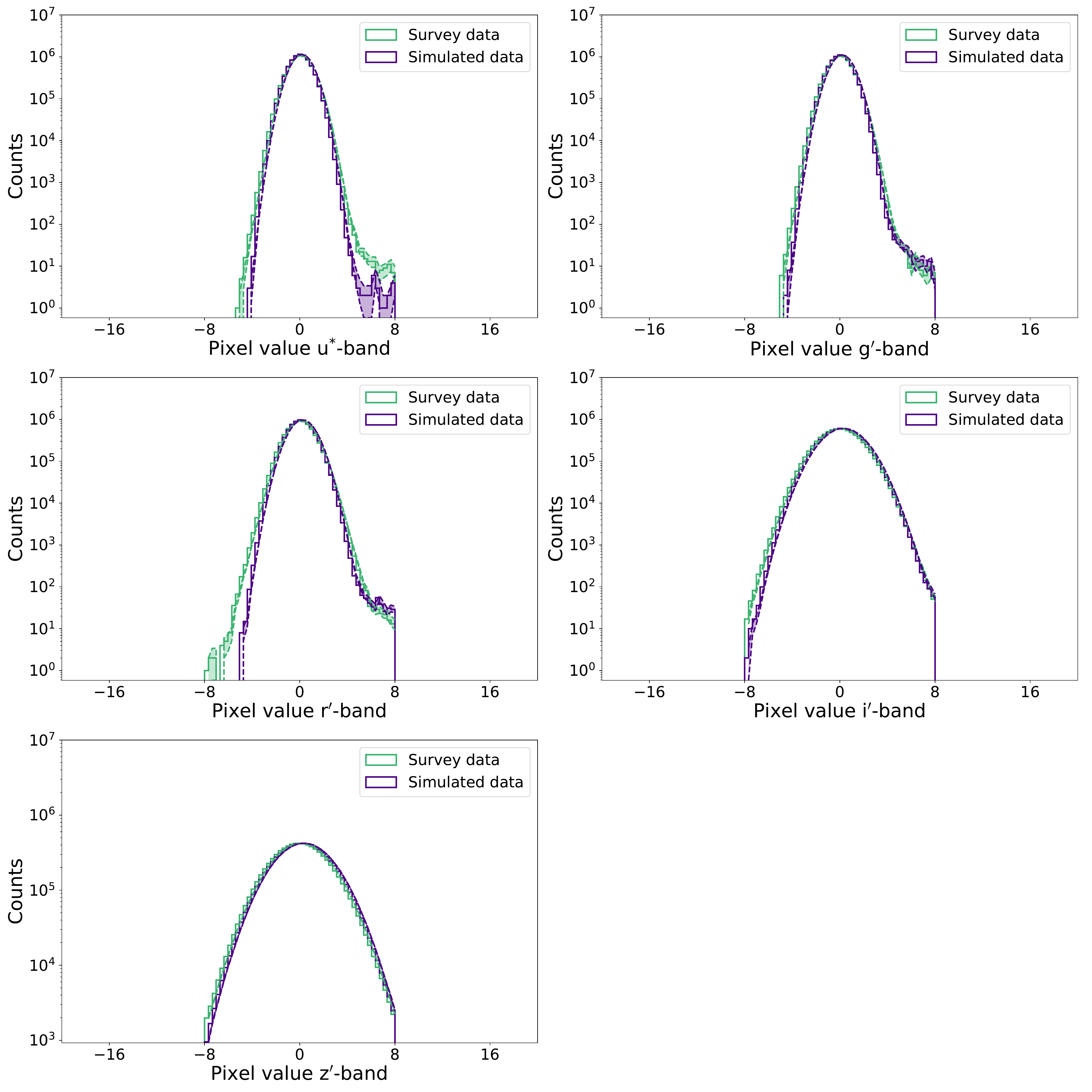}
\caption{Pixel count histograms in the 5 CFHTLS wavebands. Green and purple histograms refer to survey data and simulated data pixel counts, respectively. In order to show the background noise contribution to the pixel counts, we select only those pixels which do not belong to sources according to the SE segmentation map. Furthermore, we also exclude pixels that fall inside the star masks. The x-axis shows the pixel values in the $\mathrm{u^{*}}$, $\mathrm{g'}$, $\mathrm{r'}$, $\mathrm{i'}$, $\mathrm{z'}$ bands, while the y-axis shows the number counts. From the figures, we note that there is a slight discrepancy in the $\mathrm{u^{*}}$ and $\mathrm{r'}$ bands histograms at negative pixel values. From a visual inspection of these pixels, we find that they lie either close to the edges of the images or in the background noise, but never close to sources. Therefore, they are either artifacts of the detector or of the data reduction process. These pixels are not present in our simulations, but this does not affect the quality of our simulated data, since the photometry on survey data is not impacted by them.}
\label{fig:tortorelli_fig3}
\end{figure}

We show an example of a simulated CFHTLS image in figure \ref{fig:tortorelli_fig2}. We create the image using the parameters in table \ref{table:tortorelli_table3}. The left panel shows the survey image, while the right panel shows the simulated one. Both images represent the same patch of the sky in the $\mathrm{i'}$-band. The survey image contains $\sim 4600$ galaxies, while the simulated image $\sim 4500$ galaxies. They match within $\sim 2$ percent in terms of number of objects. They also match in terms of the background noise values as shown in figure \ref{fig:tortorelli_fig3}. The figure also shows the ability of UFig to reproduce both the population of disk galaxies and that of spheroids.

\subsection{ABC scheme}
\label{subsection:abc_scheme}

\begin{figure}[t!]
\centering
\includegraphics[width=15cm]{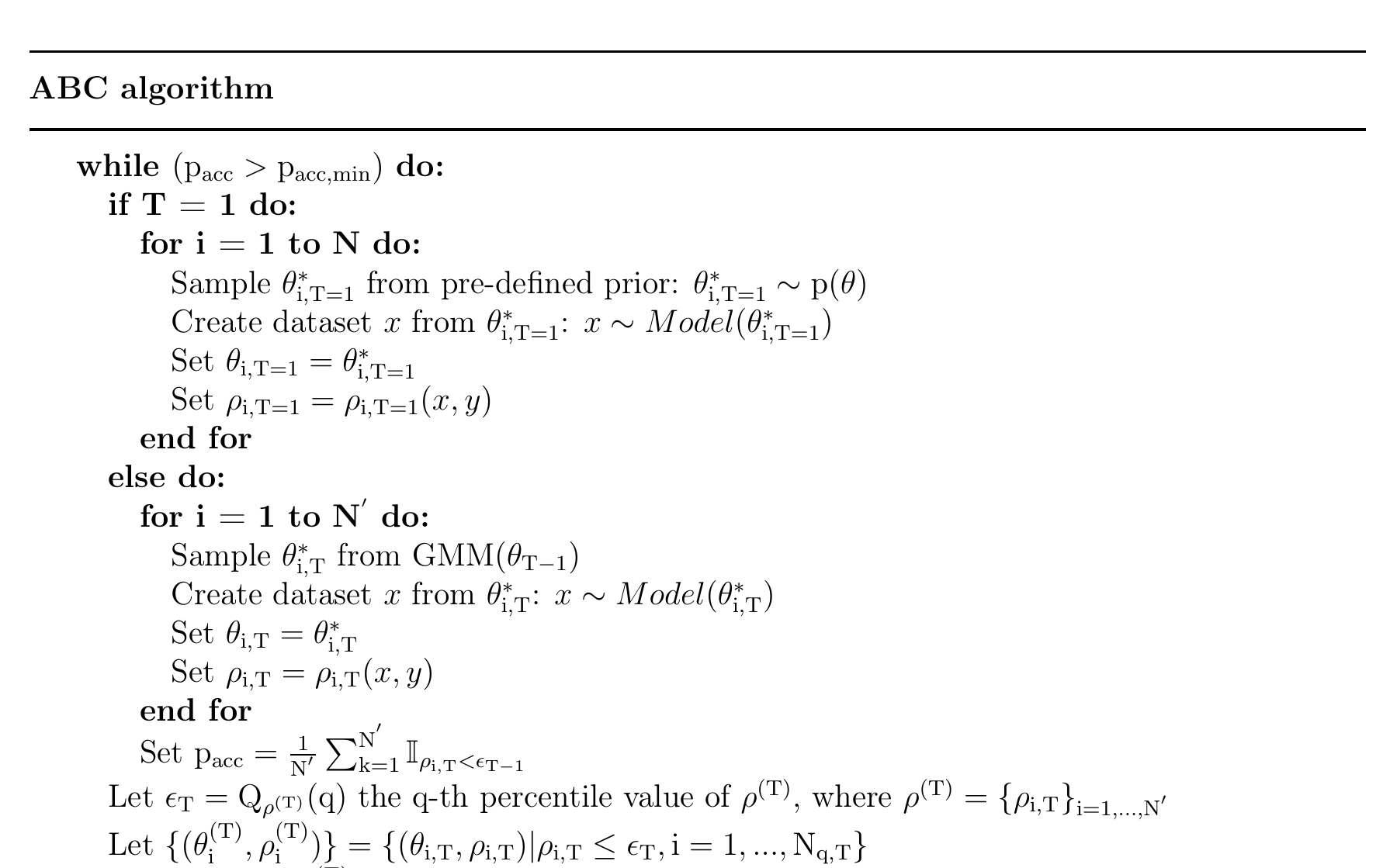}
\caption{Flowchart of the ABC inference scheme we use in our work. $\mathrm{p_{acc}}$ is the acceptance ratio of newly drawn samples. $\mathrm{N}$ and $\mathrm{N'}$ are the number of drawn samples for the $\mathrm{T = 1}$ and $\mathrm{T > 1}$ iterations of the ABC scheme. $\mathrm{GMM}$ is the Gaussian Mixture model we use to resample points between iterations. $\mathrm{p(\theta)}$ is the prior in table \ref{table:tortorelli_table2}. $Model$ refers to the galaxy population model and image rendering described in section \ref{section:galpopmodel.} $\epsilon$ is the threshold for the distance metrics.}
\label{fig:tortorelli_fig4}
\end{figure}

In standard Bayesian inference, the evaluation of the Bayesian posterior probability relies on the knowledge of the likelihood function. In many cases, however, the likelihood function is either unknown or there is no clear empirical likelihood that can be calculated for simulation-based models. Therefore, likelihood-free inference methods need to be used. We explore the use of ABC as a likelihood-free inference method. The main idea of ABC \cite{akeret15} is the approximation of the Bayesian posterior by the iterative restriction of the prior space. To do that, observed data and realistic simulations need to be compared. The Bayesian posterior, i.e. the probability of a model parametrized by a set of parameters $\theta$ given the observed data set $y$, can be approximated with
\begin{equation}
p \left ( \theta | y \right ) \simeq p \left ( \theta | \rho \left(x,y \right ) \le \epsilon \right)
\end{equation}
where $\theta$ is the set of model parameters, $x$ is the simulated data set, $\rho$ is the distance metric and $\epsilon$ is a specified threshold. This formula summarizes all the main ingredients of ABC. The algorithm samples the initial prior distribution $p(\theta)$ and a candidate set of parameters $\theta^*$ is accepted and kept as part of the approximated posterior if a defined distance metric $\rho$ between the observed data set $y$ and the simulated data set $x$ is smaller than a certain small threshold $\epsilon$.

The adopted prior distribution is described in table \ref{table:tortorelli_table2}, while the distance metrics we define are described in section \ref{subsection:distance_metrics}. We follow the prior space construction highlighted in \cite{herbel17,Kacprzak2018}, but we enlarge it by increasing the standard deviation to ensure a larger coverage of the parameter space. In short, the LF prior is built using measurements from \cite{beare15} where the uncertainties are assumed to be Gaussians and multiplied by $10$ ($5$ in \cite{herbel17}) to ensure a conservative prior. The prior on galaxy sizes is obtained from the Great-3 dataset \cite{mandelbaum14}, where the uncertainties are enlarged by a factor of $3$. The spectral coefficients are drawn from Dirichlet distributions of order five. In total, we obtain a 31-dimensional prior space.

The ABC inference scheme we adopt consists of multiple classical Rejection ABC algorithms \cite{Tavare1997}, where we increase the size of the dataset at each iteration. This is equivalent to performing a series of experiments with increasing number of data, using the posterior from the previous experiment as prior for the new one. This approach is probabilistically consistent if the dataset from the different experiments are independent. We randomly sample new patches of the full CFHTLS survey at every iteration in order the dataset to be independent.

The ABC scheme is described in the flowchart of figure \ref{fig:tortorelli_fig4}. The model is evaluated on a number of parameters sets $\theta^{*}_{\mathrm{i,T=1}}$ drawn from a pre-defined large prior distribution p($\theta$) (see table \ref{table:tortorelli_table2}), where $\mathrm{T}$ represents the iteration of the ABC algorithm. The approximate Bayesian posterior at each iteration is constituted by the samples having distance metrics less than a defined threshold $\epsilon$. During the first iteration of the ABC algorithm, we compute all the distance metrics in table \ref{table:tortorelli_table1} and we select the thresholds $\epsilon_1$ as the $\mathrm{q}=10$-th percentile value for each distance metric. We compute the approximate posterior distribution for each distance metric using the samples having $\rho_{\mathrm{i,T=1}} \le  \epsilon_1$. We keep only those $\mathrm{N_{q,1}}$ samples $\{ (\theta^{(\mathrm{1})}, \rho^{(\mathrm{1})}) \} = \{ (\theta_{\mathrm{i,T=1}}, \rho_{\mathrm{i,T=1}}) | \rho_{\mathrm{i,T=1}} \le \epsilon_{\mathrm{1}}, \mathrm{i=1,...,N_{q,1}} \}$ coming from the distance metric that provides the most stringent constraints on the LF and size parameters at $\mathrm{T = 1}$. We evaluate only this distance metric in the next $\mathrm{T > 1}$ iterations.

The following iterations use the information about the posterior sample distributions from the previous ones. We do not fix the number of total iterations $\mathrm{T}$. We keep drawing samples from the updated posterior distribution until the acceptance ratio of newly drawn samples is $\mathrm{p_{acc}} < \mathrm{p_{acc,min}} = 10\%$. At each iteration $\mathrm{T > 1}$, we generate $\mathrm{N'}$ new samples. The $\mathrm{N'}$ new parameter sets $\theta^{*}_{\mathrm{i,T}}$ are drawn from a Gaussian mixture model (GMM) to accomodate for possible non-gaussianities in the posterior distributions. At each iteration, we fit the $\mathrm{T-1}$ posterior distribution with GMMs having different number of Gaussian components. The model that best fits the data according to the Bayesian information criterion (BIC) is used to generate resampled points for the next iteration. These are then used to simulate a new data-set $x$. We select the threshold $\epsilon_{\mathrm{T}}$ as the $\mathrm{q}=10$-th percentile value for the chosen distance metric. We keep only the $\mathrm{N_{q,T}}$ samples having distance metric below $\epsilon_{\mathrm{T}}$, $\{ (\theta^{(\mathrm{T})}, \rho^{(\mathrm{T})}) \} = \{ (\theta_{\mathrm{i,T}}, \rho_{\mathrm{i,T}}) | \rho_{\mathrm{i,T}} \le \epsilon_{\mathrm{T}}, \mathrm{i=1,...,N_{q,T}} \}$

For each parameter set $\theta^{*}_{\mathrm{i,T=1}}$ drawn from the prior distribution, we simulate $10$ different CFHTLS images for a total of $\sim 0.62$ square degrees. The $10$ patches are not contiguous. To avoid clustering to impact the $\phi^{*}$ measurement and to mitigate the absence of an accurate angular correlation function in the simulations, each $\mathrm{13\ arcmin \times 13\ arcmin}$ patch is independent from the other and it is randomly drawn among all CFHTLS `Wide' patches. Furthermore, to evaluate the impact of the galaxy statistic on the approximate posterior distribution and to reduce the impact of the cosmic variance, we take new survey and simulated images and we increase their number by $10$ units at each new iteration $\mathrm{T > 1}$.

\subsection{Distance Metrics and Summary statistics}
\label{subsection:distance_metrics}

We explore the use of different distance metrics and combinations of those. All distance metrics operates on SE catalogues, either by using the full information from the distribution of properties or by using the summary statistics computed on those. Properties refer to those after the cuts described in \ref{subsection:photometry} have been applied. We define $4$ distance metrics: absolute difference, histogram distance, Maximum Mean discrepancy distance and Random Forest distance. We also combine these distance metrics to obtain additional ones. We report in table \ref{table:tortorelli_table1} the distance metrics we use. We physically motivate the use of the specific distance metric in their description.

\begin{table}
\centering
\resizebox{\textwidth}{!}{
\begin{tabular}{c c}
\hline
\hline
\textbf{Distance Metric} &  \textbf{Label}\\
\hline
Absolute difference in the number of detected galaxies & d$_1$ \\
\hline
Random Forest distance with 21 summary statistics & d$_2$ \\
\hline
Random Forest distance with 31 summary statistics & d$_3$ \\
\hline
Maximum Mean Discrepancy distance on $\mathrm{u^{*}}$, $\mathrm{g'}$, $\mathrm{r'}$, $\mathrm{i'}$, $\mathrm{z'}$ band properties & d$_{4,...,7}$ \\
\hline
Maximum Mean Discrepancy distance on $\mathrm{u^{*}}$, $\mathrm{g'}$, $\mathrm{i'}$ band properties & d$_{8,...,11}$ \\
\hline
Maximum Mean Discrepancy distance on $\mathrm{i'}$ band magnitudes and redshift distributions & d$_{12}$ \\
\hline
Magnitude histogram distance on $\mathrm{u^{*}}$, $\mathrm{g'}$, $\mathrm{r'}$, $\mathrm{i'}$, $\mathrm{z'}$ bands separately & d$_{\{13,...,17 \}}$ \\
\hline
Size histogram distance on $\mathrm{u^{*}}$, $\mathrm{g'}$, $\mathrm{r'}$, $\mathrm{i'}$, $\mathrm{z'}$bands separately & d$_{\{18,...,22 \}}$ \\
\hline
Maximum value among all previously defined rescaled distances & d$_{23} = \mathrm{max(\underline{d}_{\{1,...,22\}})}$ \\
\hline
Maximum value between the rescaled MMD distance on 5 bands and the rescaled absolute difference & $\mathrm{ d_{24,...,27} = max(\underline{d}_1,\underline{d}_{4,...,7}) }$ \\
\hline
Maximum value among the rescaled MMD distance and the rescaled magnitude histogram distance on 5 bands & $\mathrm{ d_{28,...,31} = max(\underline{d}_{4,...,7},\underline{d}_{\{13,...,17 \}} ) }$ \\
\hline
\end{tabular}
}
\caption{Table of distance metrics used in this work. $\mathrm{\underline{d_j}}$ refers to the rescaled value for the j-th distance metric.}
\label{table:tortorelli_table1}
\end{table}

\textbf{Absolute difference.} We use as a distance metric the absolute value of the difference between the number of detected galaxies in data and simulations:
\begin{equation}
\mathrm{d_{1} = |\ N_{data} - N_{sims}\ | }
\end{equation}
where $\mathrm{N_{data}}$ and $\mathrm{N_{sims}}$ refers to the number of detected galaxies by SE on real and simulated images, respectively. This distance metric is sensitive to the number of galaxies and therefore to the amplitude of the LF.

\textbf{Histogram distance.} We use the absolute values of the differences between the counts in $20$ bins of two equally binned histograms as a distance metric:
\begin{equation}
\mathrm{d_{\{ 13,...,22 \}} = \sum_i  |\ h_{data,i} - h_{sims,i}\  | }
\end{equation}
where $\mathrm{h_{data,i}}$ and $\mathrm{h_{sims,i}}$ are the counts in the i-th bin of the data and simulations histograms, respectively. The bin number choice is motivated in Appendix \ref{appendix:abc_inferece_toy_problem}. This distance metric is sensitive both to the number of galaxies and to the overall shape of the distribution. Therefore, the magnitude histogram distance can be used to constrain both the amplitude of the LF and the exponential cut-off, while the size histogram distance both the amplitude of the LF and the size distribution of galaxies. However, none of the two is sensitive to galaxy colours and therefore to the spectral coefficients distribution. We evaluate the histogram distance both for magnitudes and sizes in the $5$ CFHTLS wavebands, thereby having $10$ different distance metrics.

\textbf{Maximum Mean Discrepancy distance.} The Maximum Mean Discrepancy (hereafter, MMD \cite{gretton08}) distance measures the difference between two probability distributions and is calculated via
\begin{equation}
\mathrm{  d_{4,...,12} = \frac{1}{N(N-1)} \sum_{i,j} k(x_i,x_j) + k(y_i,y_j) - k(x_i,y_j) - k(y_i,x_j)    }
\end{equation}
where $\mathrm{N}$ is the size of the samples $\mathrm{x}$ and $\mathrm{y}$. $\mathrm{k}$ is a Gaussian kernel function of a pre-defined width $\sigma$
\begin{equation}
\mathrm{k (x_i,y_j) = \exp{\left( - \frac{\left \| x_i - y_j \right \|^2}{2\sigma} \right)}        }
\end{equation}
where $\left \| \cdot \right \|$ is the Euclidean norm and $\sigma$ is a free parameters whose choice is described in \cite{gretton08}. To apply the MMD distance, at each iteration of the ABC inference, we first transform the observed and simulated data sets to have zero mean and a standard deviation of 1. We do that by subtracting the mean and the standard deviation of the properties measured on the full CFHTLS survey to both survey data and simulations. We compute both six-dimensional ($\mathrm{d_{8,...,11}}$) and ten-dimensional MMD distances ($\mathrm{d_{4,...,7}}$) in the $3$ and $5$ CFHTLS wavebands, respectively. We compute MMD distances between PSF corrected magnitudes and sizes ($\mathrm{d_{4,8}}$), between PSF corrected magnitudes, sizes and colours ($\mathrm{d_{5,9}}$), between PSF corrected magnitudes, sizes and flux fractions ($\mathrm{d_{6,10}}$), and between PSF corrected magnitudes, sizes, colours and flux fractions ($\mathrm{d_{7,11}}$). Flux fractions are defined as the ratio between the `FLUX\_AUTO' SE property in a waveband and the sum of the latter in all available wavebands. The MMD distance captures the information from the full distribution of magnitudes, sizes and colours, thereby constraining the exponential cut-off of the LF, the size distribution and the spectral coefficients distribution. However, it is not sensitive to the number of galaxies and therefore to the amplitude of the LF.

\begin{table}
\centering
\resizebox{\textwidth}{!}{
\begin{tabular}{c c c}
\hline
\hline
\textbf{Parameter} & \textbf{Distribution} & \textbf{Prior}\\
\hline
$\alpha$ (blue) & Fixed value & -1.3 \\
\hline
$\alpha$ (red) & Fixed value & -0.5 \\
\hline
$\mathrm{M^*_{B,slope}}$ (blue) & Multivariate Normal & $\mu = -9.44 \times 10^{-1}$,  $\sigma^2 = 8.29\times10^{-1}$\\
\hline
$\mathrm{M^*_{B,slope}}$ (red) & Multivariate Normal & $\mu = -7.33 \times 10^{-1}$, $\sigma^2 = 5.30\times10^{-1}$\\
\hline
$\mathrm{M^*_{B,intcpt}} - 5 \log{h_{70}} $ (blue) & Multivariate Normal & $\mu = -2.041 \times 10^{1}$,  $\sigma^2 = 3.312\times10^{-1}$\\
\hline
$\mathrm{M^*_{B,intcpt}} - 5 \log{h_{70}} $ (red) & Multivariate Normal & $\mu = -2.035 \times 10^{-1}$, $\sigma^2 = 2.968\times10^{-1}$\\
\hline
$\mathrm{\phi^*_{exp}}$ (blue) & Multivariate Normal & $\mu = -5.66 \times 10^{-2}$,  $\sigma^2 = 9.96\times10^{-2}$\\
\hline
$\mathrm{\phi^*_{exp}}$ (red) & Multivariate Normal & $\mu = -6.97 \times 10^{-1}$, $\sigma^2 = 9.21\times10^{-1}$\\
\hline
$\mathrm{\ln{\phi^*_{amp}}}$ / 10$^{-3}$ h$^3_{70}$ Mpc$^{-3}$ mag$^{-1}$ (blue) & Multivariate Normal & $\mu = -5.28 \times 10^{0}$, $\sigma^2 = 4.1 \times 10^{-1}$ \\
\hline
$\mathrm{\ln{\phi^*_{amp}}}$ / 10$^{-3}$ h$^3_{70}$ Mpc$^{-3}$ mag$^{-1}$ (red) & Multivariate Normal & $\mu = -5.28 \times 10^{0}$, $\sigma^2 = 6.5 \times 10^{-1}$ \\
\hline
$\mathrm{r_{50,slope}^{phys}}$ & Multivariate Normal & $\mu = -2.4 \times 10^{-1}$, $\sigma^2 = 9.8\times10^{-6}$ \\
\hline
$\mathrm{r_{50,intcpt}^{phys}}$ & Uniform & [-2, 4] \\
\hline
$\mathrm{\sigma_{phys}}$ & Multivariate Normal & $\mu = 5.7 \times 10^{-1}$, $\sigma^2 = 1.9 \times 10^{-5}$ \\
\hline
$\mathrm{a}_{i,0}$ & Dirichlet $\times$ Uniform & [1., 1., 1., 1., 1.] $\times$ [5, 15] \\
\hline
$\mathrm{a}_{i,1}$ & Dirichlet $\times$ Uniform & [1., 1., 1., 1., 1.] $\times$ [5, 15] \\
\hline
\end{tabular}
}
\caption{Prior range of the parameters used to simulate CFHTLS `Wide' images for red and blue galaxies. The LF and size parameters are drawn from two Multivariate normals, except for $\mathrm{r_{50,intcpt}^{phys}}$ which has a uniform distribution. The spectral coefficients are drawn from Dirichlet distributions of order five.}
\label{table:tortorelli_table2}
\end{table}

\textbf{Random Forest distance.} So far we have defined distance metrics based on the information coming from the full distribution of properties. We also explore the use of a distance metric based on summary statistics, the Random Forest (RF) distance. Different summary statistics can be defined, e.g., the number of detected galaxies, the mean magnitudes and so on, but adding more does not necessarily mean gaining more information about the data set. Indeed, different summary statistics may correlate with each other and as a result be redundant. We define two sets of summary statistics and therefore two distance metrics $\mathrm{d_2}$ and $\mathrm{d_3}$: one consisting of $21$ and the other of $31$ summary statistics. The summary statistics we initially define are:  the number of total detected galaxies `$\mathrm{N_{gal}}$', the mean magnitude `$\mathrm{m}$', the mean size `$\mathrm{r_{50}}$', the magnitude variance `$\sigma^2_{\mathrm{m}}$', the size variance `$\sigma^2_{\mathrm{r_{50}}}$', the elements of the covariance matrix between magnitudes and sizes `$\mathrm{cov}$' and the median of the colour distributions. The difference between the $21$ and $31$ elements sets is that in the first case the properties are computed only for the $\mathrm{u^{*}}$, $\mathrm{g'}$, $\mathrm{i'}$ bands, while in the second case for the $\mathrm{u^{*}}$, $\mathrm{g'}$, $\mathrm{r'}$, $\mathrm{i'}$, $\mathrm{z'}$ bands. The magnitudes are the ones measured by the PSF corrected `$\mathrm{MAG\_ISO\_CORR}$' parameter, while sizes refer to the SE `$\mathrm{FLUX\_RADIUS}$' parameters. The choice of the specific summary statistics is meant to select properties that could constrain both the LF, the size and the spectral coefficients distributions.

We also decide to evaluate new distance metrics by combining the already defined ones. The idea is to exploit their full constraining power by obtaining distance metrics which are sensitive to the LF, to the size and spectral coefficients distributions. In order to do that, we rescale them to have the same numerical range. We follow the prescription in \cite{Kacprzak2018}. We divide each original distance metric $\mathrm{d_j}$ by a factor $\mathrm{d^{10}_j}$, corresponding to the 10-th percentile value found using the first $500$ samples we simulate. The rescaled distance metric $\mathrm{\underline{d_j}}$ is then $\mathrm{\underline{d_j} = d_j / d^{10}_j}$. We define three new distance metrics.

\textbf{Maximum distance.} We compute the maximum value among all the rescaled distance metrics:
\begin{equation}
\mathrm{ d_{23} = max(\underline{d}_{\{1,...,22\}}) } 
\end{equation}
where $\mathrm{\underline{d_j}}$ is the rescaled value for the j-th distance.

\textbf{Maximum between MMD and Absolute difference distances.} We compute the maximum value between the rescaled MMD distance on $5$ bands and the rescaled absolute difference in the number of detected galaxies:
\begin{equation}
\mathrm{ d_{24,...,27} = max(\underline{d}_1,\underline{d}_{4,...,7}) } 
\end{equation}
We do this separately for the different defined MMD distances ($\mathrm{d_{4,...,7}}$).

\textbf{Maximum between MMD and magnitude histogram distances.} We compute the maximum value between the rescaled MMD distance on 5 bands and the rescaled magnitude histogram distances in all $5$ bands:
\begin{equation}
\mathrm{ d_{28,...,31} = max(\underline{d}_{4,...,7},\underline{d}_{\{13,...,17 \}} ) }
\end{equation}
We do this separately for the different defined MMD distances ($\mathrm{d_{4,...,7}}$).

\begin{figure}[t!]
\centering
\includegraphics[width=16cm]{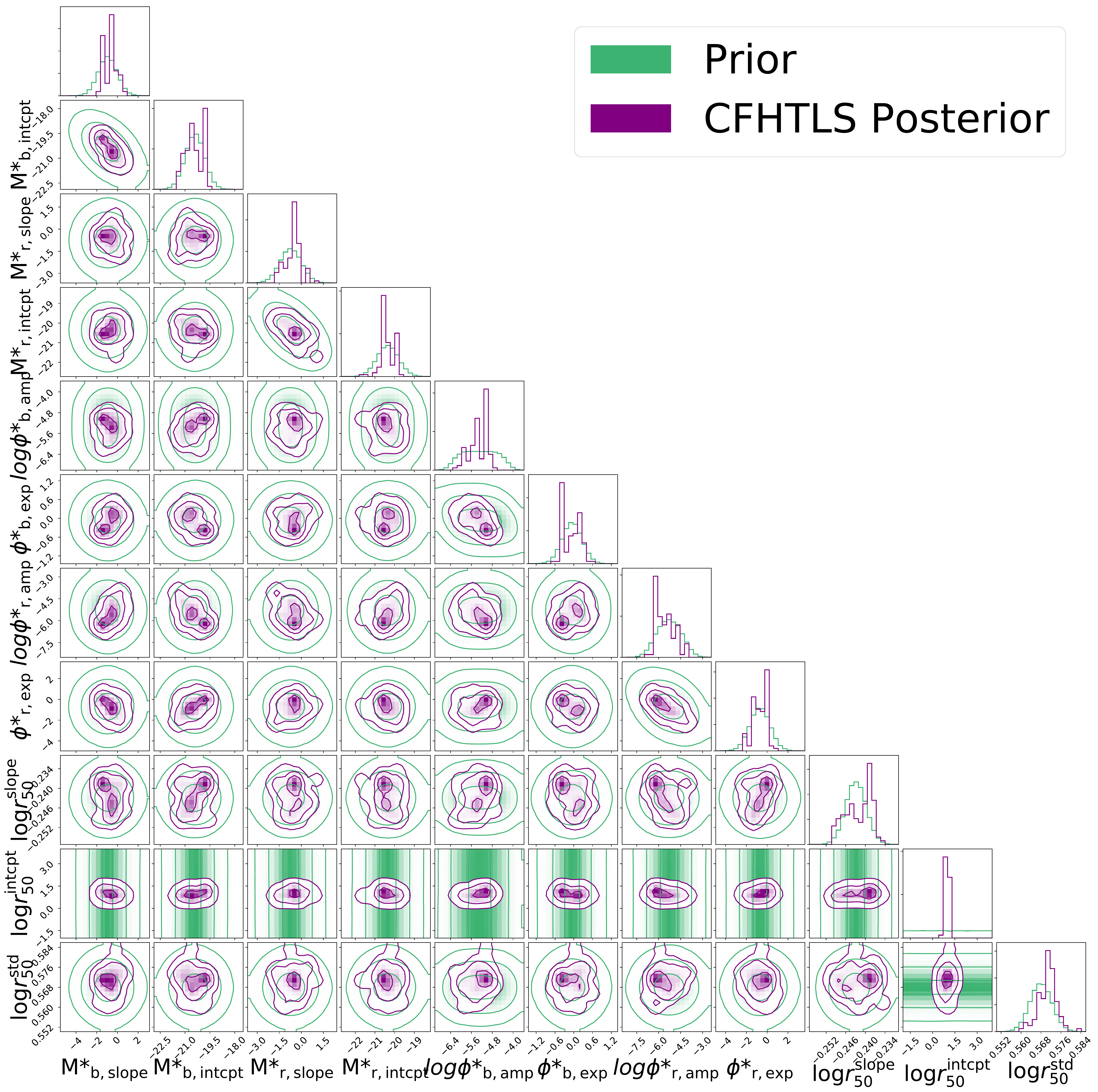}
\caption{Posterior distributions of the LF and size parameters for the ABC inference on survey data. Green contours show to the prior distributions, while purple contours the approximate Bayesian posterior. The subscripts `b' and `r' refer to parameters for blue and red galaxies, respectively. The contour levels refer to the $\sim 39\%$, $\sim 86\%$ and $\sim 98\%$ confidence levels for a 2D Gaussian distribution.}
\label{fig:tortorelli_fig5}
\end{figure}

\section{Results}
\label{section:results}

This section presents our results. We describe the ABC inference we perform on CFHTLS `Wide' survey data and we show the approximate Bayesian posterior contours with $31$ free parameters ($8$ for the LF, $20$ for the spectral coefficients and $3$ for the sizes). We also show the galaxy population properties resulting from the approximate posterior. We plot the LFs for blue and red galaxies, together with the full galaxy sample LFs and we compare them with low-redshift and high-redshift literature works. To check the validity of our approach for cosmology applications, we perform the same selection cuts in colours as in the VIPERS survey and we compare our set of redshift distributions with that from the official VIPERS catalogue.

\subsection{ABC inference on CFHTLS survey data}
\label{subsection:abc_data}

\begin{figure}[t!]
\centering
\includegraphics[width=16cm]{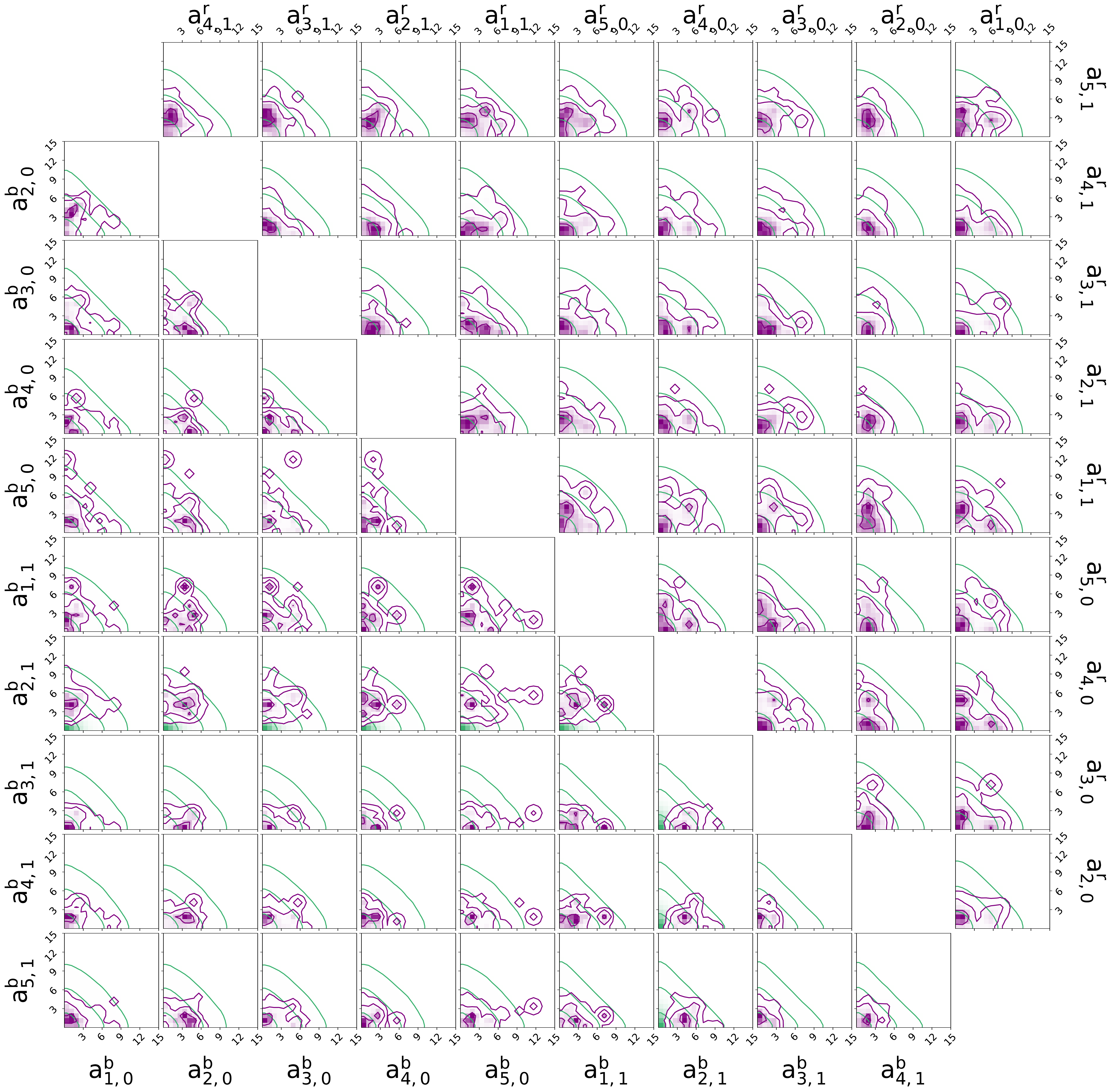}
\caption{Posterior distributions of the spectral coefficients for blue galaxies (lower corner plot) and red galaxies (upper corner plot) for the ABC inference on survey data. Green contours refer to the prior distributions, while purple contours to the approximate Bayesian posterior. The subscript `b' and `r' refer to parameters for blue and red galaxies, respectively. The subscripts `$0$' and `$1$' for the five components refer to the coefficients at redshift $\mathrm{z = 0}$ and $\mathrm{z = 1}$. The contour levels refer to the $\sim 39\%$, $\sim 86\%$ and $\sim 98\%$ confidence levels for a 2D Gaussian distribution.}
\label{fig:tortorelli_fig6}
\end{figure}

The prior space used for our inference on survey data is in table \ref{table:tortorelli_table2}. In the $\mathrm{T = 1}$ iteration, we draw $\mathrm{N = 10^5}$ samples from the prior distribution. For each set of parameters, we simulate $10$ randomly chosen CFHTLS images. Each simulated image has the same instrumental parameters of the corresponding CFHTLS survey image and different noise seed. We measure galaxy properties for each image and we stack the $10$ different catalogues. We then evaluate the $31$ distance metrics in table \ref{table:tortorelli_table1}.

Upon inspection of the posterior distribution at $\mathrm{T = 1}$ obtained with different distance metrics, the maximum between the MMD distance on PSF magnitudes and sizes in 5 bands and the absolute difference $\mathrm{d_{24}}$ is the distance metric that provides the most constrained LF and size parameters. This is not surprising since it provides both information about the magnitudes, sizes and colours distributions (MMD) and on the number of detected galaxies (absolute difference). Furthermore, in the inference on mock observations (see Appendix \ref{appendix:simonsim_run}), this distance metric is also the one that gives the most constrained result centered on the true input value. We compute the threshold $\epsilon_1$ as the $\mathrm{q}=10$-th percentile value for the distribution of the $\mathrm{d_{24}}$ distance metric. The $\mathrm{T = 1}$ approximate posterior consists of the $\mathrm{N_{q,1}}$ samples having distance $\mathrm{d_{24}}$ less than $\epsilon_1$, $\{ (\theta^{(\mathrm{1})}, \mathrm{d}_{24}^{(\mathrm{1})}) \} = \{ (\theta_{\mathrm{i,T=1}}, \mathrm{d}_{24,\mathrm{i,T=1}}) | \mathrm{d}_{24,\mathrm{i,T=1}} \le \epsilon_{\mathrm{1}}, \mathrm{i=1,...,N_{q,1}} \}$.

Each following iteration can be considered as a new experiment where we increase the size of the dataset per sample. We fit the $\mathrm{N_{q,1}}$ samples with GMMs having different number of Gaussian components. We then select the model that best fit the data using the BIC and we use this model to perform the resampling. We draw $\mathrm{N'} = 10^4$ new samples. For each set of parameters, we simulate new $\mathrm{n = 10 \times T}$ randomly chosen CFHTLS images and we evaluate the $\mathrm{d_{24}}$ distance metric only. We then compute the threshold $\mathrm{\epsilon_T}$ as the $\mathrm{q}=10$-th percentile value for the distribution of the $\mathrm{d_{24}}$ distance metric. The approximate posterior at each iteration $\mathrm{T>1}$ consists of the $\mathrm{N_{q,T}}$ samples having $\mathrm{d_{24}}$ less than $\mathrm{\epsilon_T}$, $\{ (\theta^{(\mathrm{T})}, \mathrm{d}_{24}^{(\mathrm{T})}) \} = \{ (\theta_{\mathrm{i,T}}, \mathrm{d}_{24,\mathrm{i,T}}) | \mathrm{d}_{24,\mathrm{i,T}} \le \epsilon_{\mathrm{T}}, \mathrm{i=1,...,N_{q,T}} \}$. We repeat the same scheme until the acceptance ratio drops below $\mathrm{p_{acc,min}} = 10\%$.

The computed acceptance ratio drops below $10\%$ at the $\mathrm{T = 6}$ iteration, where we simulate $60$ CFHTLS images corresponding to roughly $3$ square degrees. Figure \ref{fig:tortorelli_fig5} and \ref{fig:tortorelli_fig6} show the approximate Bayesian posterior distributions (purple contours) for blue and red galaxies of the galaxy population model parameters we vary in our ABC inference on CFHTLS `Wide' data. This approximate posterior contains $10^3$ samples. Given the small number and sparsity of the samples, we smooth the contours with a Gaussian kernel for a better visualization of the results. We also over-plot the prior distributions with green contours. The 50-th percentile values (medians) and the upper (86-th - 50th percentile values) and lower (50-th - 16th percentile values) errors of the different model parameters are reported in table \ref{table:tortorelli_table3}.

\begin{table}
\centering
\begin{tabular}{c c c}
\hline
\hline
& \textbf{Blue} & \textbf{Red} \\
\hline
$\alpha$ & -1.3 & -0.5 \\
\hline
$\mathrm{M^*_{B,slope}}$ & -0.565$^{+0.394}_{-0.789}$ & -0.537$^{+0.337}_{-0.585}$ \\
\hline
$\mathrm{M^*_{B,intcpt}} - 5 \log{h_{70}} $ & -20.475$^{+0.629}_{-0.539}$ & -20.482$^{+0.572}_{-0.189}$ \\
\hline
$\mathrm{\ln{\phi^*_{amp}}}$ / 10$^{-3}$ h$^3_{70}$ Mpc$^{-3}$ mag$^{-1}$ & -5.326$^{+0.312}_{-0.344}$ & -5.683$^{+0.920}_{-0.463}$ \\
\hline
$\mathrm{\phi^*_{exp}}$ & -0.093$^{+0.308}_{-0.303}$ & -0.661$^{+0.683}_{-0.664}$ \\
\hline
$\mathrm{r_{50,slope}^{phys}}$ & -0.241$^{+0.003}_{-0.005}$ & -0.241$^{+0.003}_{-0.005}$ \\
\hline
$\mathrm{r_{50,intcpt}^{phys}}$ & 0.986$^{+0.070}_{-0.143}$ & 0.986$^{+0.070}_{-0.143}$ \\
\hline
$\mathrm{\sigma_{phys}}$ & 0.571$^{+0.003}_{-0.004}$ & 0.571$^{+0.003}_{-0.004}$ \\
\hline
$\mathrm{a}_{1,0}$ & 1.171$^{+0.693}_{-1.322}$ & 1.316$^{+0.544}_{-4.014}$ \\ 
\hline
$\mathrm{a}_{2,0}$ & 3.055$^{+1.325}_{-1.778}$ & 1.936$^{+1.073}_{-1.570}$ \\
\hline
$\mathrm{a}_{3,0}$ & 1.394$^{+1.019}_{-3.675}$ & 1.683$^{+1.336}_{-2.866}$ \\
\hline
$\mathrm{a}_{4,0}$ & 1.669$^{+1.459}_{-1.181}$ & 1.281$^{+0.582}_{-3.332}$ \\
\hline
$\mathrm{a}_{5,0}$ & 1.855$^{+0.867}_{-2.420}$ & 1.844$^{+1.269}_{-2.507}$ \\
\hline
$\mathrm{a}_{1,1}$ & 2.385$^{+1.303}_{-3.499}$ & 2.644$^{+1.438}_{-1.701}$ \\ 
\hline
$\mathrm{a}_{2,1}$ & 4.294$^{+1.329}_{-1.412}$ & 1.8760$^{+0.964}_{-0.742}$ \\
\hline
$\mathrm{a}_{3,1}$ & 0.898$^{+0.488}_{-1.464}$ & 1.421$^{+0.973}_{-1.420}$ \\
\hline
$\mathrm{a}_{4,1}$ & 1.895$^{+1.101}_{-0.265}$ & 1.404$^{+0.695}_{-0.861}$ \\
\hline
$\mathrm{a}_{5,1}$ & 1.459$^{+0.818}_{-1.009}$ & 2.566$^{+1.224}_{-1.562}$ \\
\hline
\end{tabular}
\caption{50-th percentile values and errors (86-th - 50th percentile and 50-th - 16th percentile values) of the model parameters from the ABC inference on survey data for red and blue galaxies.}
\label{table:tortorelli_table3}
\end{table}

\subsection{Galaxy population properties from the Approximate Bayesian posterior}

\begin{figure}[t!]
\centering
\includegraphics[width=16cm]{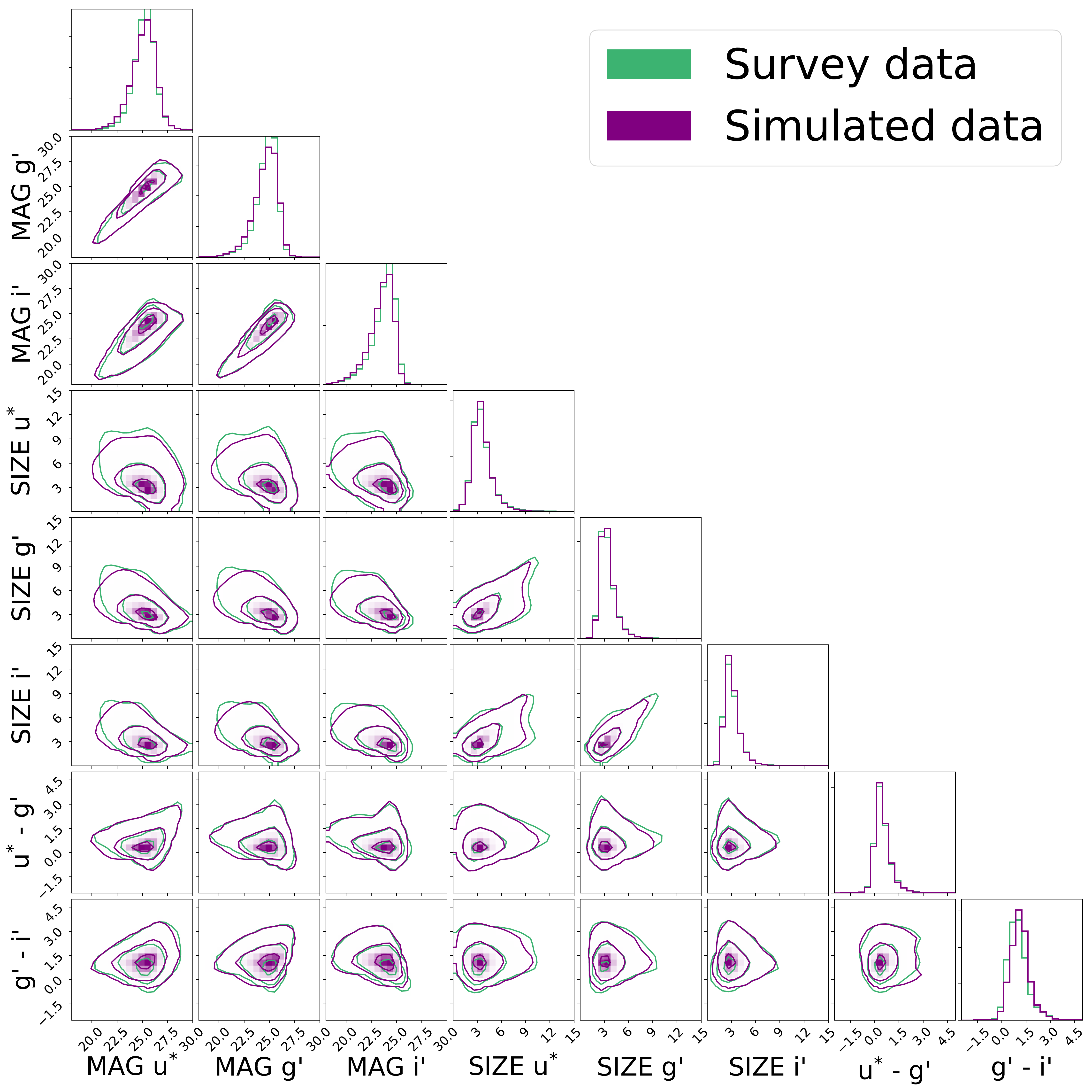}
\caption{Comparison of the galaxy population properties between survey data and simulations. Green contours refer to properties measured on observed CFHTLS `Wide' data, while purple contours refer to the simulated ones. The magnitude distributions in the $\mathrm{u^{*}}$, $\mathrm{g'}$, $\mathrm{i'}$ bands are the PSF corrected magnitudes described in \ref{subsection:photometry}, while sizes refer to SE `$\mathrm{FLUX\_RADIUS}$' parameter. The $\mathrm{u^{*} - g'}$ and $\mathrm{g' - i'}$ colours are given by the difference between the PSF corrected magnitudes.}
\label{fig:tortorelli_fig7}
\end{figure}

We compare different galaxy population properties between survey data and simulations from the approximate posterior. We show only a subset of these properties to avoid an overcrowded plot, but the same conclusions apply to all of them. The properties we compare are the magnitude distributions and the size distributions in the $\mathrm{u^{*}}$, $\mathrm{g'}$, $\mathrm{i'}$ bands and the $\mathrm{u^{*} - g'}$ and $\mathrm{g' - i'}$ colours. The comparison is shown in figure \ref{fig:tortorelli_fig7}. Purple contours show the properties of the simulated galaxies from a subset of $100$ approximate posterior distribution samples. Green contours show the properties of the observed galaxies in the same CFHTLS `Wide' fields as the simulated ones. The properties refer to $11.58 \times 10^{6}$ and $11.57 \times 10^{6}$ observed and simulated galaxies, respectively.

The properties from the observed and simulated images are in good visual agreement. The only minor discrepancies are in the $\mathrm{u^{*}}$ and $\mathrm{g'}$ band sizes, where observations show a slightly higher number of objects with sizes greater than 10 pixels, and in the peak of the simulated $\mathrm{i'}$-band magnitude distribution that is slightly offset towards brighter objects. This, in turns, is also reflected in the $\mathrm{g'-i'}$ colour. When we compare the 50-th percentile values of the galaxy properties in table \ref{table:tortorelli_table4}, we find that they are all consistent within errors. Overall, the galaxy properties are in agreement between survey data and simulations.

\begin{table}
\centering
\begin{tabular}{c c c}
\hline
\hline
&\textbf{Observations} & \textbf{Posterior simulations} \\
\hline
MAG\_$\mathrm{u^{*}}$ & 25.25 $^{+0.92}_{-1.10}$ & 25.21 $^{+1.02}_{-1.34}$ \\ 
\hline
MAG\_$\mathrm{g'}$ & 24.85$^{+0.79}_{-1.13}$ & 24.75 $^{+0.95}_{-1.34}$ \\ 
\hline
MAG\_$\mathrm{i'}$ & 23.92 $^{+0.83}_{-1.41}$ & 23.66 $^{+0.91}_{-1.44}$ \\ 
\hline
SIZE\_$\mathrm{u^{*}}$ & 3.38 $^{+1.39}_{-0.92}$ & 3.40 $^{+1.23}_{-0.89}$ \\ 
\hline
SIZE\_$\mathrm{g'}$ & 3.19 $^{+1.10}_{-0.68}$ & 3.23 $^{+1.01}_{-0.67}$ \\ 
\hline
SIZE\_$\mathrm{i'}$ & 2.89 $^{+1.09}_{-0.71}$ & 2.94 $^{+1.02}_{-0.64}$ \\ 
\hline
$\mathrm{u^{*}}$ - $\mathrm{g'}$ & 0.41 $^{+0.52}_{-0.34}$ & 0.42 $^{+0.49}_{-0.32}$ \\ 
\hline
$\mathrm{g'}$ - $\mathrm{i'}$ & 0.95 $^{+0.59}_{-0.51}$ & 1.12 $^{+0.53}_{-0.53}$ \\ 
\hline
\end{tabular}
\caption{50-th percentile values and upper (86-th - 50th percentile values) and lower (50-th - 16th percentile values) errors of magnitude, size and color distributions for observations and posterior simulations. The magnitude distributions in the $\mathrm{u^{*}}$, $\mathrm{g'}$, $\mathrm{i'}$ bands are the PSF corrected magnitudes described in \ref{subsection:photometry}, while sizes refer to SE `$\mathrm{FLUX\_RADIUS}$' parameter. The $\mathrm{u ^{*}- g'}$ and $\mathrm{g' - i'}$ colours are given by the difference between the PSF corrected magnitudes. The properties refer to $11.58 \times 10^{6}$ and $11.57 \times 10^{6}$ observed and simulated galaxies, respectively.}
\label{table:tortorelli_table4}
\end{table}

\subsection{The Luminosity Function measurement}

\begin{figure}[t!]
\centering
\includegraphics[width=7.65cm]{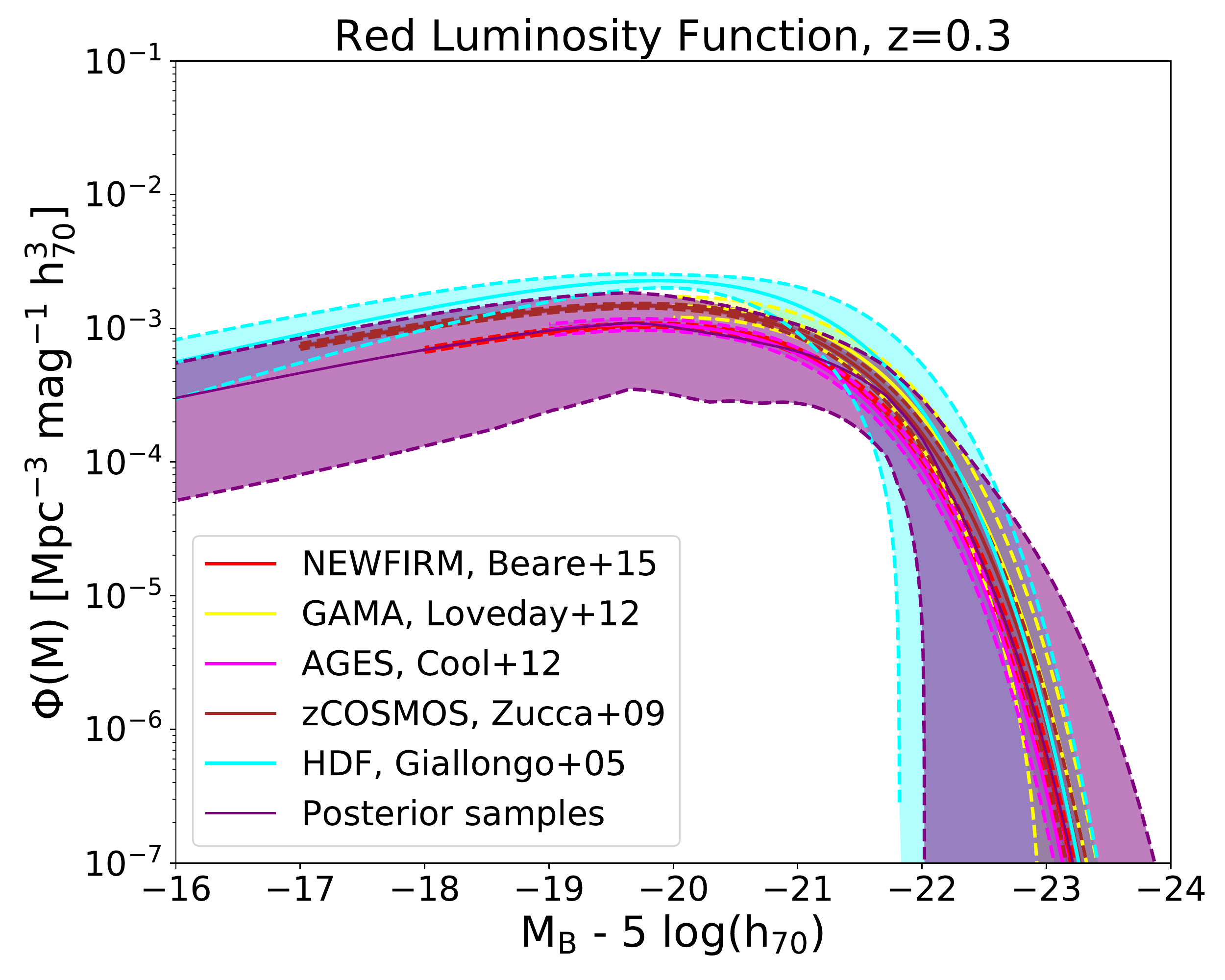}
\includegraphics[width=7.65cm]{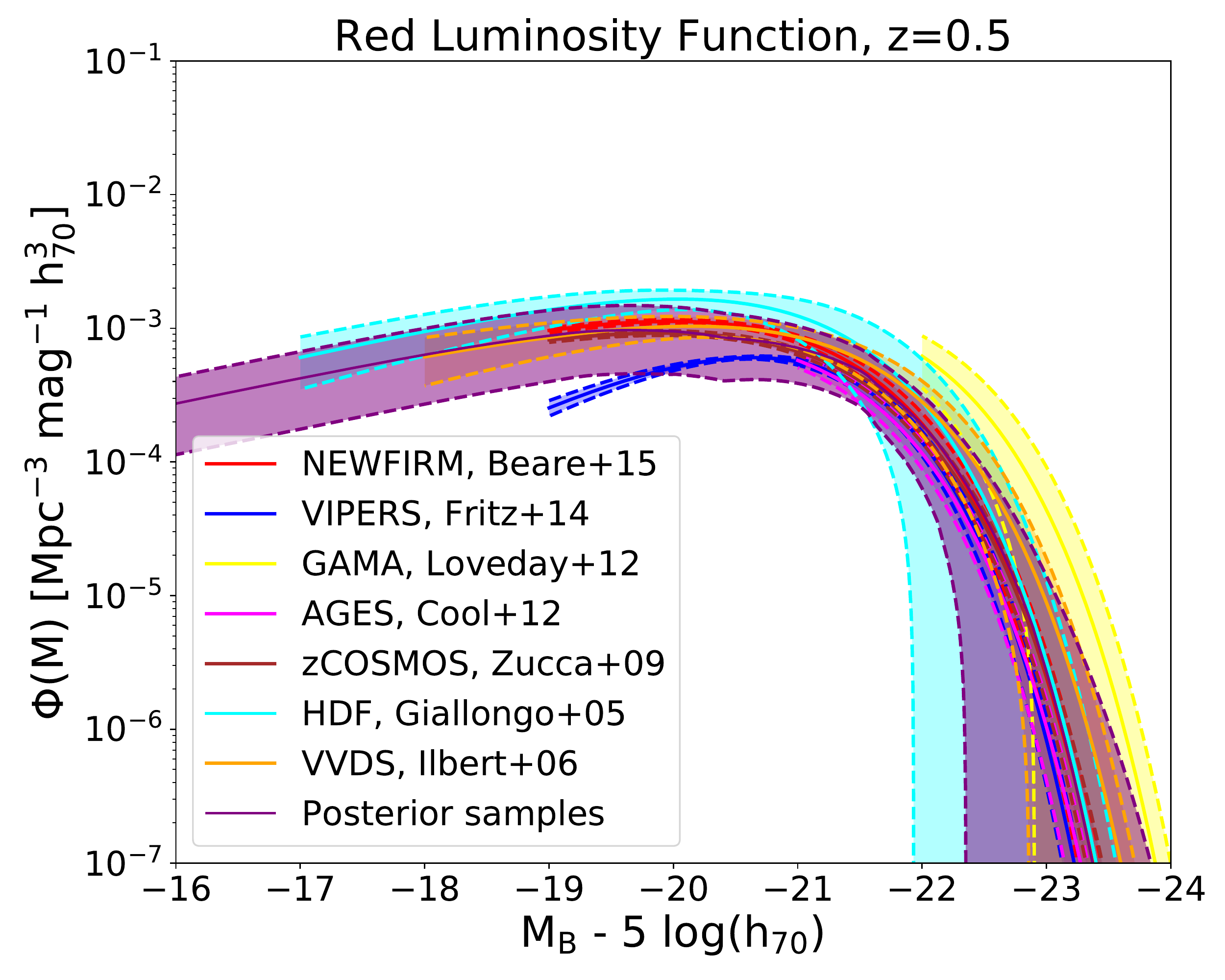}
\includegraphics[width=7.65cm]{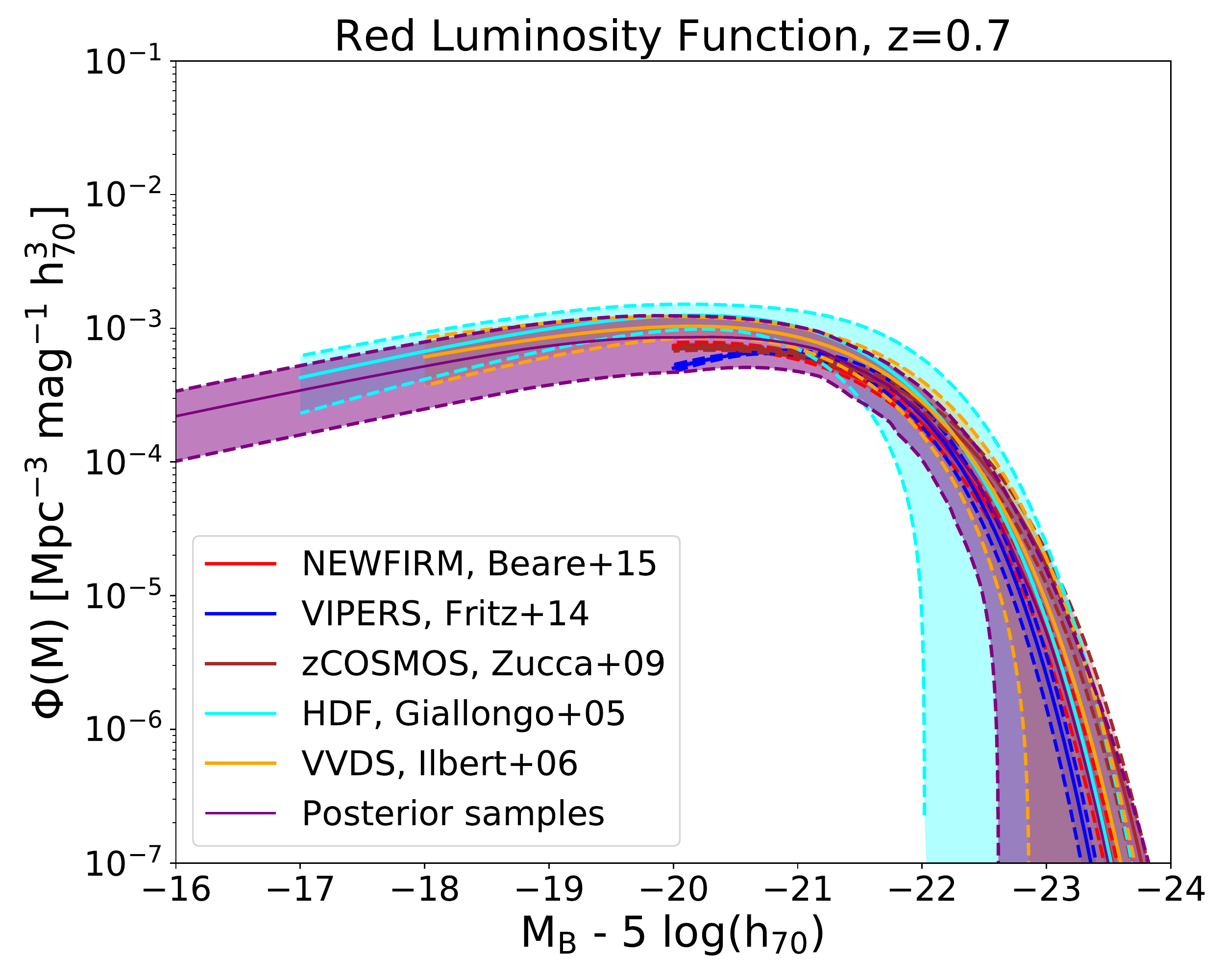}
\includegraphics[width=7.65cm]{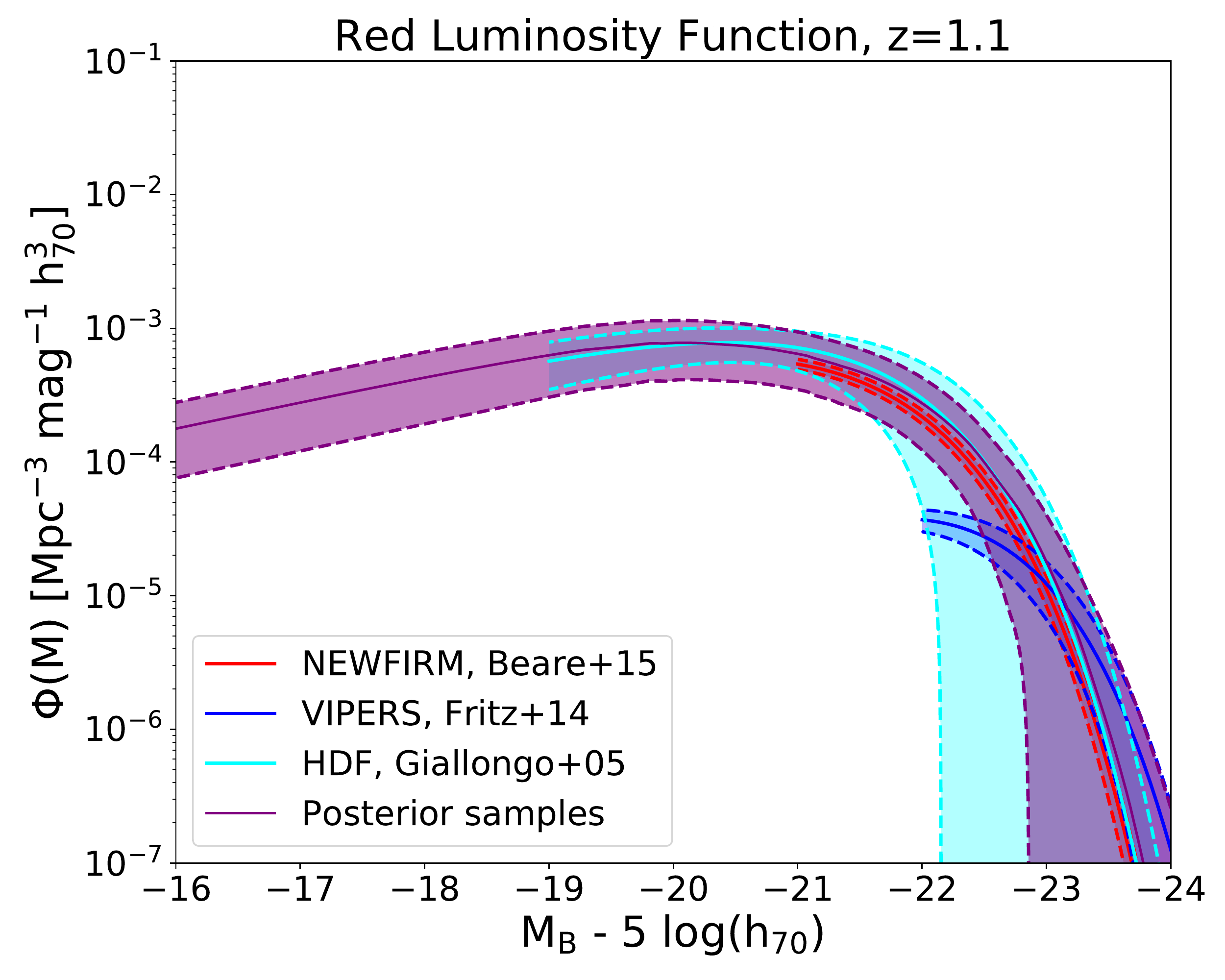}
\caption{Comparison between the LFs built from the posterior distributions (purple bands) and the works of \cite{giallongo05,ilbert06,zucca09,loveday12,cool12,fritz14,beare15} (cyan, orange, brown, yellow, fuchsia, blue and red bands, respectively) for red galaxies. The purple bands refer to the median and the one standard deviation error of our set of LFs as a function of absolute magnitude. $1 \sigma$ errors on the literature LFs are represented as coloured bands. Comparisons are shown for redshifts $\mathrm{z = 0.3}$, $0.5$, $0.7$ and $1.1$ in the upper left, upper right, lower left, lower right panels, respectively. The absolute magnitude in the B-band is in units of $\mathrm{M_B - 5 \log{h_{70}}}$, while the number density of galaxies is in units of $\mathrm{Mpc^{-3}\ mag^{-1}\ h^3_{70}}$.}
\label{fig:tortorelli_fig8}
\end{figure}

\begin{figure}[t!]
\centering
\includegraphics[width=7.65cm]{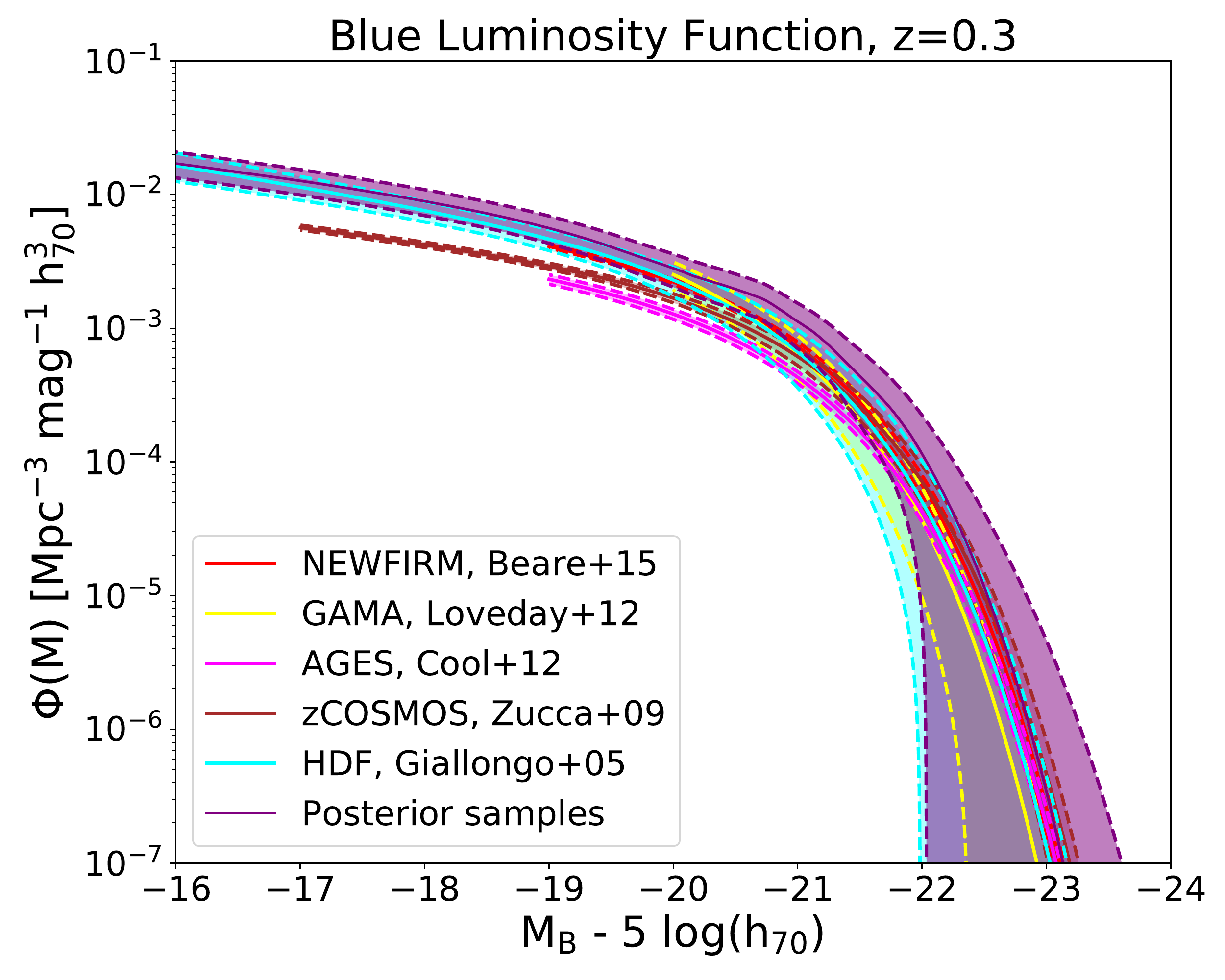}
\includegraphics[width=7.65cm]{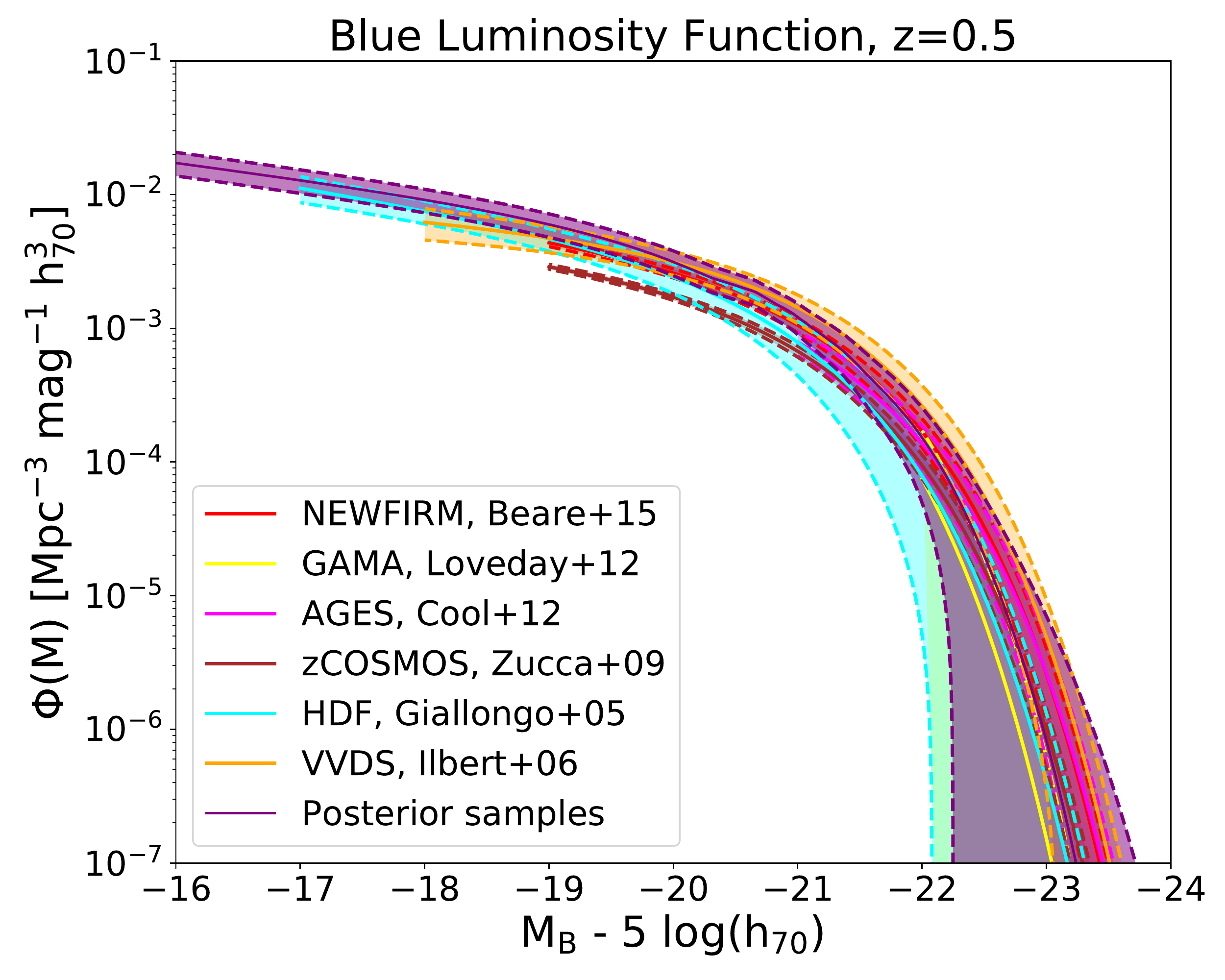}
\includegraphics[width=7.65cm]{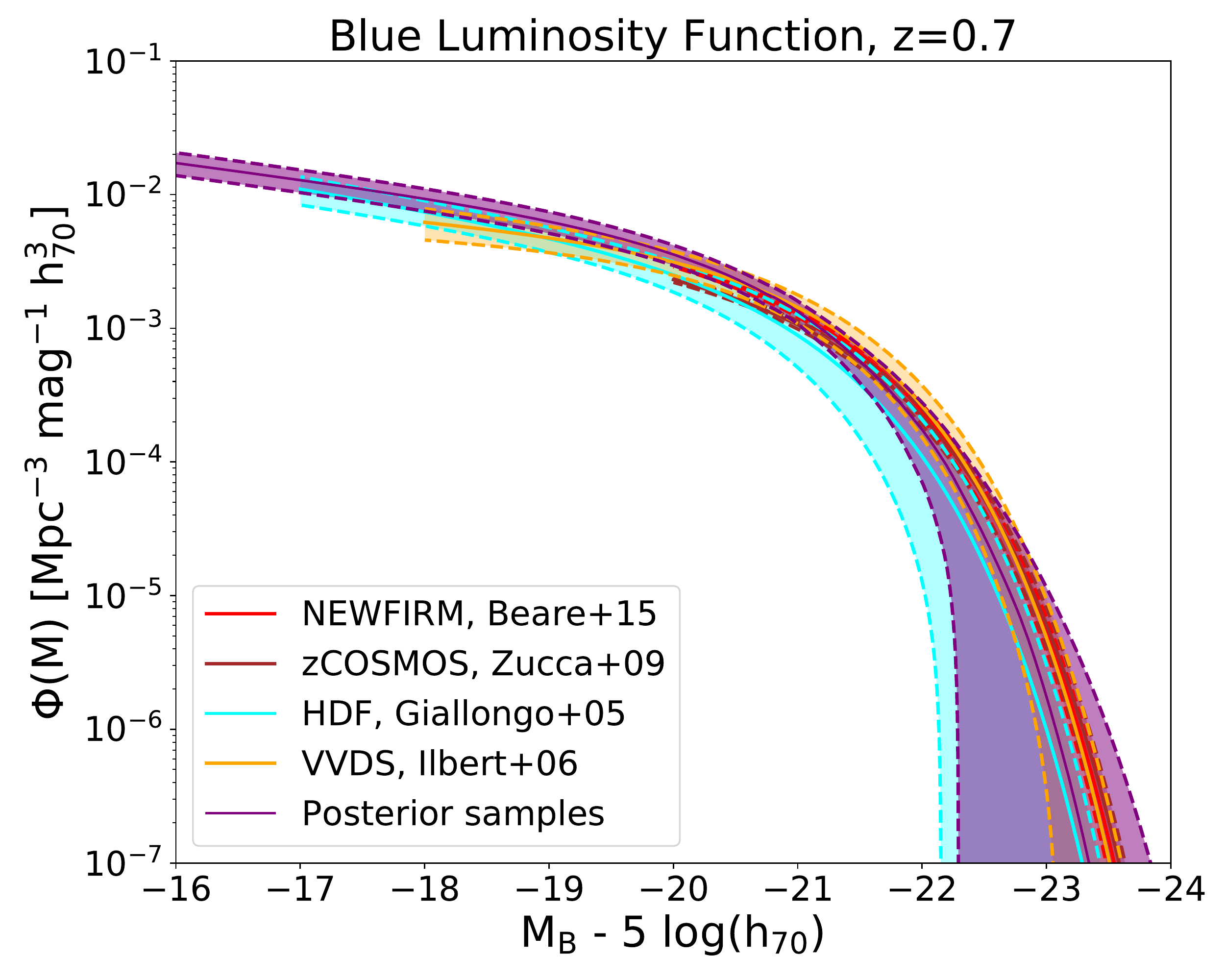}
\includegraphics[width=7.65cm]{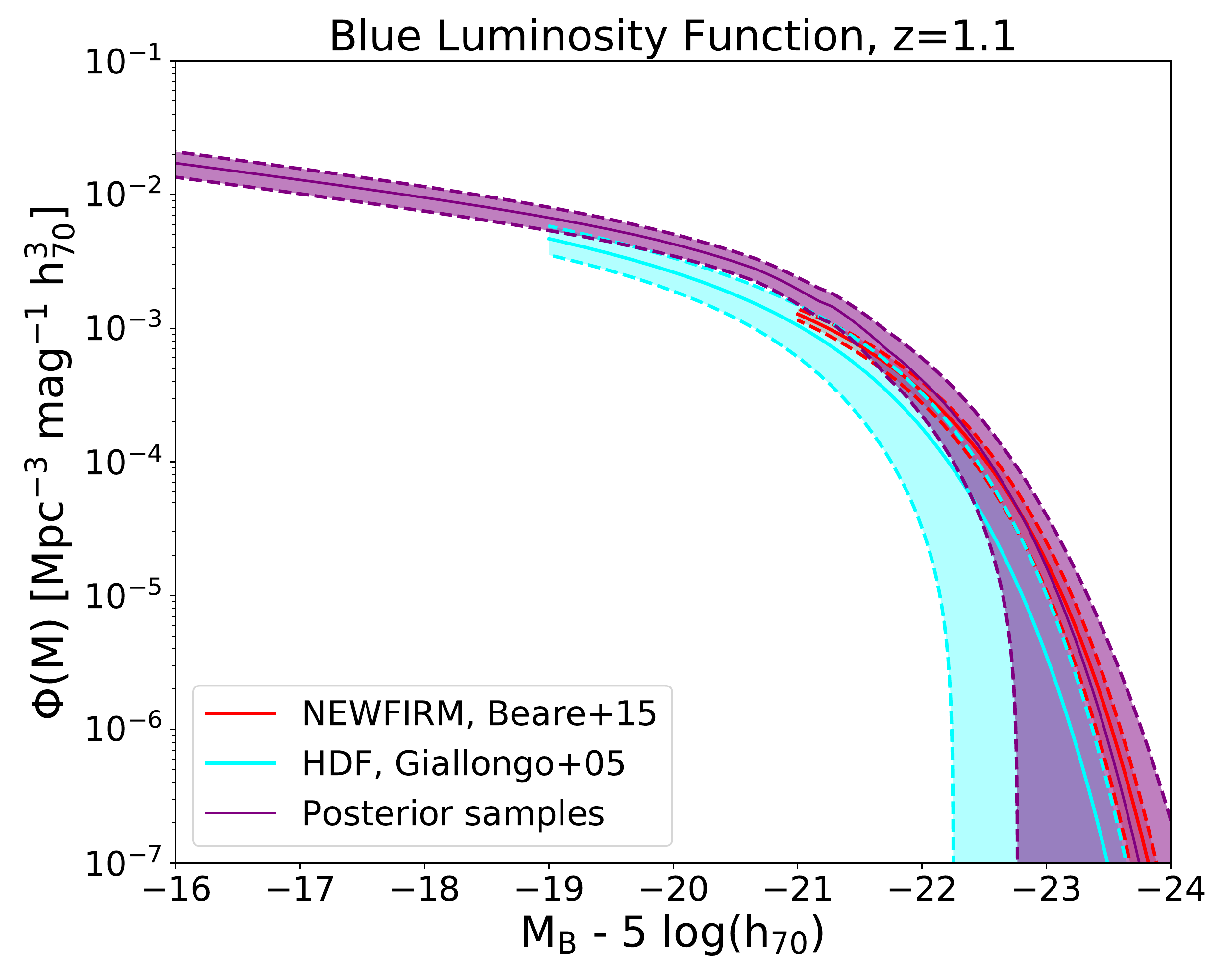}
\caption{Comparison between the LFs built from the posterior distributions (purple bands) and the works of \cite{giallongo05,ilbert06,zucca09,loveday12,cool12,beare15} (cyan, orange, brown, yellow, fuchsia and red bands, respectively) for blue galaxies. The purple bands refer to the median and the one standard deviation error of our set of LFs as a function of absolute magnitude. $1\sigma$ errors on the literature LFs are represented as coloured bands. Comparisons are shown for redshifts $\mathrm{z = 0.3}$, $0.5$, $0.7$ and $1.1$ in the upper left, upper right, lower left, lower right panels, respectively. The absolute magnitude in the B-band is in units of $\mathrm{M_B - 5 \log{h_{70}}}$, while the number density of galaxies is in units of $\mathrm{Mpc^{-3}\ mag^{-1}\ h^3_{70}}$.}
\label{fig:tortorelli_fig9}
\end{figure}

Using the posterior distributions resulting from the ABC inference, we build the sets of LFs for blue and red galaxies as a function of redshift. We also build the LFs for the full sample of galaxies as a function of redshift (hereafter, global LF). We compare our sets of LFs with those estimated in other literature studies, both at low and high-redshifts. In particular, we compare our LFs with those of \cite{giallongo05,ilbert06,zucca09,loveday12,cool12,fritz14,beare15} at redshifts $\mathrm{z = 0.3}$, $0.5$, $0.7$, $1.1$, spanning different selection criteria, areas and depths. We choose these 4 redshifts on the basis of the number of literature studies available to compare our results with. The LFs in \cite{fritz14} are measured only for red galaxies and for the total sample of objects. To be able to compare the different LFs, we rescale them to have the same units, i.e. the absolute magnitude in units of $\mathrm{M_B - 5 \log{h_{70}}}$ and the number density of galaxies in units of $\mathrm{Mpc^{-3}\ mag^{-1}\ h^3_{70}}$. We compute the median and the one standard deviation error of our set of LFs as a function of absolute magnitude and we plot them as purple bands. We also plot the $1 \sigma$ errors on the literature LFs as coloured bands. The literature LFs are restricted to the magnitude range given by their measured points. The LFs built from the prior distribution in table \ref{table:tortorelli_table2} are not shown since they span a range in number densities and absolute magnitudes that is larger than the plotted range.

Figure \ref{fig:tortorelli_fig8} and \ref{fig:tortorelli_fig9} show the comparison between our LFs (purple bands) and the literature studies for red and blue galaxies, respectively. Our uncertainties on the LF measurements are comparable in size or slightly larger than earlier measurements, depending on the sample and the redshift probed. The errors on our LF measurements are smaller at the faint-end and larger at the bright-end, given the larger number of galaxies at faint magnitudes. Furthermore, they are generally smaller at $\mathrm{z = 0.5}$ and $\mathrm{z = 0.7}$, where the peak of the CFHTLS redshift distribution resides for galaxies with $\mathrm{m_i \le 22.5}$ \cite{coupon09}.

Our red galaxies LFs are consistent within $1 \sigma$ with all studies at $\mathrm{z = 0.3}$, $\mathrm{z = 0.7}$, $\mathrm{z = 1.1}$, and with all studies at $\mathrm{z = 0.5}$, except for the VIPERS field study \cite{fritz14} at $\mathrm{M_B - 5 \log{h_{70}}} > -20$. At the bright-end ($\mathrm{M_B - 5 \log{h_{70}}} < -22$), our uncertainties are of the same order as those of the GAMA field study \cite{loveday12} at $\mathrm{z = 0.5}$, of the VVDS field study \cite{ilbert06} at $\mathrm{z = 0.7}$ and of the HDF field study \cite{giallongo05} at $\mathrm{z = 0.3}$ and at $\mathrm{z = 0.5}$. Furthermore, the uncertainties at the bright-end are smaller than those of the HDF field at $\mathrm{z = 0.7}$ and $\mathrm{z = 1.1}$, while they are slightly larger than the other studies. At the faint-end ($\mathrm{M_B - 5 \log{h_{70}}} > -21$), instead, our uncertainties are of the same order as those of the HDF field at $\mathrm{z = 0.7}$ and $\mathrm{z = 1.1}$. At the faint-end ($\mathrm{M_B - 5 \log{h_{70}}} > -21$) of $\mathrm{z = 0.3}$, our LFs tend to favour a lower number density of objects with respect to other literature studies. Despite having slightly larger errors, our results encapsulate the dispersions among the different results.

Our blue galaxies LFs are consistent within $1\sigma$ with all studies at $\mathrm{z = 0.7}$ and $\mathrm{z = 1.1}$, with all studies at $\mathrm{z = 0.3}$, except for the zCOSMOS field study \cite{zucca09} and the AGES field study \cite{cool12} at $\mathrm{M_B - 5 \log{h_{70}}} > -21$, and with all studies at $\mathrm{z = 0.5}$, except for the zCOSMOS field study \cite{zucca09} at $\mathrm{M_B - 5 \log{h_{70}}} > -21$. The uncertainties on our measurements are comparable to those of the HDF field study \cite{giallongo05} at $\mathrm{z = 0.3}$, $\mathrm{z = 0.5}$ and $\mathrm{z = 0.7}$, both at the faint and bright-end, while they are smaller than the HDF field study at $\mathrm{z = 1.1}$. Furthermore, the uncertainties are of the same order as those of the GAMA field study \cite{loveday12} at the bright-end of $\mathrm{z = 0.5}$ and of the VVDS field study \cite{ilbert06} at the faint-end of $\mathrm{z = 0.7}$. The integrated number density of blue galaxies is larger than that of red galaxies at all redshifts. The resulting larger galaxy statistics causes the blue galaxies LFs to be more constrained than the red ones.

\begin{figure}[t!]
\centering
\includegraphics[width=7.65cm]{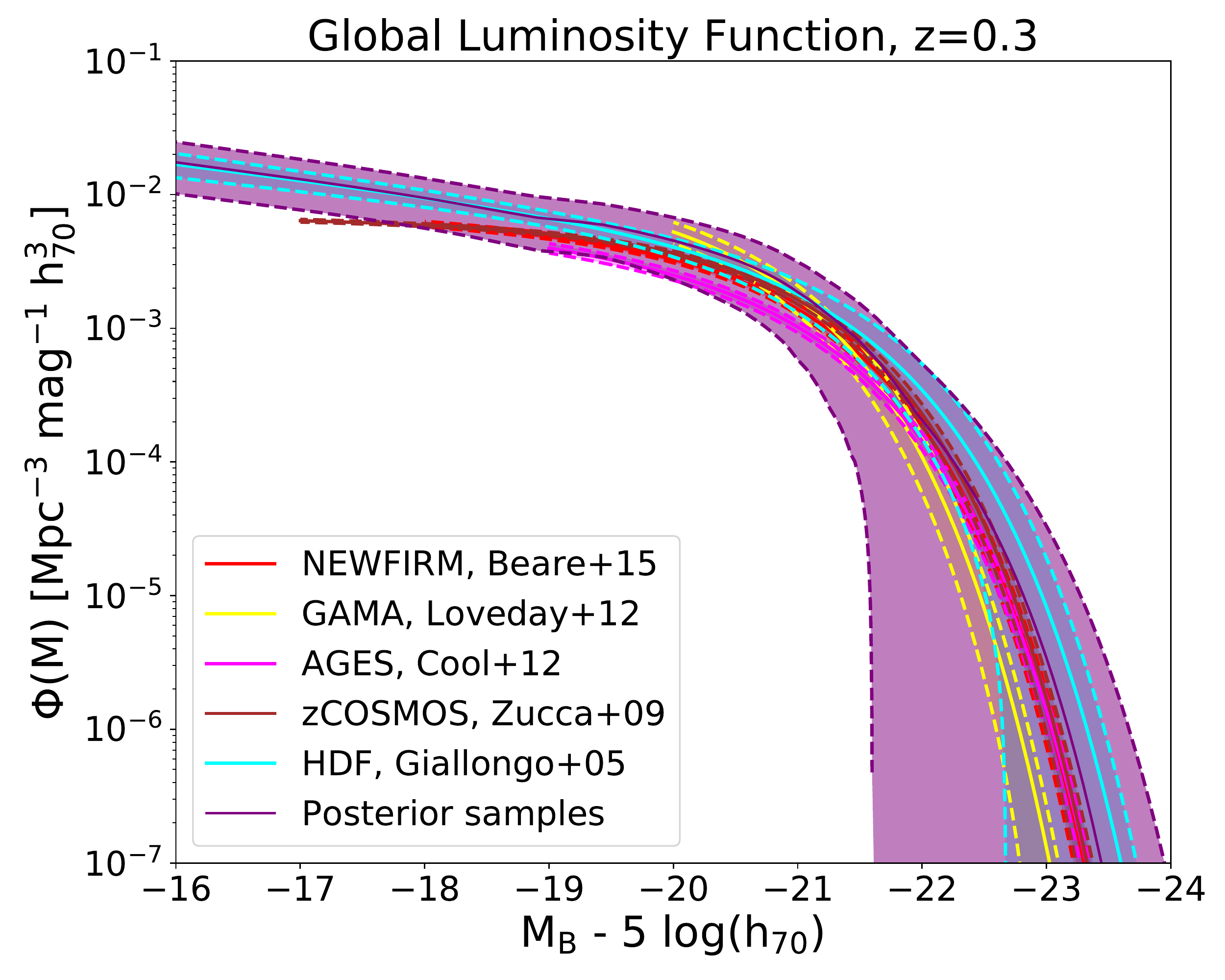}
\includegraphics[width=7.65cm]{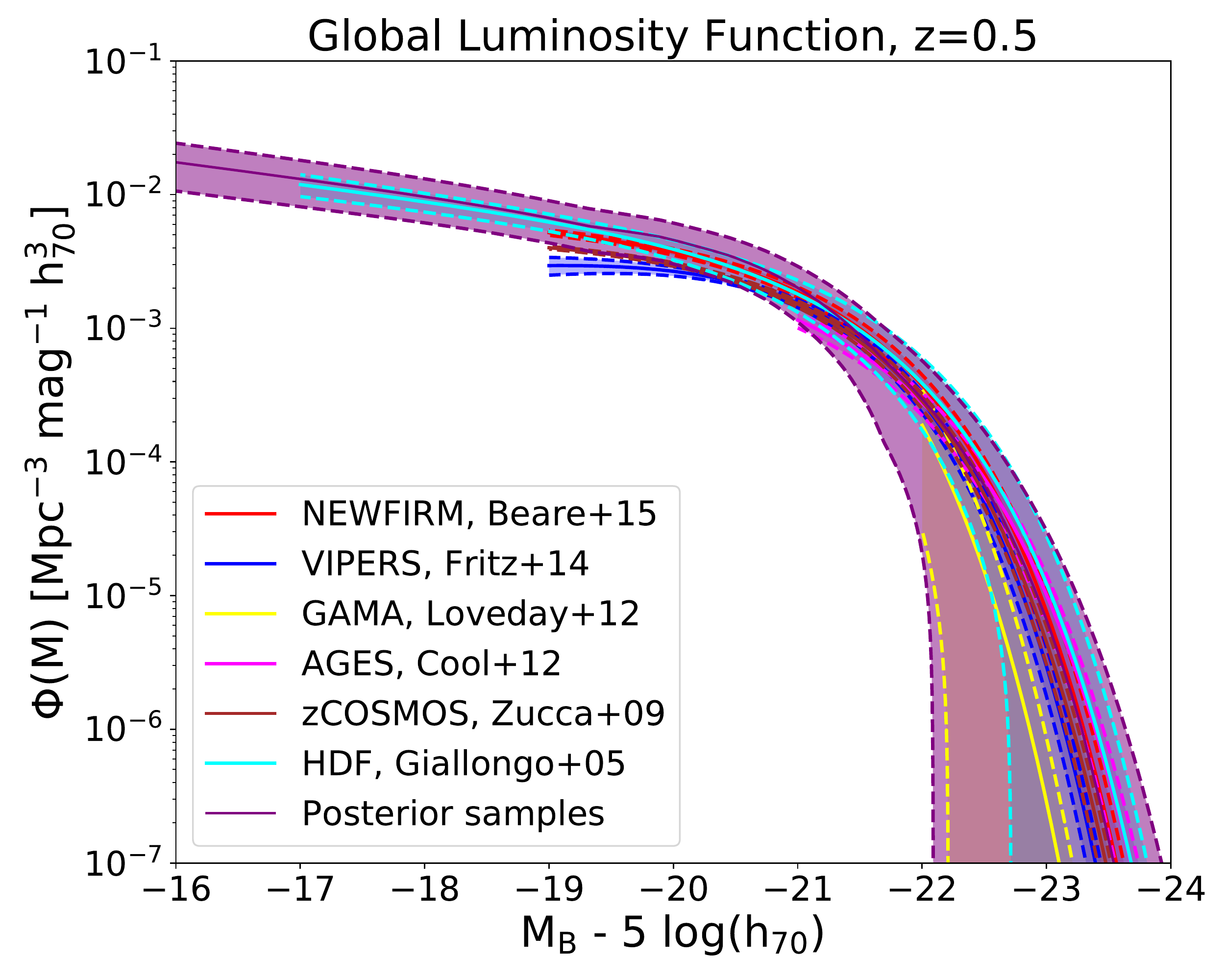}
\includegraphics[width=7.65cm]{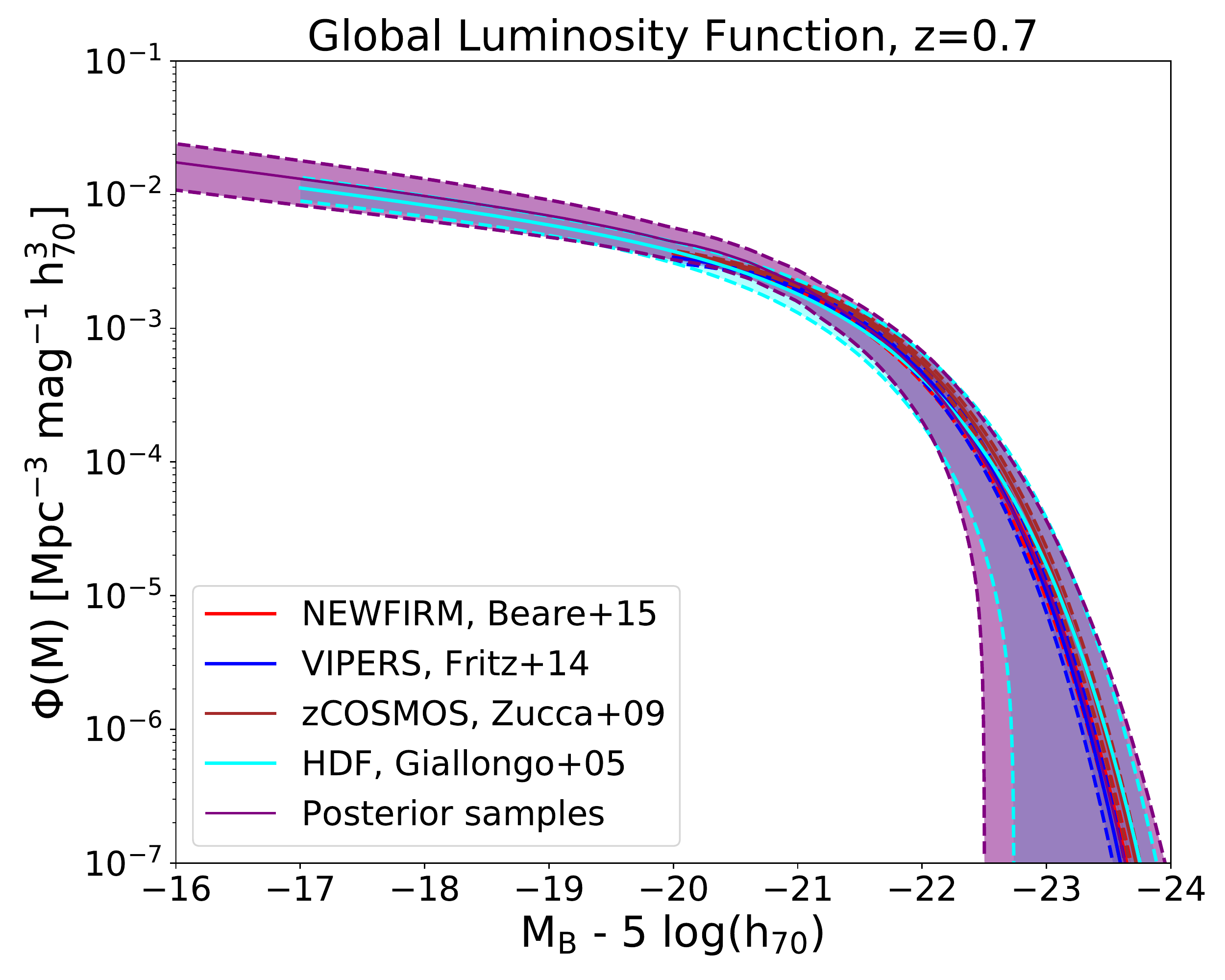}
\includegraphics[width=7.65cm]{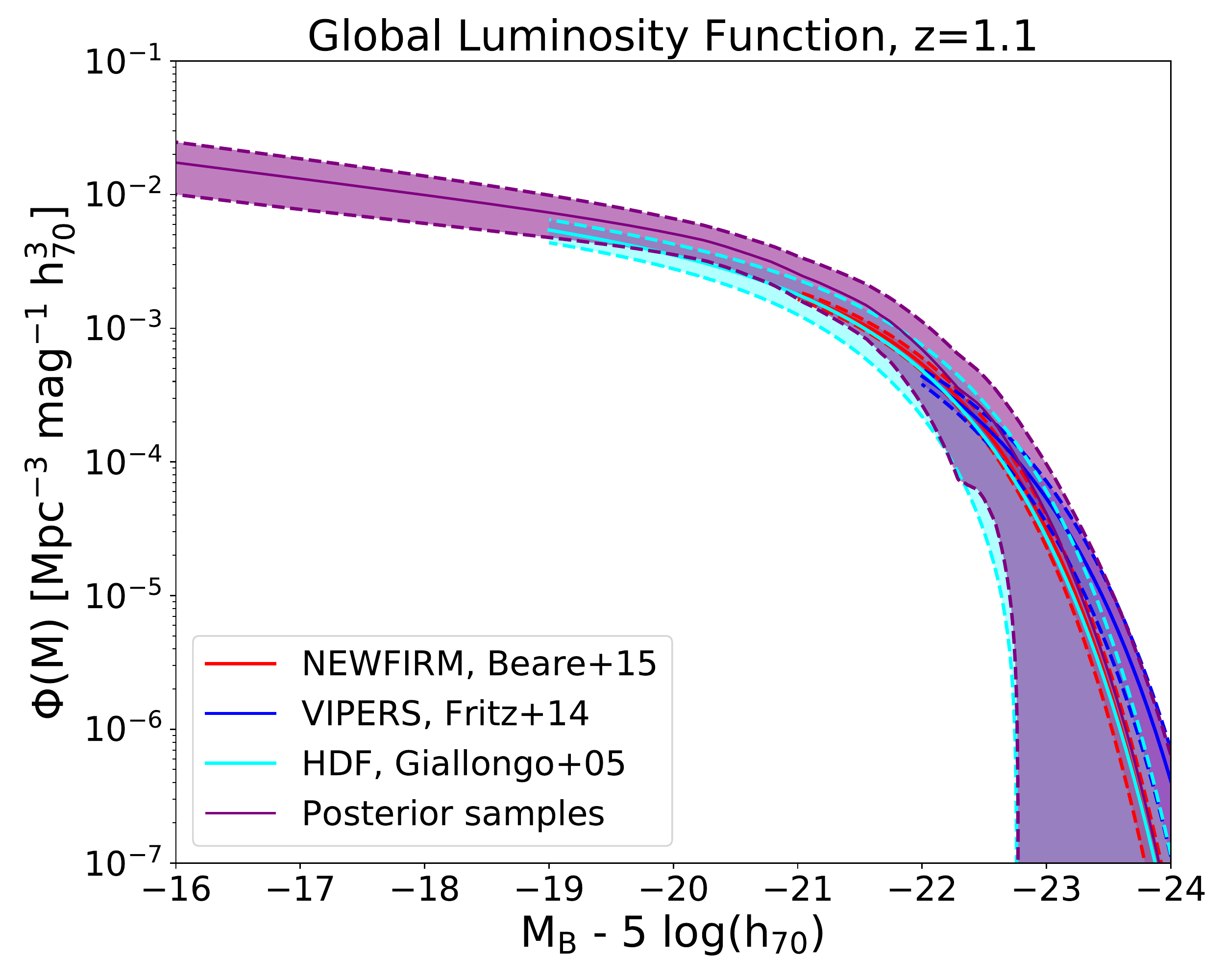}
\caption{Comparison between the LFs built from the posterior distributions (purple bands) and the works of \cite{giallongo05,ilbert06,zucca09,loveday12,cool12,fritz14,beare15} (cyan, orange, brown, yellow, fuchsia, blue and red bands, respectively). The purple bands refer to the median and the one standard deviation error of our set of LFs as a function of absolute magnitude. $1\sigma$ errors on the literature LFs are represented as coloured bands. We build our global LFs as sum of blue and red LFs from the same posterior sample.  Comparisons are shown for redshifts $\mathrm{z = 0.3}$, $0.5$, $0.7$ and $1.1$ in the upper left, upper right, lower left, lower right panels, respectively. The absolute magnitude in the B-band is in units of $\mathrm{M_B - 5 \log{h_{70}}}$, while the number density of galaxies is in units of $\mathrm{Mpc^{-3}\ mag^{-1}\ h^3_{70}}$.}
\label{fig:tortorelli_fig10}
\end{figure}

We also show the LFs for the full sample of galaxies in figure \ref{fig:tortorelli_fig10}. Similarly to red and blue galaxies, the global LFs are consistent within $1 \sigma$ with all studies at all redshifts, except for the $\mathrm{z = 0.5}$ VIPERS field study \cite{fritz14} at $\mathrm{M_B - 5 \log{h_{70}}} > -20$. The uncertainties on the LFs measurements are slightly larger than those of all the other studies. However, they are comparable with those of the HDF field study at the bright-end ($\mathrm{M_B - 5 \log{h_{70}}} < -21$) of $\mathrm{z = 0.7}$ and $\mathrm{z = 1.1}$. For $\mathrm{M_B - 5 \log{h_{70}}} > -21$, our results show that at all probed redshifts the number density at fixed absolute magnitude of blue galaxies is greater than that of red galaxies. 

\subsection{The redshift evolution of $\mathrm{M^*}$ and $\phi^*$}

\begin{figure}[t!]
\centering
\includegraphics[width=7.65cm]{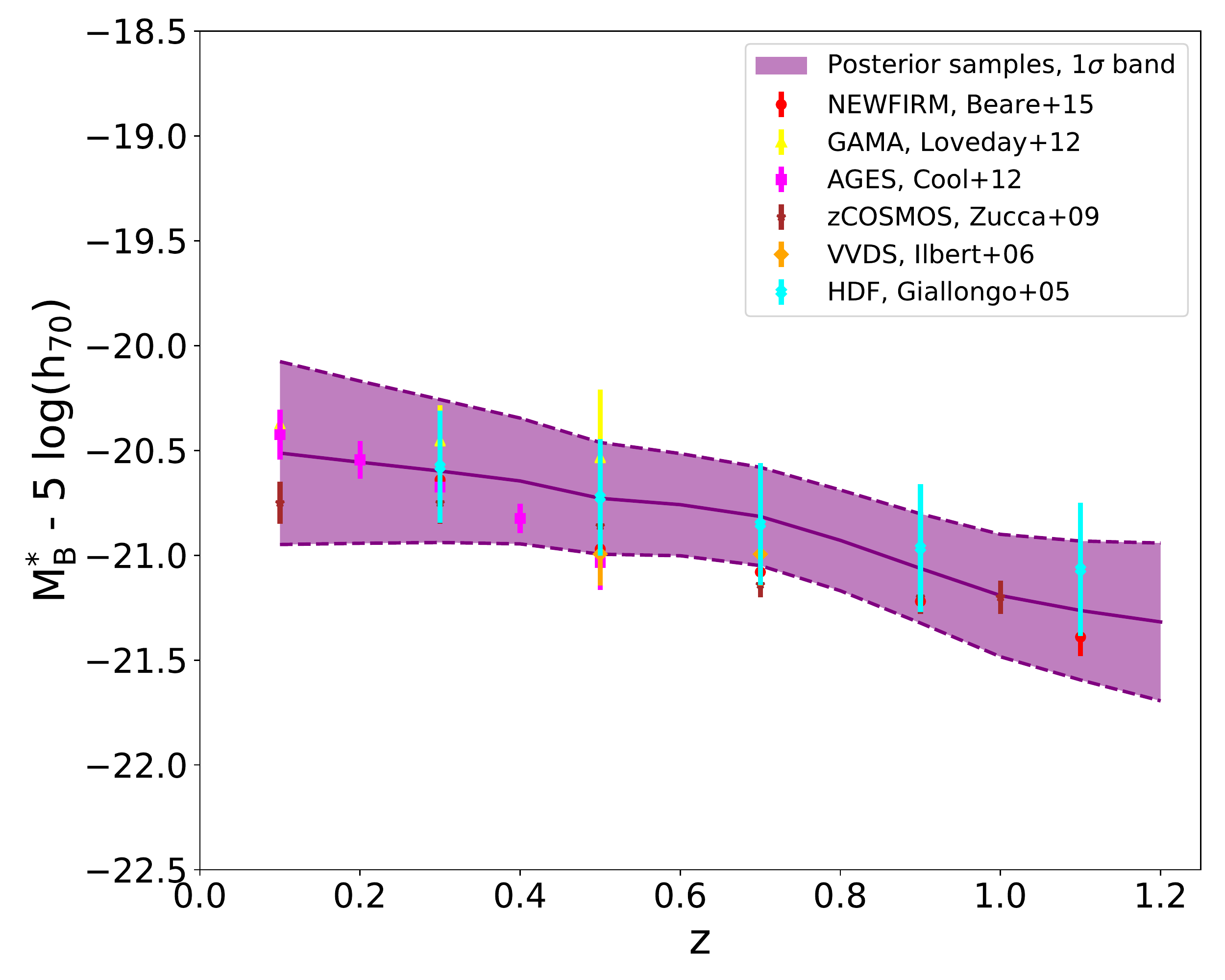}
\includegraphics[width=7.65cm]{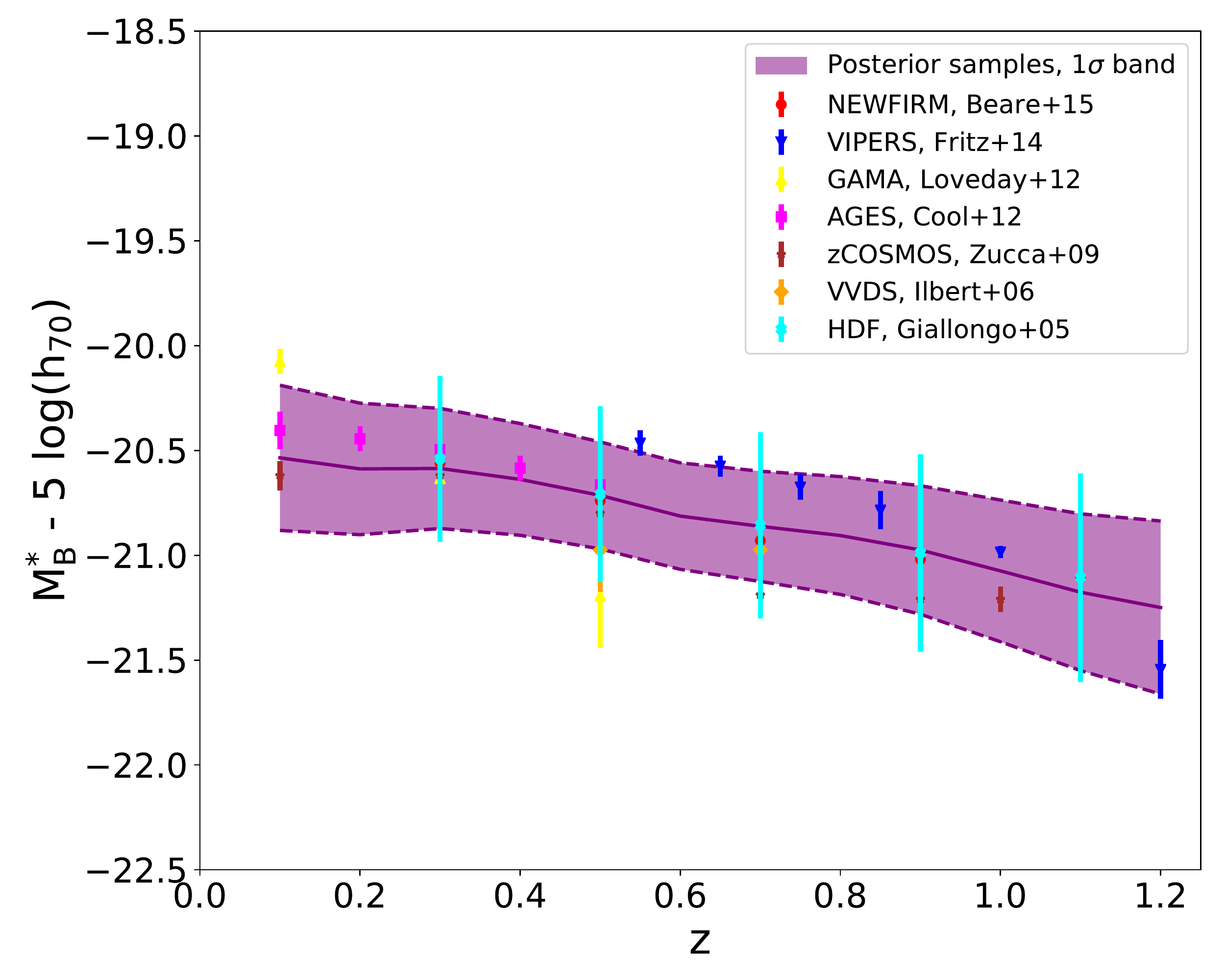}
\caption{The left panel shows the redshift evolution of $\mathrm{M^*}$ for blue galaxies, while the right panel the evolution for red galaxies. The purple line and band refer to the 50-th percentile values and one standard deviation errors, respectively, of our approximate Bayesian posterior. The coloured points and relative $1 \sigma$ error bars refer to different works in literature.}
\label{fig:tortorelli_fig11}
\end{figure}

\begin{figure}[t!]
\centering
\includegraphics[width=7.65cm]{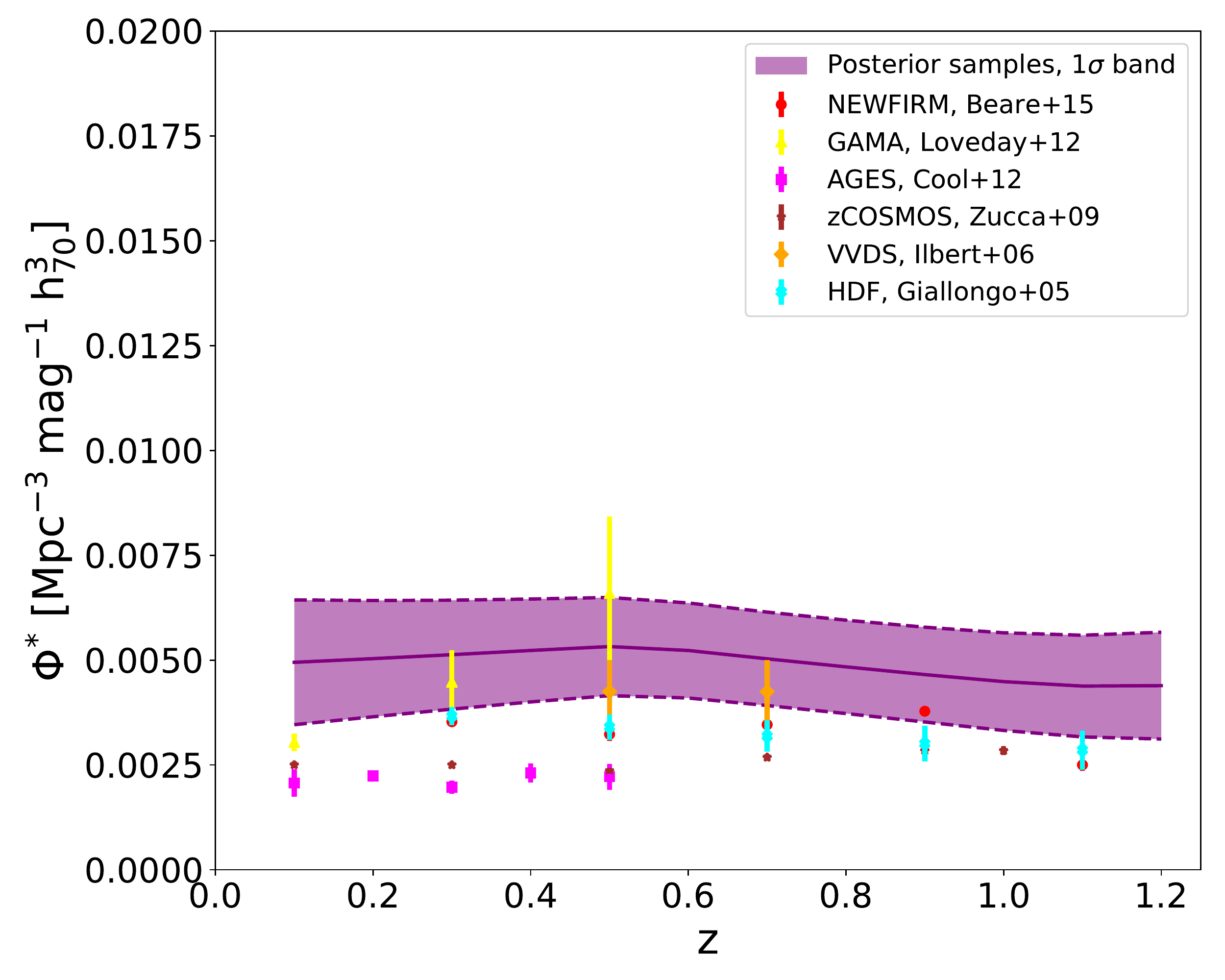}
\includegraphics[width=7.65cm]{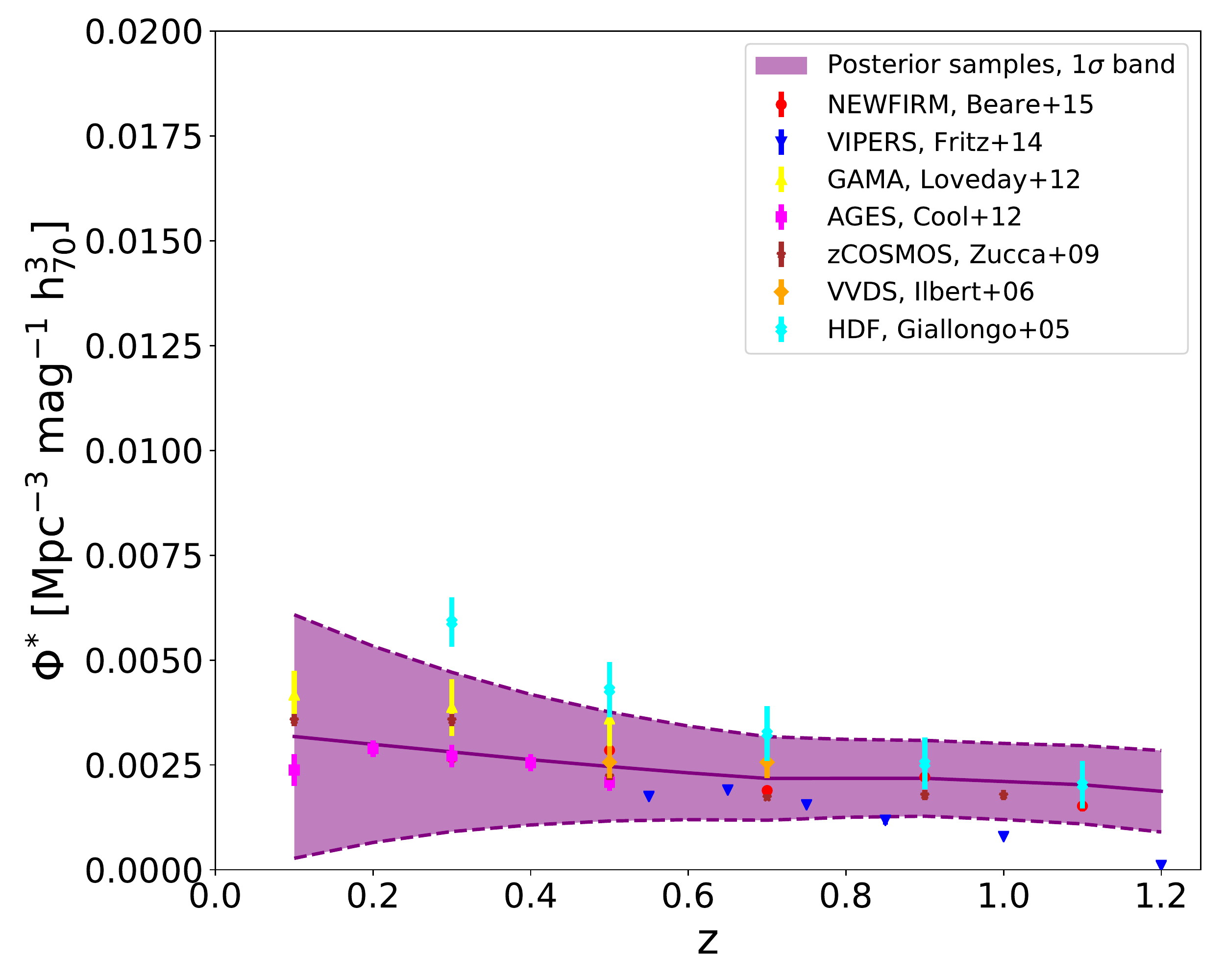}
\caption{The left panel shows the redshift evolution of $\phi^*$ for blue galaxies, while the right panel the evolution for red galaxies. The purple line and band refer to the 50-th percentile values and one standard deviation errors, respectively, of our approximate Bayesian posterior. The coloured points and relative $1 \sigma$ error bars refer to different works in literature.}
\label{fig:tortorelli_fig12}
\end{figure}

In this section, we compare the $\mathrm{M^*}$ and $\phi^*$ values from the approximate Bayesian posterior with those from other literature studies \cite{giallongo05,ilbert06,zucca09,loveday12,cool12,fritz14,beare15}. We show in figures \ref{fig:tortorelli_fig11} and \ref{fig:tortorelli_fig12} the redshift evolution of $\mathrm{M^*}$ and $\phi^*$ for red and blue galaxies. We compute the medians of $\mathrm{M^*}$ and $\phi^*$ and their one standard deviation errors in bins of $\Delta \mathrm{z} = 0.1$ and we plot them as purple bands.

The redshift evolution of the characteristic galaxy luminosity $\mathrm{M^*}$ from our results is consistent within $1 \sigma$ with all the quoted literature studies, both for blue galaxies (left panel of figure \ref{fig:tortorelli_fig11}) and for red galaxies (right panel of figure \ref{fig:tortorelli_fig11}). The only marginal inconsistency is with the GAMA field study at $\mathrm{z = 0.1}$. We find that $\mathrm{M^*}$ fades by $\Delta \mathrm{M}^*_{\mathrm{0.1-1.0,b}} = 0.68 \pm 0.52$ and $\Delta \mathrm{M}^*_{\mathrm{0.1-1.0,r}} = 0.54 \pm 0.48$ magnitudes between redshift $\mathrm{z = 1}$ and $\mathrm{z = 0.1}$ for blue and red galaxies, respectively. The subscript `b' and `r' refer to blue and red galaxies. By extrapolating our measurements to redshift $\mathrm{z = 2}$, we find that $\mathrm{M^*}$ fades by $\Delta \mathrm{M}^*_{\mathrm{0.1-2.0,b}} = 1.21 \pm 0.92$ and $\Delta \mathrm{M}^*_{\mathrm{0.1-2.0,r}} = 1.21 \pm 0.85$ magnitudes for blue and red galaxies, respectively. These values are 0.8 magnitudes smaller than what has been found in literature (see section \ref{section:introduction}), although they are consistent within errors with those results. The $\mathrm{M^*}$ median trend with redshift follows that of the HDF field study \cite{giallongo05} for blue galaxies, and that of the HDF \cite{giallongo05}, AGES \cite{cool12} and NEWFIRM \cite{beare15} field studies for red galaxies. Our results imply that $\mathrm{M^*}$ for blue galaxies fades more than that for red galaxies from $\mathrm{z = 1}$ to $\mathrm{z = 0.1}$, while from $\mathrm{z = 2}$ to $\mathrm{z = 0.1}$ the amount of fading is of the same order.

Our results for the LF amplitude $\phi^*$ at the characteristic galaxy luminosity show different behaviours for blue (left panel of figure \ref{fig:tortorelli_fig12}) and red (right panel of figure \ref{fig:tortorelli_fig12}) galaxies. Both for blue and red galaxies, the uncertainties are larger than most of the other studies at all redshifts, except for the GAMA field study \cite{loveday12} at $\mathrm{z=0.5}$. $\phi^*$ for blue galaxies stays roughly constant between $\mathrm{z = 0.1}$ and  $\mathrm{z = 1}$, $\Delta \mathrm{\phi}^*_{\mathrm{0.1-1.0,b}} = 0.0004 \pm 0.0019$, confirming what has already been found in literature (see section \ref{section:introduction}). $\phi^*$ for blue galaxies is consistent within $1 \sigma$ with the GAMA \cite{loveday12} field study at $\mathrm{z = 0.3}$ and $\mathrm{z = 0.5}$, with the VVDS \cite{ilbert06} field study at $\mathrm{z = 0.5}$ and $\mathrm{z = 0.7}$, with the HDF \cite{giallongo05} field study at $\mathrm{z = 0.3}$, $\mathrm{z = 0.9}$ and $\mathrm{z = 1.1}$, and with the NEWFIRM \cite{beare15} field study at $\mathrm{z = 0.9}$. Except for the GAMA field result \cite{loveday12} at $\mathrm{z = 0.5}$, the median values of $\phi^*$ with redshift for blue galaxies are larger than all the other studies. This implies that the number density of blue galaxies that we find in our work is higher than that of the other studies. This trend was already visible at all redshifts in figure \ref{fig:tortorelli_fig9}. $\phi^*$ for red galaxies decreases by $\sim 35 \%$ between $\mathrm{z = 0.1}$ and  $\mathrm{z = 1}$, $\Delta \mathrm{\phi}^*_{\mathrm{0.1-1.0,r}} = 0.001 \pm 0.003$, a value 15\% smaller than what has been found in literature (see section \ref{section:introduction}). $\phi^*$ for red galaxies is consistent within $1 \sigma$ with all studies at all redshifts, except for the HDF field study at $\mathrm{z = 0.3}$ and the VIPERS field study at $\mathrm{z = 1.0}$ and $\mathrm{z = 1.2}$. The median trend of $\phi^*$ for red galaxies with redshift follows that of the AGES field study at $\mathrm{z \le 0.5}$ and that of the NEWFIRM field study \cite{beare15} at $\mathrm{z \ge 0.7}$. The errors on the $\phi^*$ estimate for red galaxies are larger at $\mathrm{z \le 0.3}$, an effect already seen in the LFs of red galaxies in figure \ref{fig:tortorelli_fig8}. Our results suggest that the number density $\phi^*$ of blue galaxies at the characteristic luminosity $\mathrm{M^*}$ is larger than that of red galaxies at all redshifts.

These results demonstrate that the approximate Bayesian posterior we obtain with our method gives comparable results and it is consistent with most of the literature studies of the LF. In addition, our method provides probabilistically consistent error bars and it captures cosmic variance and part of the limitations from selection effects, such as incorrect redshift estimation and blue/red galaxies separation.

\subsection{Redshift distribution comparison with VIPERS}
\label{subsection:red_distr}

\begin{figure}[t!]
\centering
\includegraphics[width=14cm]{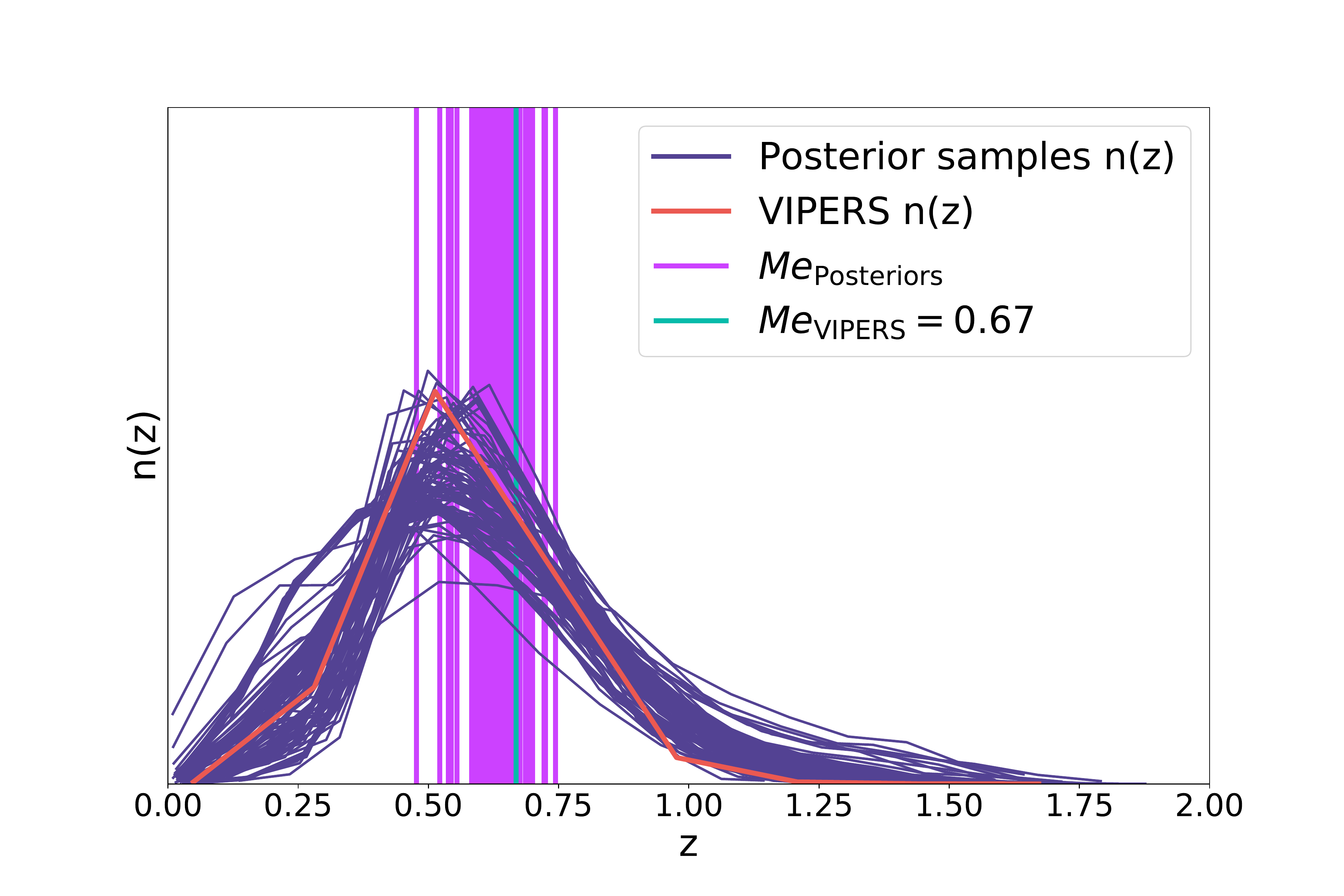}
\caption{Comparison between the set of simulated n(z) from the posterior distribution (purple lines) and the observed n(z) (orange line) from the VIPERS survey. The green and fuchsia vertical lines are the medians of the VIPERS n(z) and posterior n(z), respectively. $Me$ stands for median.}
\label{fig:tortorelli_fig13}
\end{figure}

The redshift distribution n(z) of galaxy samples is an important ingredient for different cosmological probes, including weak lensing \cite{Kacprzak2019}. To further validate our result and show its application in cosmology studies, we compare the n(z) from the VIPERS survey with those built from our posterior samples. 

The target selection for VIPERS spectra is based on a pure magnitude and color selection. Spectra are selected as those of galaxies having 
\begin{equation}
\begin{split}
17.5 \le &i' \le 22.5 \\
(r' - i') > 0.5 \times (u^{*} - g') \quad O&R \quad (r' - i') > 0.7
\end{split}
\end{equation}
The magnitudes in VIPERS are measured in the CFHTLS `W1' and `W4' fields. We apply the same cuts on our posterior samples catalogues. Furthermore, we select only those VIPERS spectra with redshifts flagged as measured with $> 95\%$ confidence (see section 4 in \cite{Scodeggio2018}).

In figure \ref{fig:tortorelli_fig13}, we show the comparison of the redshift distribution n(z) from the VIPERS survey (orange line) with the set of redshift distributions built with the approximate Bayesian posterior samples (purple lines). The figure shows that both the full VIPERS n(z) and the median of the VIPERS n(z) (green vertical line) are in good agreement with our posterior redshift distributions and medians (fuchsia vertical lines) of the posterior redshift distributions. In particular, the mean of the medians distribution is $0.63 \pm 0.05$ and it is consistent with the VIPERS median $Me_{\mathrm{VIPERS}} = 0.67$. This further validates the results and the use of UFig for cosmology measurements.

\section{Conclusions}
\label{section:conclusions}

The Luminosity Function of galaxies is a key observational ingredient for galaxy evolution studies and for cosmology. Different LF studies exist in the literature, but disagreement is still present on the value of its parameters and many systematic effects may bias its estimate. To overcome some of these limitations, we develop a forward modeling approach to measure galaxy population properties as a function of redshift and specifically the B-band Luminosity Function of galaxies.

In an earlier work, \cite{herbel17} developed a galaxy population model that needs to be constrained using existing data. We constrain the model using the Approximate Bayesian computation applied to CFHTLS images, since the likelihood of the model is not empirically tractable. We define a large prior space and we use an iterative approach of Rejection ABC algorithms to constrain the model parameters. For each set of parameters, we simulate CFHTLS images and we stack the catalogues of measured galaxy properties. We then define a number of distance metrics between survey data and simulations. By thresholding the distance metrics based on the $\mathrm{q}=10$-th percentile value, we derive the approximate Bayesian posterior for the LF and the size parameters and for the spectral coefficients for blue and red galaxies separately. 

We test our results by comparing galaxy population properties between survey data and simulations and we find a good agreement between the two. We then plot the set of LFs built from the posterior samples against different literature studies spanning different ranges in redshift and with different sample selections. We find a very good agreement for the red, blue and global LFs with the LF studies in literature at all probed redshifts.  For $\mathrm{M_B - 5 \log{h_{70}}} > -21$, our results show that at all probed redshifts the number density at fixed absolute magnitude of blue galaxies is greater than that of red galaxies. 

We find that our $\mathrm{M^*}$ measurements are consistent with all the quoted literature studies, both for blue and red galaxies. We find that $\mathrm{M^*}$ fades by $\Delta \mathrm{M}^*_{\mathrm{0.1-1.0,b}} = 0.68 \pm 0.52$ and $\Delta \mathrm{M}^*_{\mathrm{0.1-1.0,r}} = 0.54 \pm 0.48$ magnitudes between redshift $\mathrm{z = 1}$ and $\mathrm{z = 0.1}$ for blue and red galaxies, respectively. Furthermore, it fades by $\Delta \mathrm{M}^*_{\mathrm{0.1-2.0,b}} = 1.21 \pm 0.92$ and $\Delta \mathrm{M}^*_{\mathrm{0.1-2.0,r}} = 1.21 \pm 0.85$ magnitudes between redshift $\mathrm{z = 2}$ and $\mathrm{z = 0.1}$. These values are 1.2 magnitudes smaller than what has been found in literature, although consistent within errors. These results imply that $\mathrm{M^*}$ for blue galaxies fades more than that for red galaxies from $\mathrm{z = 1}$ to $\mathrm{z = 0.1}$, while from $\mathrm{z = 2}$ to $\mathrm{z = 0.1}$ the amount of fading is of the same order.

$\phi^*$ for blue galaxies stays roughly constant between $\mathrm{z = 0.1}$ and $\mathrm{z = 1}$, while for red galaxies it decreases by about $\sim 35 \%$. The first result is consistent with what has been already found in literature, while the second value is $15 \%$ smaller than other studies. The $\phi^*$ trend with redshift for red galaxies is consistent within $1 \sigma$ with all the quoted literature results, while for blue galaxies it is consistent mainly with \cite{giallongo05,ilbert06,loveday12}. Our results point to a number density of blue galaxies that is larger than what other studies found. Furthermore, the number density $\phi^*$ of blue galaxies at the characteristic galaxy luminosity $\mathrm{M^*}$ is always higher than that of red galaxies.

We also validate our results for cosmology applications comparing the redshift distributions from the approximate Bayesian posterior with that from the VIPERS survey. We perform the same magnitude and colours cuts as in the survey data and we compare the VIPERS n(z) with the set of n(z) from the posterior samples. We find a good agreement both for the distributions and for the median of the distributions, validating the method application for cosmology measurements.

In the spirit of the forward-modeling approach, the galaxy population model we calibrate in this work can be extended and complexified, if needed. For instance, we do not provide an accurate angular correlation function, but we position galaxies randomly on the sky and this might affect the completeness of faint sources, especially in highly clustered areas. Furthermore, in the current implementation, the size of galaxies does not explicitly evolve with redshift, but it does so due to its coupling with the LF evolution. Therefore, a possible extension could be the introduction of a redshift-evolving size relation, where the functional form is still log-normal, but the mean and standard deviation explicitly evolve with redshift. This and other possible extensions are left to future work.

Despite the simple yet realistic galaxy population model, we are able to provide the first end-to-end measurement of galaxy population properties using ABC. With our method, we therefore obtain a measurement of the LF with realistic uncertainties that takes into account part of the systematics affecting those measurements and is consistent with other LF measurements in literature. Furthermore, the method can be applied to other photometric and spectroscopic surveys that can help to better constrain the different parameters of the galaxy population model.

\acknowledgments

We acknowledge support by Swiss National Science Foundation (SNF) grant 200021\_169130. AR is grateful for the hospitality of KIPAC at Stanford University/SLAC where part of his contribution was made. LT is grateful to Uwe Schmitt and Ritabrata Dutta for useful discussions. Based on observations obtained with MegaPrime/MegaCam, a joint project of CFHT and CEA/IRFU, at the Canada-France-Hawaii Telescope (CFHT) which is operated by the National Research Council (NRC) of Canada, the Institut National des Science de l'Univers of the Centre National de la Recherche Scientifique (CNRS) of France, and the University of Hawaii. This work is based in part on data products produced at Terapix available at the Canadian Astronomy Data Centre as part of the Canada-France-Hawaii Telescope Legacy Survey, a collaborative project of NRC and CNRS. This work has made use of data from the European Space Agency (ESA) mission {\it Gaia} (\url{https://www.cosmos.esa.int/gaia}), processed by the {\it Gaia} Data Processing and Analysis Consortium (DPAC, \url{https://www.cosmos.esa.int/web/gaia/dpac/consortium}). Funding for the DPAC has been provided by national institutions, in particular the institutions participating in the {\it Gaia} Multilateral Agreement. This paper uses data from the VIMOS Public Extragalactic Redshift Survey (VIPERS). VIPERS has been performed using the ESO Very Large Telescope, under the "Large Programme" 182.A-0886. The participating institutions and funding agencies are listed at \url{http://vipers.inaf.it}.

\appendix

\section{\textsc{Source Extractor} Configuration}
\label{appendix:sexconfig}

\begin{table}
\centering
\begin{tabular}{c|c}
\hline
\textbf{\textsc{Source Extractor} parameter name} & \textbf{Value} \\
\hline
CATALOG\_TYPE & FITS\_1.0\\
DETECT\_TYPE & CCD\\
DETECT\_MINAREA & 10\\
THRESH\_TYPE & RELATIVE\\
DETECT\_THRESH & 2.0\\
ANALYSIS\_THRESH & 2.0\\
FILTER & Y\\
FILTER\_NAME & gauss\_3.0\_5x5.conv\\
DEBLEND\_NTHRESH & 32\\
DEBLEND\_MINCONT & 0.001\\
CLEAN & Y\\
CLEAN\_PARAM & 1.0\\
MASK\_TYPE & CORRECT\\
PHOT\_APERTURES & 5, 10, 15, 20, 25\\
PHOT\_AUTOPARAMS & 2.5, 3.5 \\
PHOT\_FLUXFRAC & 0.5\\
SATUR\_LEVEL & tile- \& band-dependent \\
MAG\_ZEROPOINT & 30 \\
GAIN & tile- \& band-dependent \\
PIXEL\_SCALE & 0.186 \\
SEEING\_FWHM & tile- \& band-dependent\\
STARNNW\_NAME & default.nnw\\
BACK\_SIZE & 64\\
BACK\_FILTERSIZE & 3\\
BACKPHOTO\_TYPE & LOCAL\\
BACKPHOTO\_THICK & 24\\
WEIGHT\_TYPE, MAP\_RMS & NONE \\
CHECKIMAGE\_TYPE & SEGMENTATION\\
\hline
\end{tabular}
\caption{\textsc{Source Extractor} configuration used in this work.}
\label{table:tortorelli_table5}
\end{table}

We report in table \ref{table:tortorelli_table5} the \textsc{Source Extractor} configuration that we use to analyze the CFHTLS survey and simulated images. The same configuration is used for \textsc{Source Extractor} single image and dual-image mode.

\begin{figure}[t!]
\centering
\includegraphics[width=16cm]{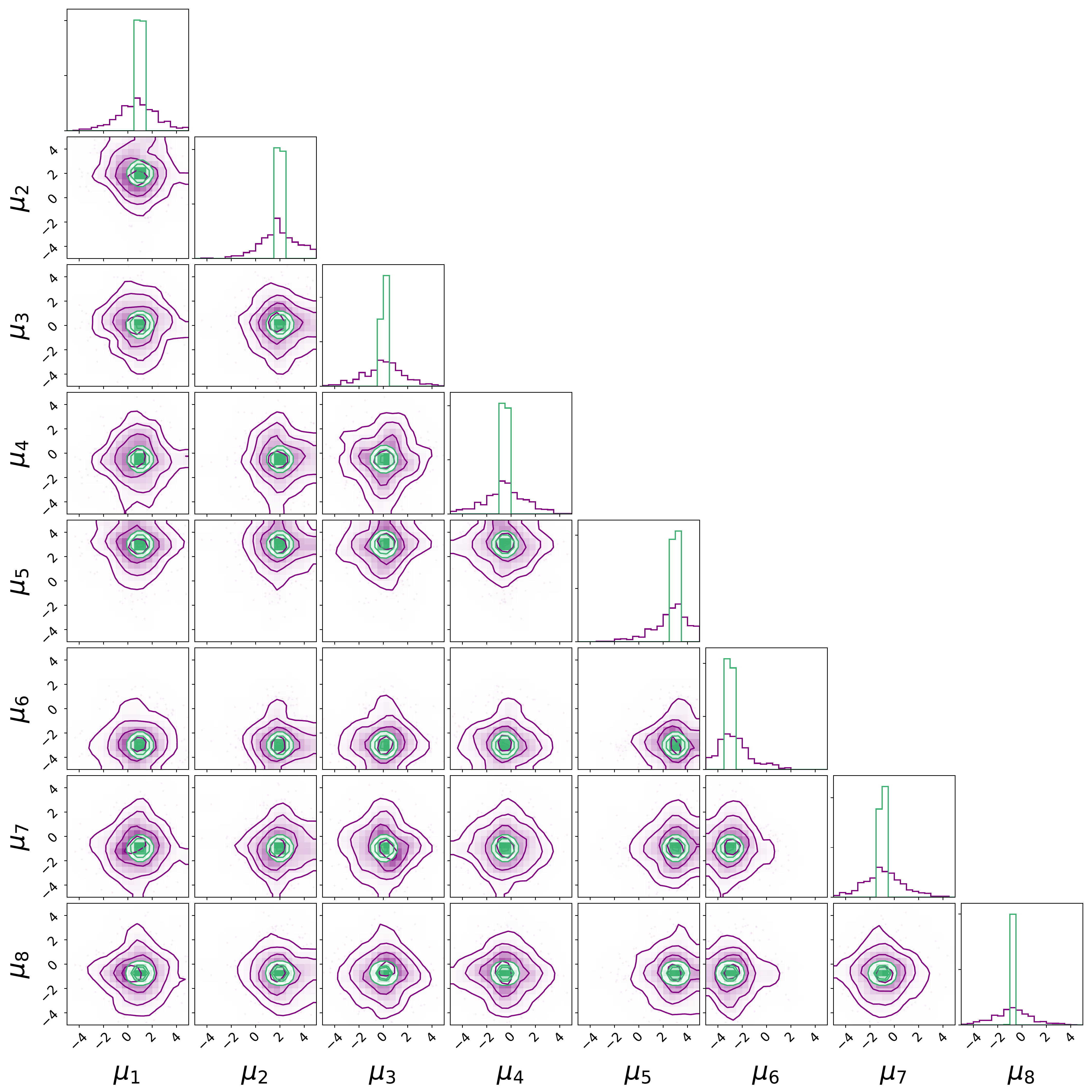}
\caption{Posterior distribution resulting from the $\mathrm{T = 1}$ iteration of our ABC algorithm on the 8-dimensional toy model . The parameters $\mu_i$ refer to the means of the distribution. Purple contours show the posterior distribution from the Euclidean distance, while green contours show the true Bayesian posterior.}
\label{fig:tortorelli_fig14}
\end{figure}

\section{ABC inference for Toy problem}
\label{appendix:abc_inferece_toy_problem}

To test our ABC inference scheme, we consider an 8-dimensional multivariate Gaussian case where we try to constrain its mean.

\begin{figure}[t!]
\centering
\includegraphics[width=15cm]{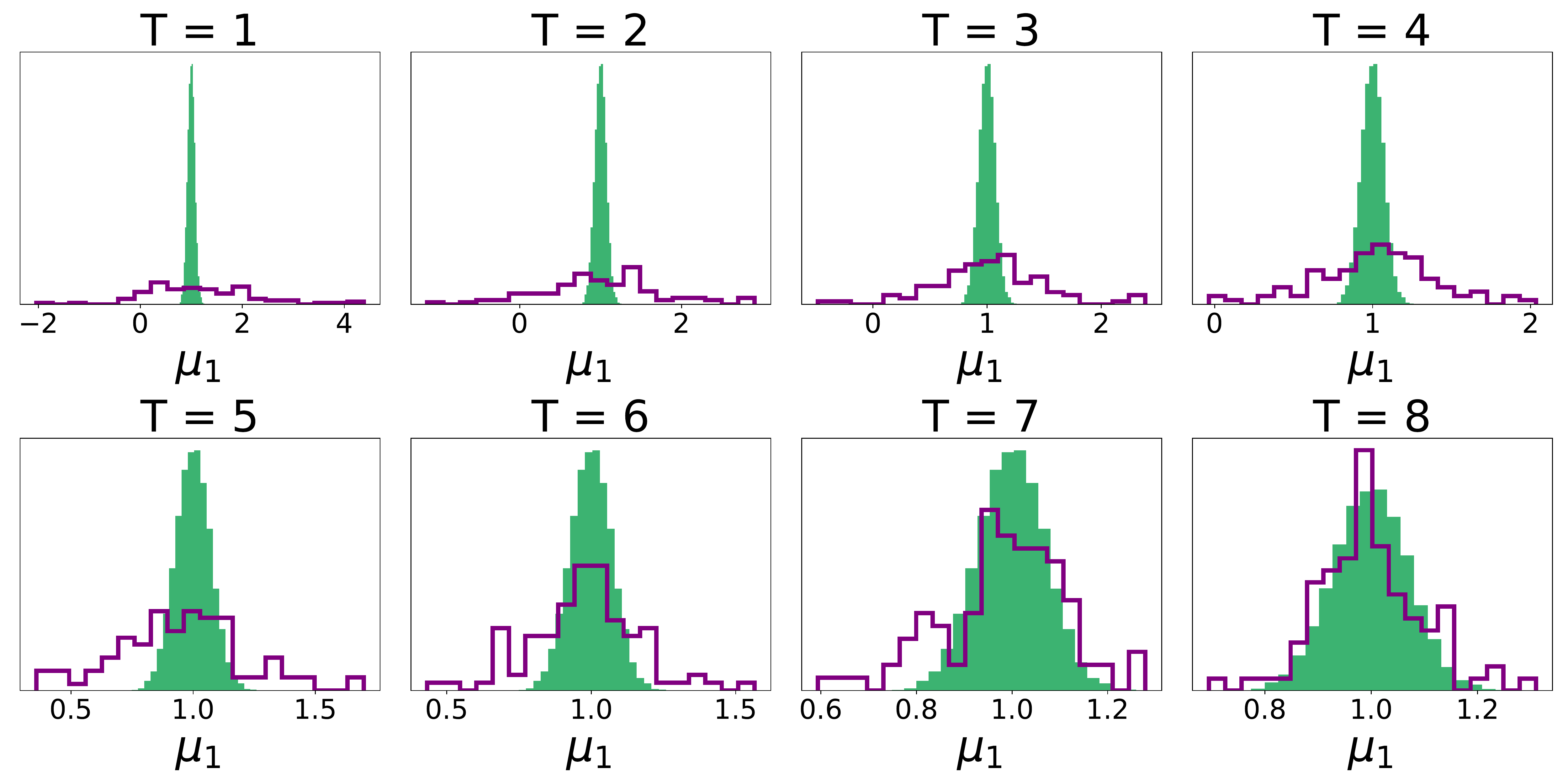}
\caption{The panels show the evolution of a 1-dimensional projection of the approximate Bayesian posterior as a function of the iteration $\mathrm{T}$ of the ABC algorithm. Purple and green histograms refer to the approximate and true Bayesian posteriors. The iteration number is shown at the top of each panel. The true Bayesian posterior is the same function in all panels, but the numerical range on the x-axis changes according to the approximate posterior size.}
\label{fig:tortorelli_fig15}
\end{figure}

We create the observed data by drawing $10^4$ samples from a Multivariate Normal distribution having mean $\mu_{1,..,8} = [1,2,0,-0.5,3,-3,-1,-0.8]$ and a diagonal covariance matrix where each diagonal element $\sigma_{\mathrm{ii}}^2=0.5$. We compute the true Bayesian posterior from the observed data. The probability distribution function (PDF) of a single sample $\mathrm{y_i}$ is given by
\begin{equation}
\mathrm{p(y_i|\theta) = \frac{ e^{ -(y_i-\theta)^2 / 2 \sigma^2}}{\sigma \sqrt{2\pi}} }
\end{equation}
where $\theta$ is the mean. Since the variables $\mathrm{y_i}$ are independent, the likelihood is given by the joint probability
\begin{equation}
\mathrm{p(y|\theta) = \prod_{i=1}^N  p(y_i|\theta) = \left( \sigma \sqrt{2 \pi} \right)^{-n} e^{-\sum_{i=1}^N \frac{(y_i-\theta)^2}{2\sigma^2}} }
\end{equation}
Since we assume a flat distribution for our prior ($\mathrm{p(\theta) \propto constant}$), then the normalized posterior probability is
\begin{equation}
\mathrm{ p(\theta|y) = \left( \frac{n}{2\pi\sigma^2} \right)^{\frac{1}{2}} e^{-\frac{n(\theta - \bar{y})^2}{2\sigma^2}} }
\end{equation} 
where $\mathrm{\bar{y} = \sum_{i=1}^N y_i / n }$ is the mean of the data points. Therefore, the true Bayesian posterior distribution of $\theta$ is a Multivariate Normal having means given by the means computed on the observed data and covariance given by the fixed covariance matrix rescaled for the square root of the number of samples.

The $\mathrm{T = 1}$ iteration of our ABC inference consists in sampling $6 \times 10^4$ 8-dimensional sets of means from an 8-dimensional flat prior distribution in the range $[-5, 5]$. Each prior sample is used to generate a Multivariate Normal with mean given by the prior point and covariance matrix as in the observed data. We draw $10^4$ samples from each Multivariate Normal. We evaluate $7$ different distance metrics for each prior point: RF distance on distribution means $\mathrm{d'_1}$, RF distance on distribution means and variances $\mathrm{d'_2}$, MMD distance between observed and simulated data $\mathrm{d'_3}$, histogram distance between observed and simulated data $\mathrm{d'_4}$, maximum among all rescaled distances $\mathrm{d'_{5}}$, maximum among rescaled MMD and histogram distance $\mathrm{d'_{6}}$ and a new defined $\mathrm{d'_{7}}$ distance metric as the absolute value of the differences between the observed and simulated means (hereafter, Euclidean distance). For the $\mathrm{d'_{4,6}}$ distances, we try different number of bins. The one that is giving the most constrained results is 20, so we use this value for this analysis and also for the inference on CFHTLS survey and mock data. By inspecting the posterior distributions resulting from the $\mathrm{q}=10$-th quantile for each distance metric, we find that $\mathrm{d'_{7}}$ provides the most stringent constraints. Therefore we use this posterior distribution to draw samples for the next iteration and we evaluate $\mathrm{d'_{7}}$ only for $\mathrm{T > 1}$. We apply the ABC algorithm described in section \ref{subsection:abc_scheme} enlarging the number of observations and simulations at each iteration. We stop our algorithm when the acceptance ratio drops below $10\%$.

\begin{figure}[t!]
\centering
\includegraphics[width=16cm]{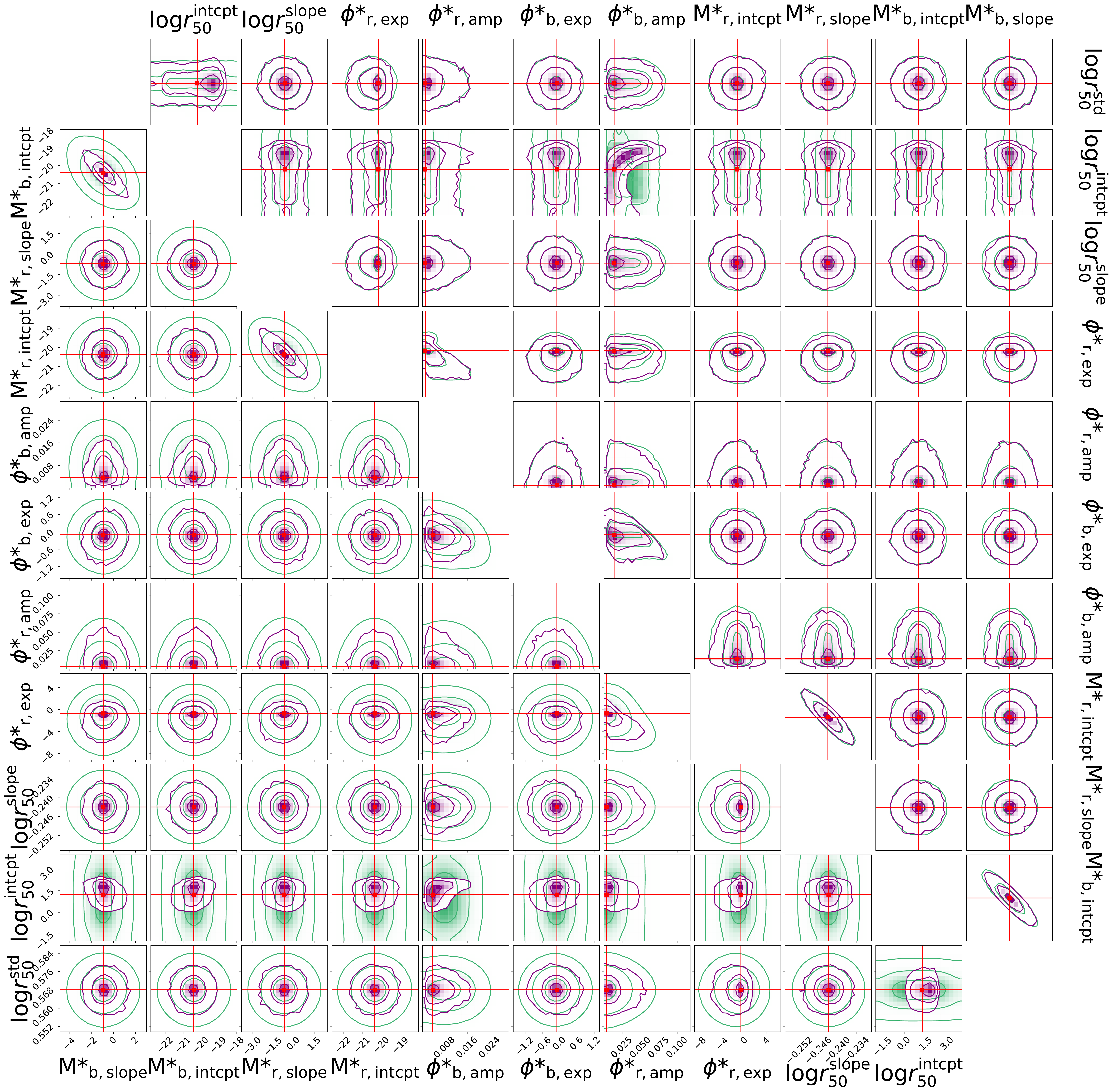}
\caption{LF and size parameters prior (green contours) and posterior (purple contours) distributions from the $\mathrm{T=1}$ iteration of the ABC inference on mock data. Upper and lower corner plots show the results when using $\mathrm{d_1}$ and $\mathrm{d_{24}}$ as distance metrics, respectively. The red crosses represent the true input values. The subscript `b' and `r' refer to parameters for blue and red galaxies, respectively.}
\label{fig:tortorelli_fig16}
\end{figure}

We show in figure \ref{fig:tortorelli_fig14} the posterior distribution (purple contours) resulting from the $\mathrm{T = 1}$ iteration using the Euclidean distance. The posterior samples are those for which the Euclidean distance between real and simulated means is smaller than the $\mathrm{q}=10$-th percentile value. We also over-plot the true Bayesian posterior in green. The $\mathrm{T = 1}$ iteration already gives a posterior distribution that contains the true Bayesian posterior within the $1\sigma$ error. We show in figure \ref{fig:tortorelli_fig15} a 1-dimensional projection of the evolution of the constraints on the means for the $\mathrm{T > 1}$ iterations. Purple and green histograms refer to the approximate and true Bayesian posteriors, respectively. The algorithm stops at the $\mathrm{T = 8}$ iteration. The resulting approximate Bayesian posterior is both centered on the true mean and larger than the true Bayesian posterior. Same conclusions apply to the full 8-dimensional posterior constraints.

\section{ABC inference on mock CFHTLS observations}
\label{appendix:simonsim_run}

\begin{figure}[t!]
\centering
\includegraphics[width=16cm]{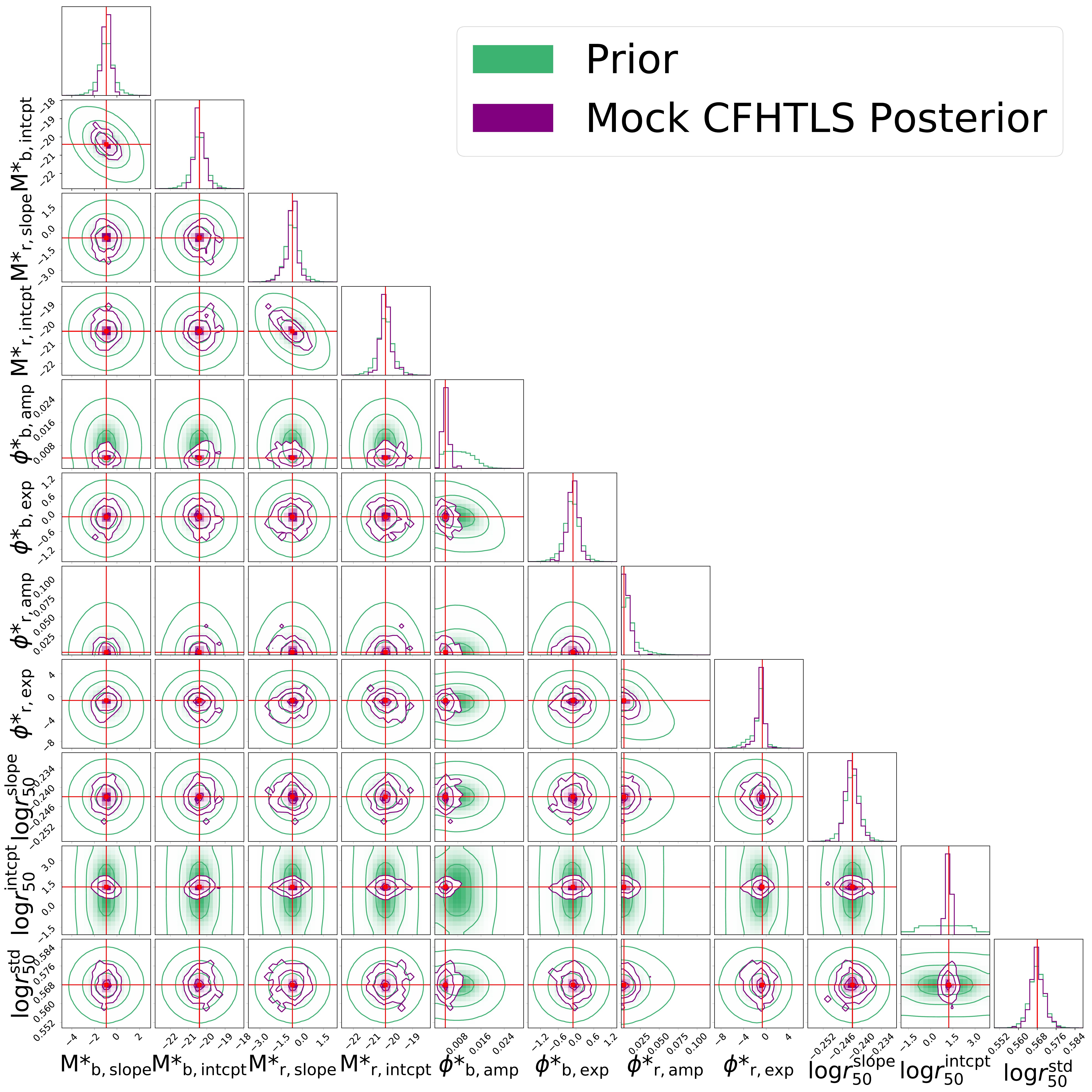}
\caption{LF and size parameters prior (green contours) and posterior (purple contours) distributions from the ABC inference on mock data. The red crosses represent the true input values. The subscripts `b' and `r' refer to parameters for blue and red galaxies, respectively.}
\label{fig:tortorelli_fig17}
\end{figure}

In order to test the performance of our method and to test the impact of the distance metrics on the final result, we perform an ABC inference on mock CFHTLS observations. The mock observations consists of a simulated CFHTLS `Wide' survey generated with a defined set of LF, spectral coefficient and size parameters for UFig. This set of UFig parameters is reported in Appendix B of \cite{tortorelli18b}. In \cite{tortorelli18b}, the LF parameters for the blue and the red populations are taken from \cite{beare15}, sizes are from the Great-3 challenge \cite{mandelbaum14} and the spectral coefficients are taken from \cite{herbel17}. Each mock image has the same instrumental parameters of the corresponding CFHTLS survey image and it has also a different noise seed to mimic the noise variations between different observations. We use the prior space in table \ref{table:tortorelli_table2} and we evaluate the $31$ distance metrics in table \ref{table:tortorelli_table1}.

\begin{figure}[t!]
\centering
\includegraphics[width=7.65cm]{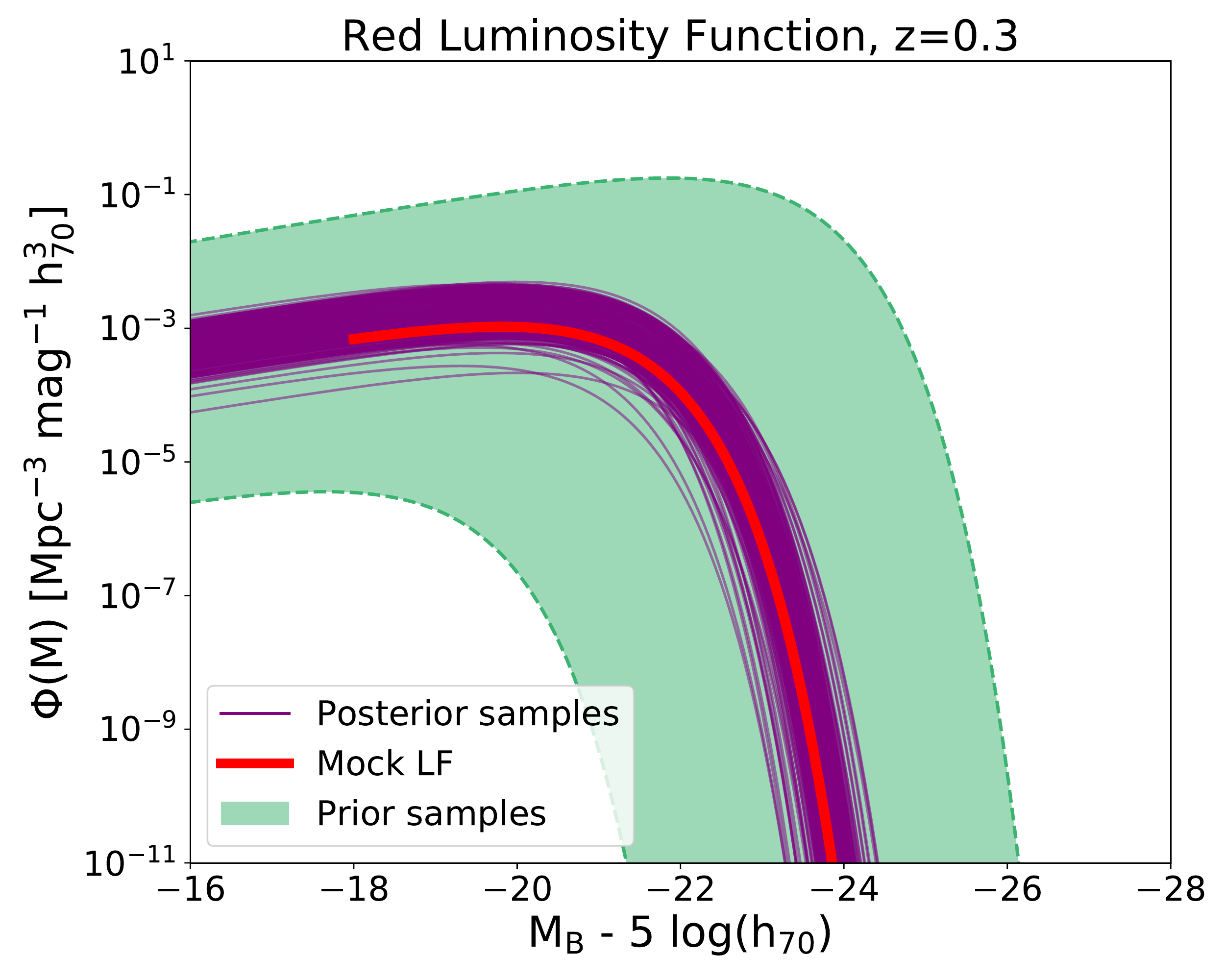}
\includegraphics[width=7.65cm]{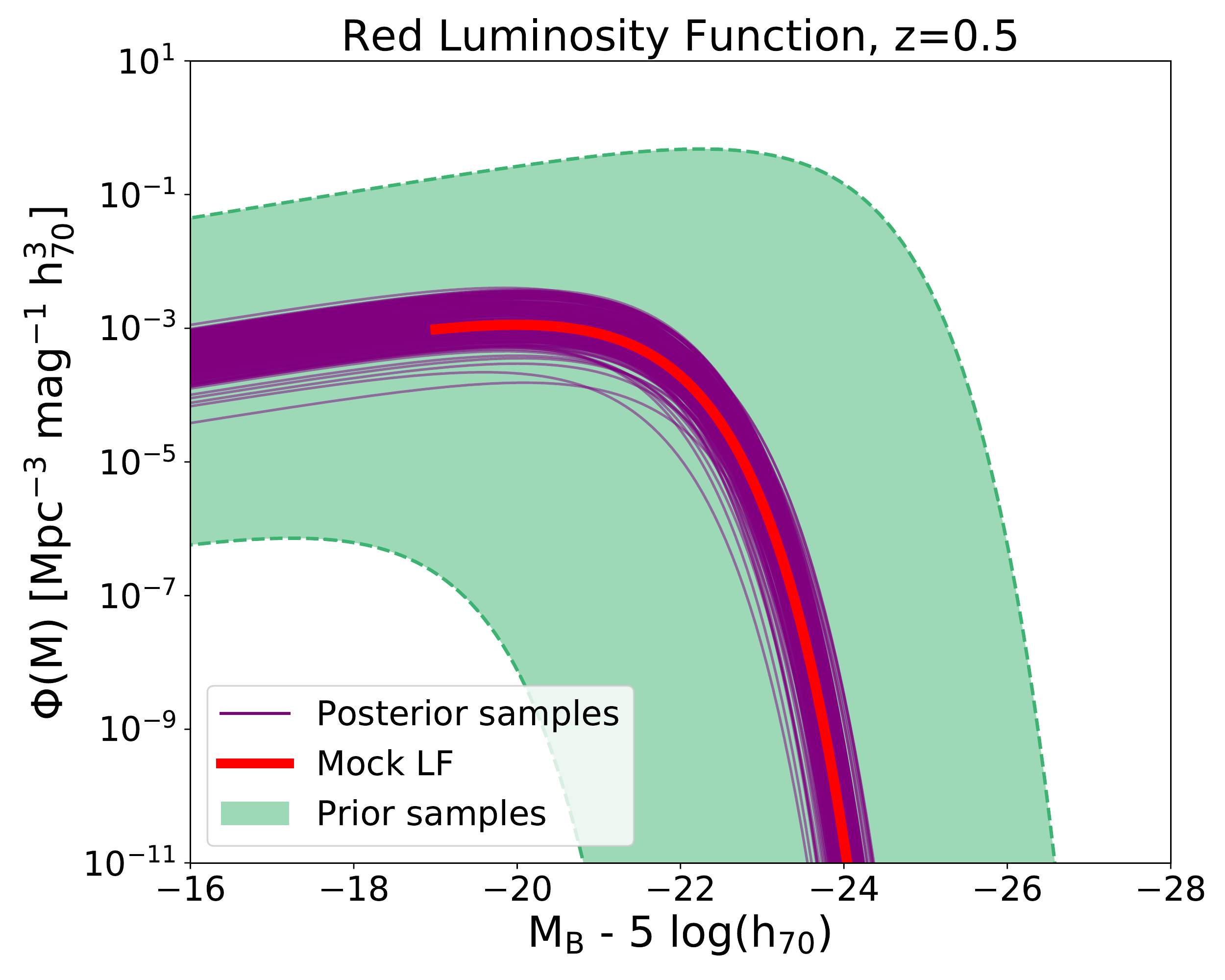}
\includegraphics[width=7.65cm]{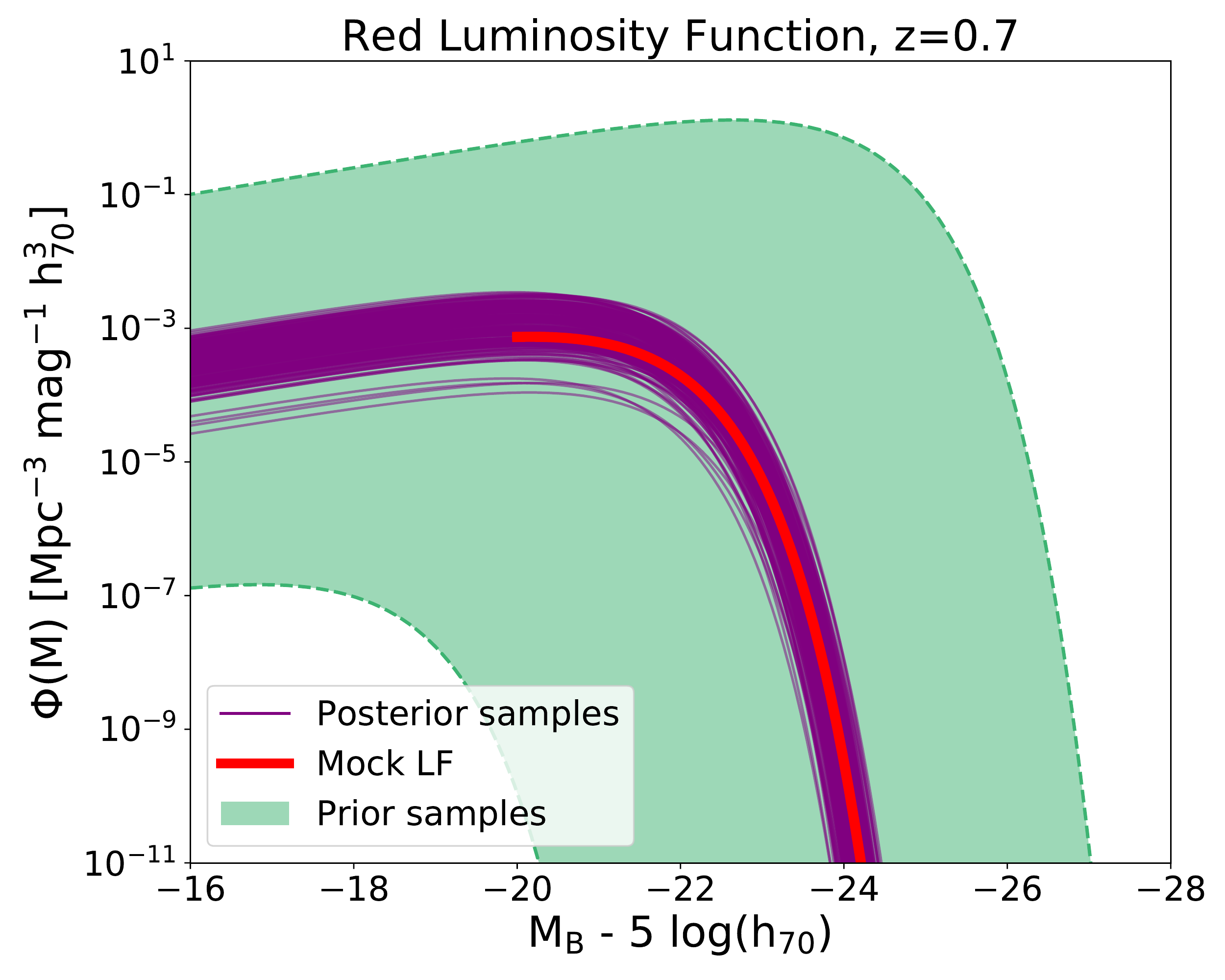}
\includegraphics[width=7.65cm]{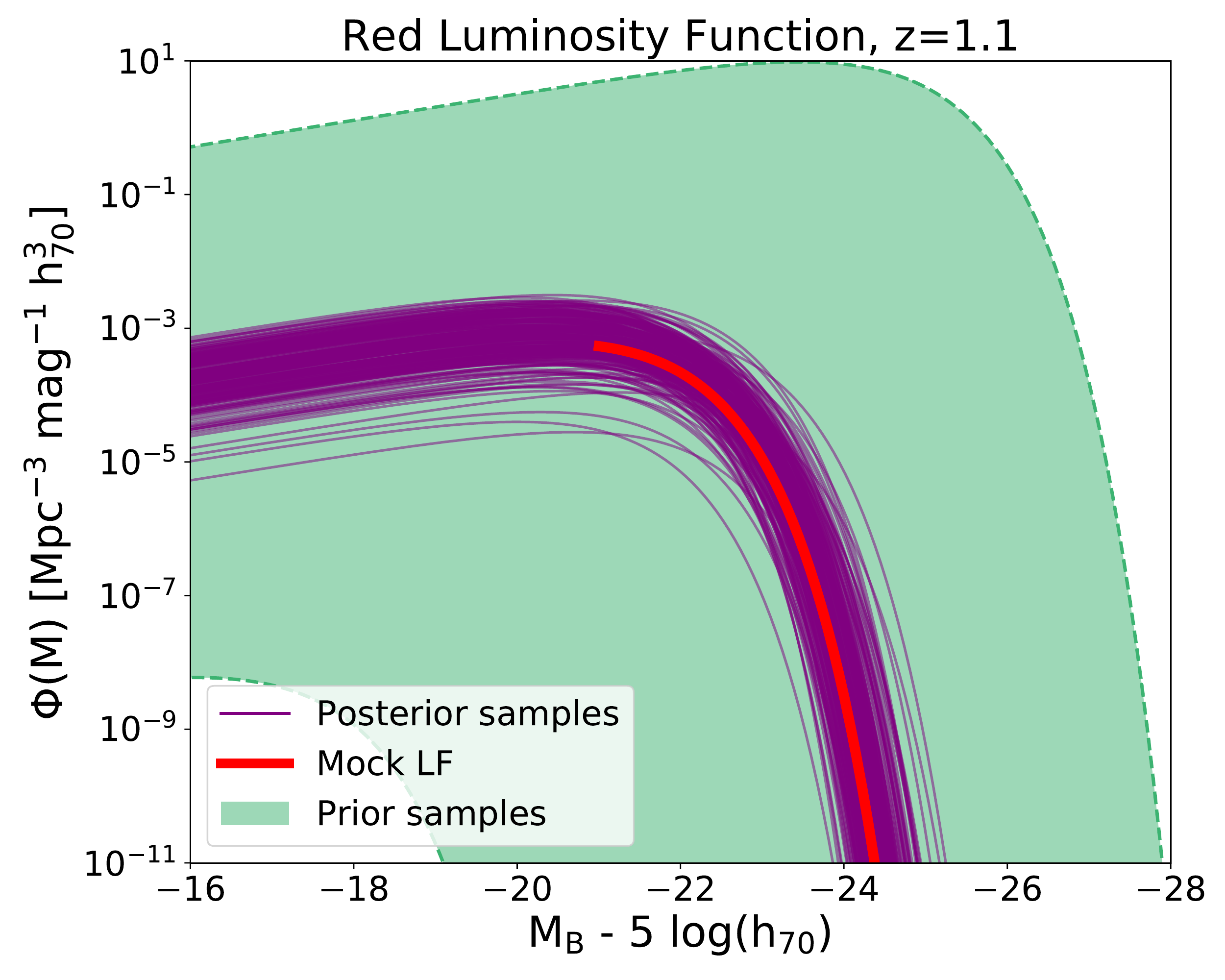}
\caption{Comparison between the set of LFs built from the posterior distributions (purple curves in the background) and the LF from the mock CFHTLS survey for red galaxies (red curves). The green band in the background represents LFs from the prior distribution. We plot comparisons for redshifts $\mathrm{z = 0.3}$, $0.5$, $0.7$ and $1.1$ in the upper left, upper right, lower left, lower right panels, respectively. The absolute magnitude in the B-band is in units of $\mathrm{M_B - 5 \log{h_{70}}}$, while the number density of galaxies is in units of $\mathrm{Mpc^{-3}\ mag^{-1}\ h^3_{70}}$.}
\label{fig:tortorelli_fig18}
\end{figure}

\begin{figure}[t!]
\centering
\includegraphics[width=7.65cm]{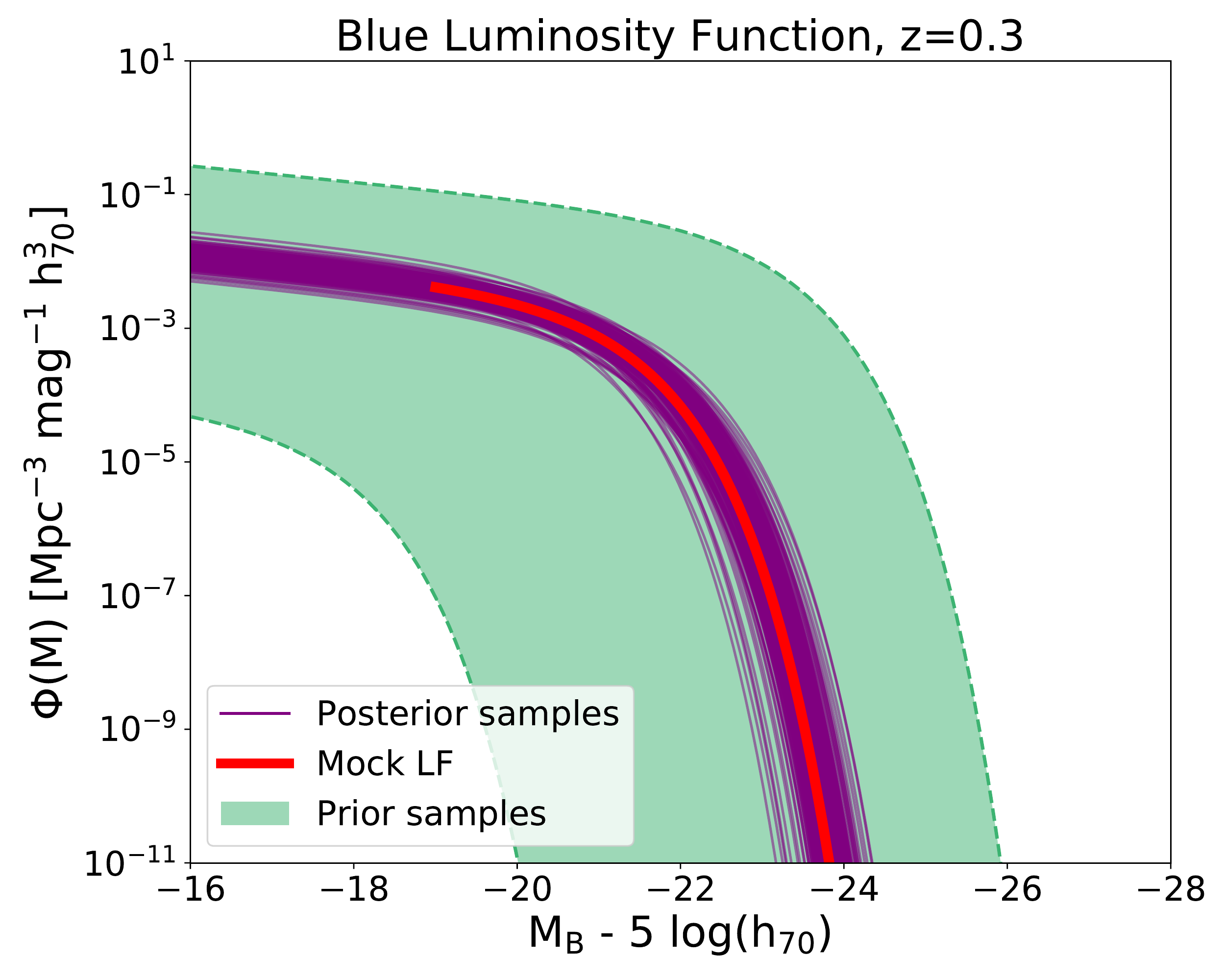}
\includegraphics[width=7.65cm]{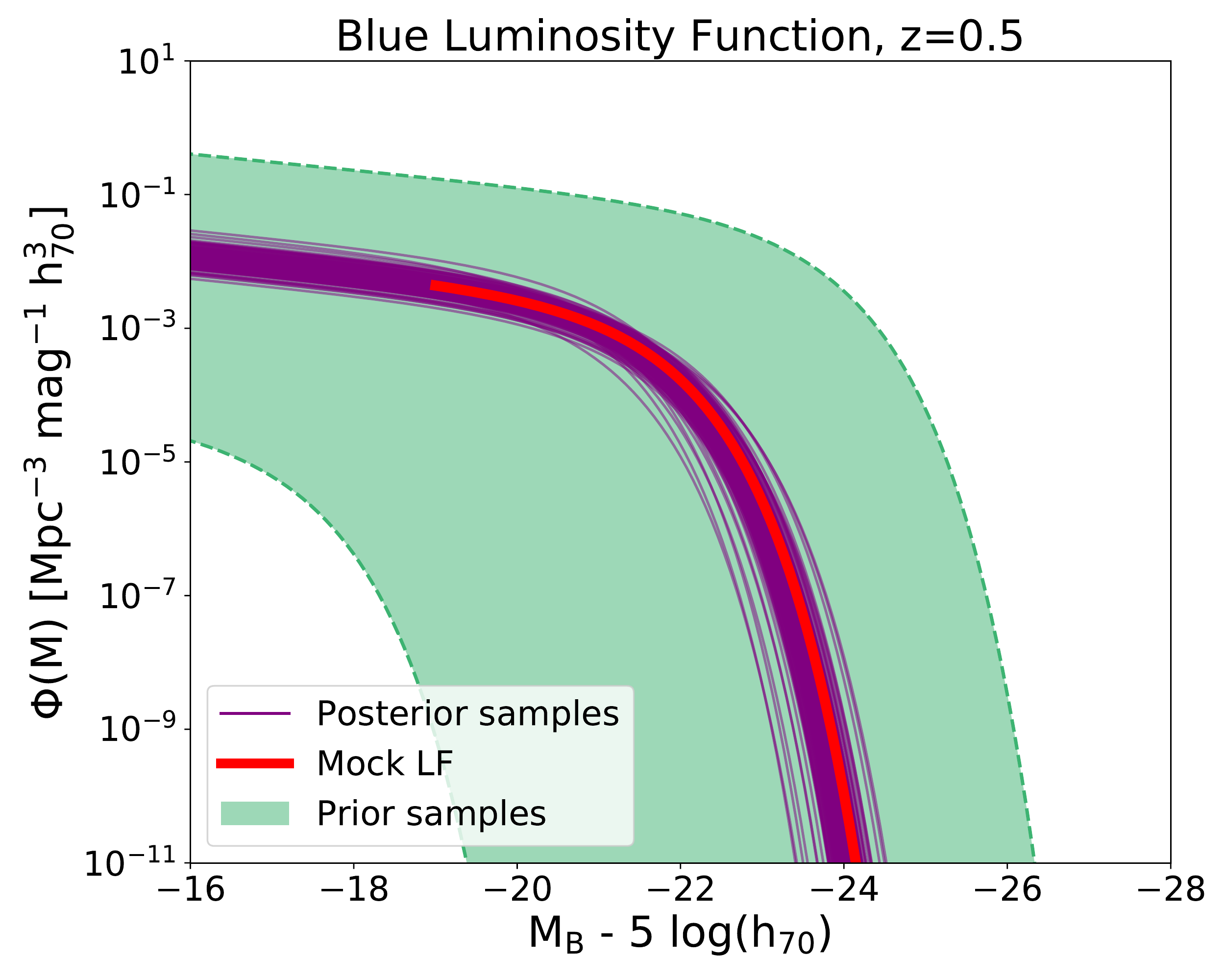}
\includegraphics[width=7.65cm]{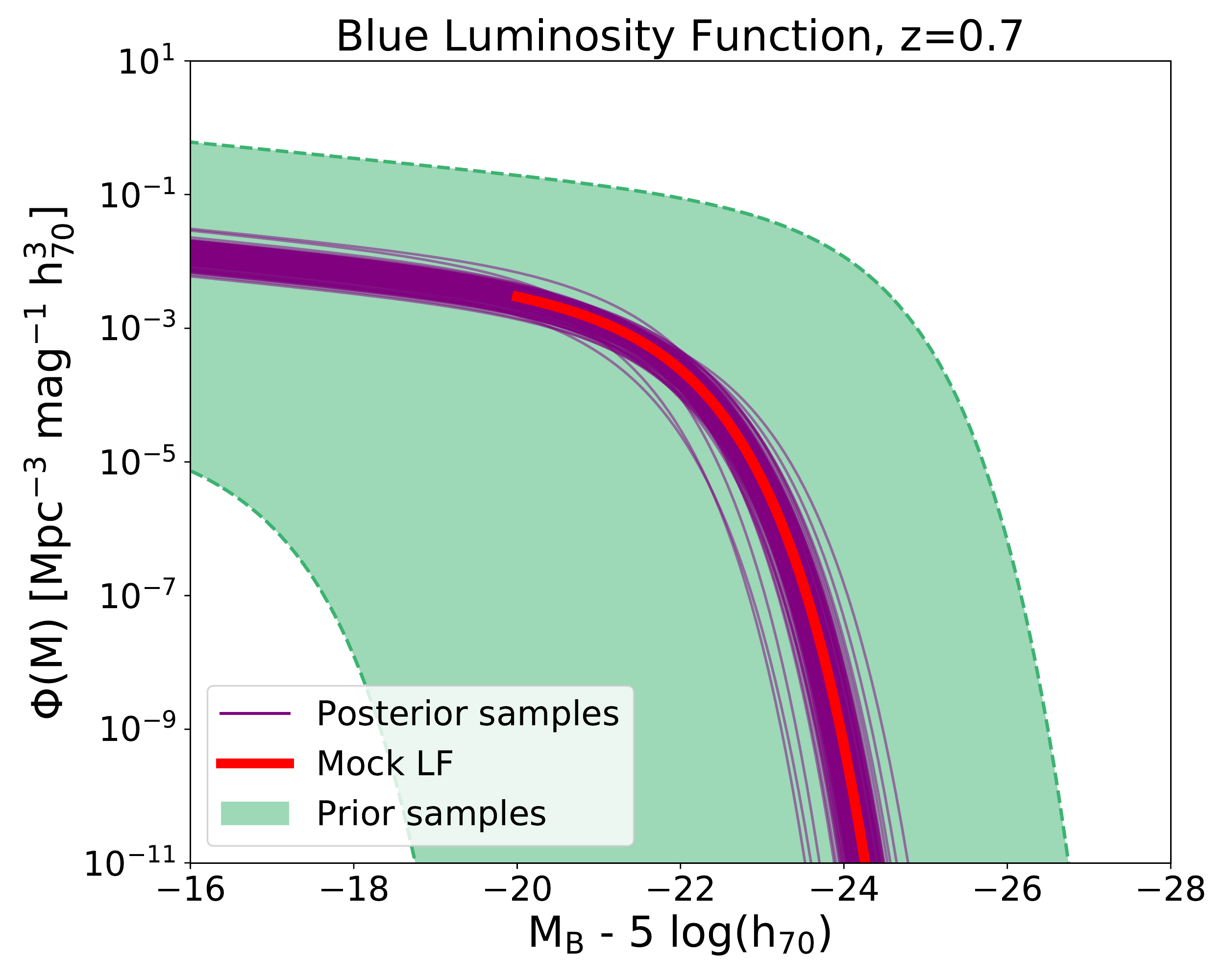}
\includegraphics[width=7.65cm]{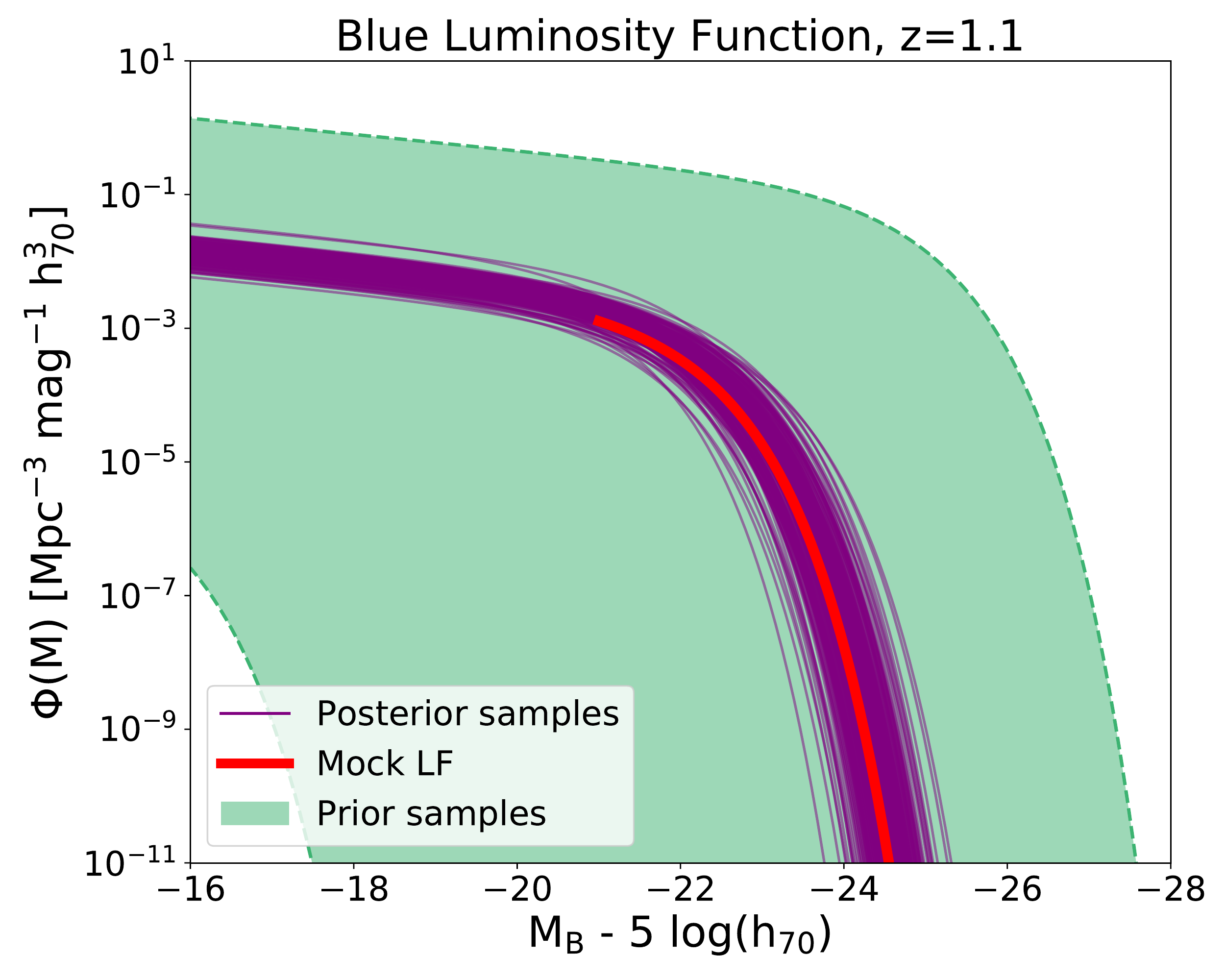}
\caption{Comparison between the set of LFs built from the posterior distributions (purple curves in the background) and the LF from the mock CFHTLS survey for blue galaxies (red curves). The green band in the background represents LFs from the prior distribution. We plot comparisons for redshifts $\mathrm{z = 0.3}$, $0.5$, $0.7$ and $1.1$ in the upper left, upper right, lower left, lower right panels, respectively. The absolute magnitude in the B-band is in units of $\mathrm{M_B - 5 \log{h_{70}}}$, while the number density of galaxies is in units of $\mathrm{Mpc^{-3}\ mag^{-1}\ h^3_{70}}$.}
\label{fig:tortorelli_fig19}
\end{figure}

We perform the same ABC inference as described in section \ref{subsection:abc_data}. At $\mathrm{T = 1}$, we evaluate $10^5$ samples from the prior distribution. For each sample, we simulate $10$ randomly chosen CFHTLS `Wide' patches and we evaluate all the distance metrics. The posterior distributions resulting from the different distance metric already contain the true input value of the mock data at the $\mathrm{T = 1}$ iteration. However, not all of them offer the same constraining power. We show in figure \ref{fig:tortorelli_fig16} the approximate Bayesian posterior at $\mathrm{T = 1}$ for the LF and size parameters for two different distance metrics. The upper corner plot shows the posterior distribution (purple contours) when using the absolute difference in detected galaxies $\mathrm{d_1}$ as distance metric. Its constraining power is not optimal compared to the initial prior distribution (green contours), specifically for size parameters. Nonetheless, the posterior distribution contains the true input value (red crosses). The lower corner plot shows the posterior distribution (purple contours) when using $\mathrm{d_{24}}$. The constraining power is better than $\mathrm{d_1}$, especially for the size parameters, and the distribution is centered on the true input value. 

The distance metric that provides the best constraints on LF and size parameters at $\mathrm{T} = 1$ is $\mathrm{d_{24}}$. Therefore, we only evaluate this distance metric at $\mathrm{T > 1}$. At each new iteration, we increase the number of simulated patches by $10$. We show in figure \ref{fig:tortorelli_fig17}, \ref{fig:tortorelli_fig18} and \ref{fig:tortorelli_fig19} the resulting approximate Bayesian posterior and the LFs for red and blue galaxies from the inference on mock data. The approximate Bayesian posterior distribution is very well constrained and centered on the true input value of the mocks. The LFs for blue and red galaxies are plotted as set of purple curves. Each curve represents a sample of the approximate posterior. These LFs are also very well constrained and centered on the true input LFs (red curves) at all redshifts.

\section{Blue and Red galaxies in the UVJ diagram}
\label{appendix:blue_red_uvj}

\begin{figure}[t!]
\centering
\includegraphics[width=7.6cm]{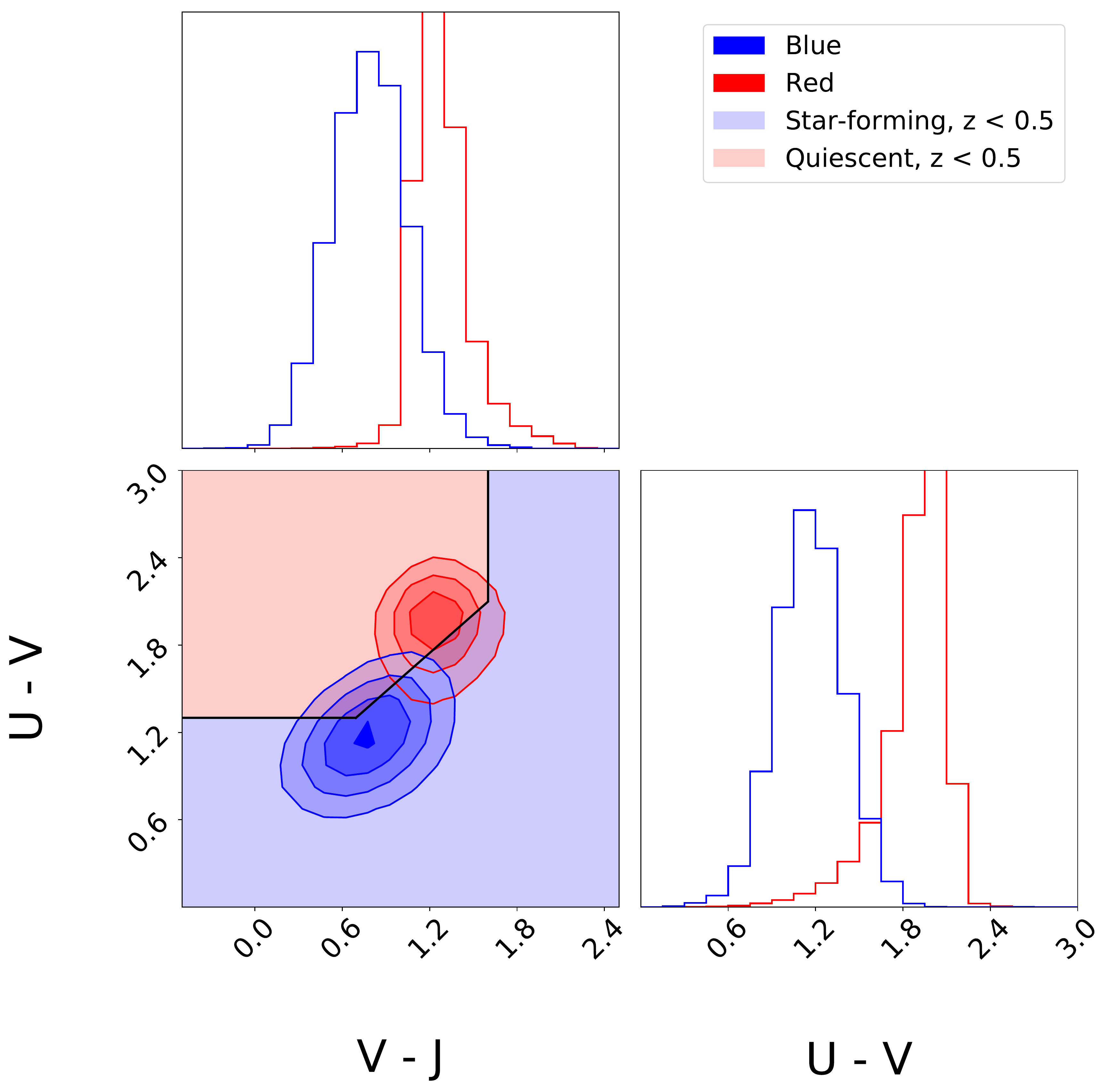}
\includegraphics[width=7.6cm]{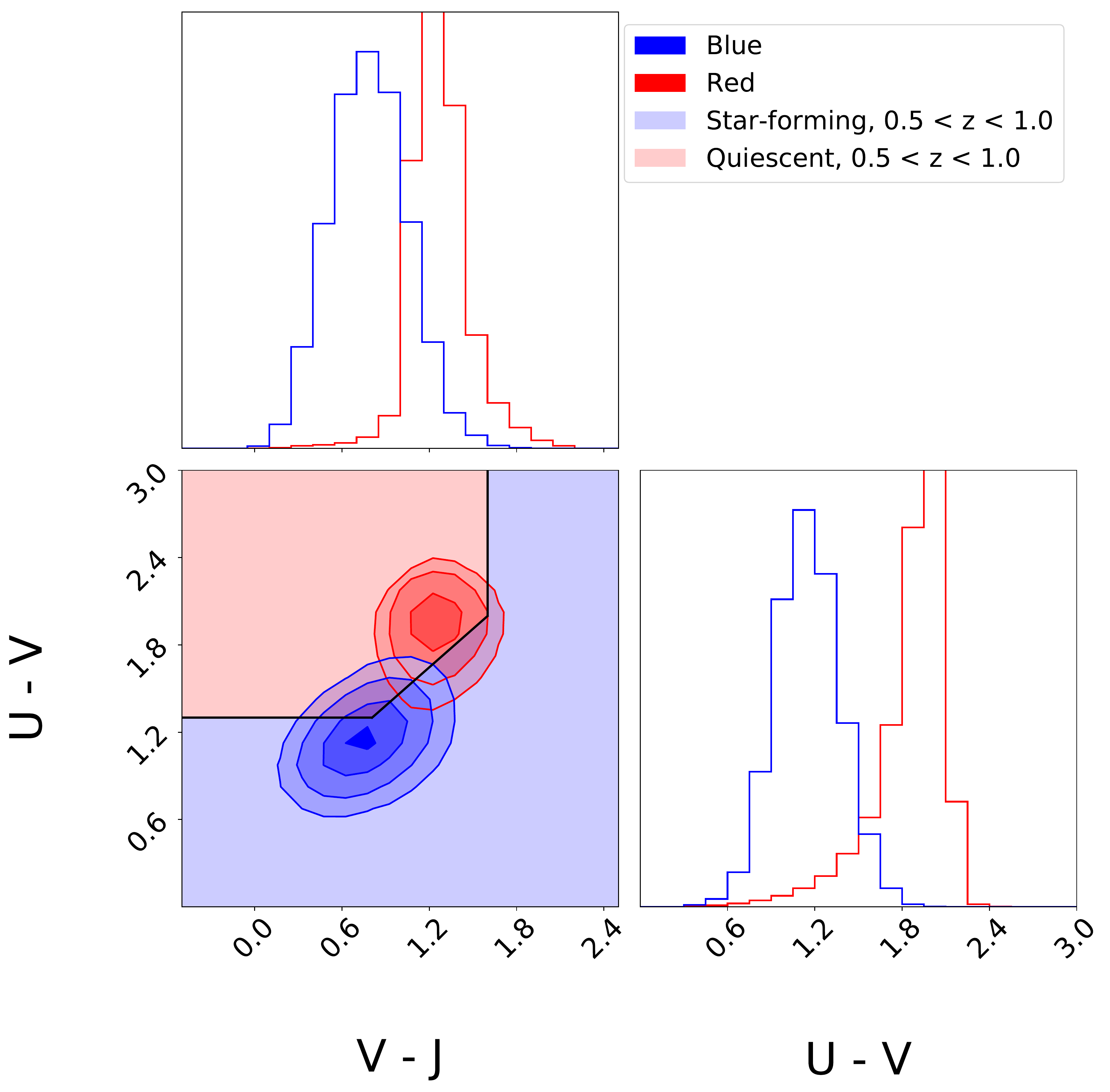}
\includegraphics[width=7.6cm]{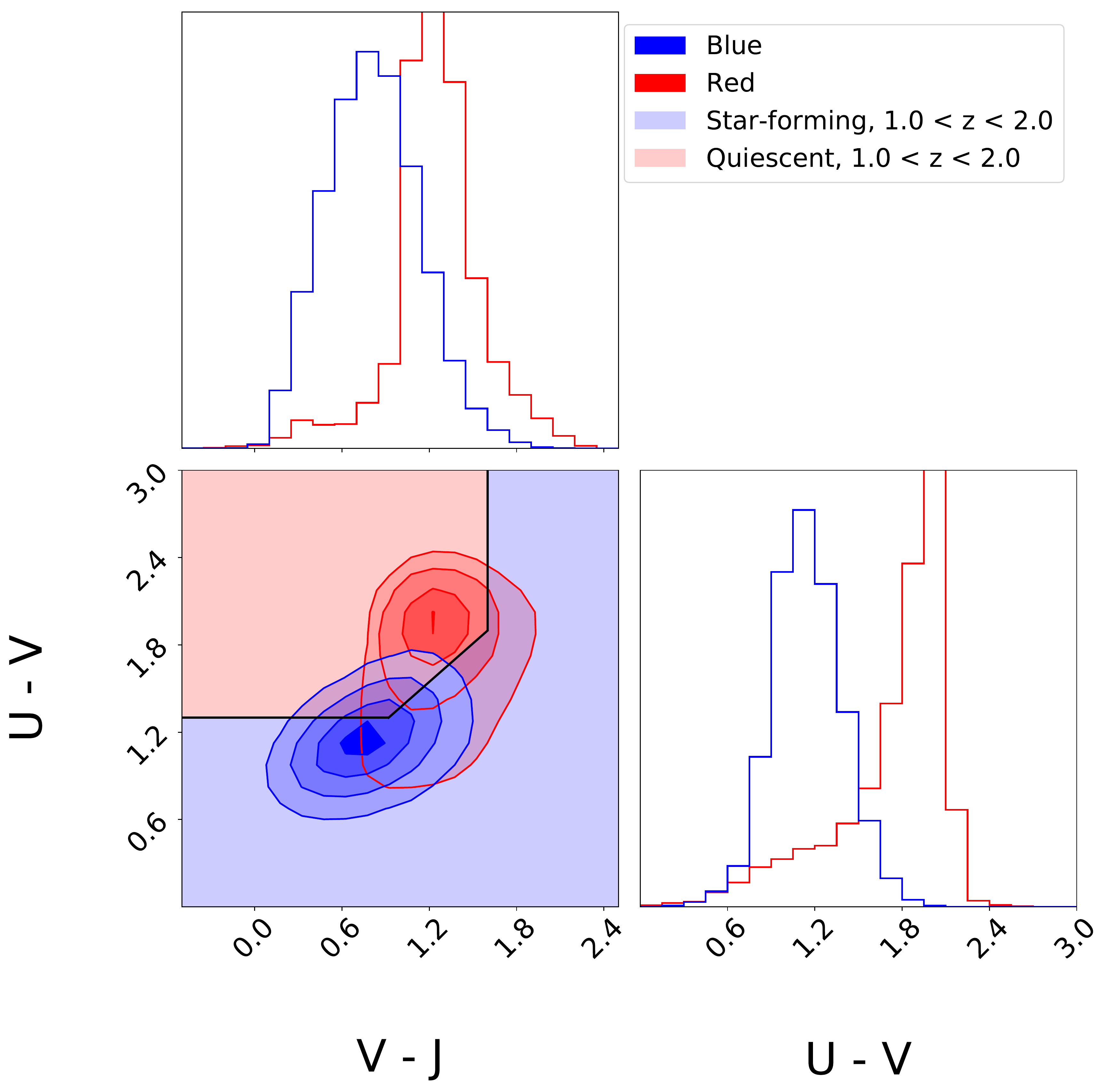}
\caption{Top left, top right, bottom panels show the UVJ diagram in the redshift ranges $\mathrm{z < 0.5}$, $\mathrm{0.5 < z < 1.0}$, $\mathrm{1.0 < z < 2.0}$, respectively. Red and blue galaxies are classified as such according to the intrinsic classification in our galaxy population model. Red and blue contours refer to the distribution of our red and blue galaxies in the UVJ diagram. The black lines separate regions of quiescent (light red) and star-forming (light blue) galaxies according to the empirical relation in \cite{Williams2009}. The distributions of red and blue galaxies are peaked in the quiescent and star-forming galaxies regions, respectively, at all redshifts.}
\label{fig:tortorelli_fig20}
\end{figure}

The galaxy population model we use in our work draws galaxies from two redshift-dependent Luminosity Functions for blue and red galaxies. Furthermore, two different rest-frame spectra are assigned to blue and red objects. In order to check that our definition of blue and red galaxies is consistent with what it is usually found in literature, we check the location of our galaxies in the rest-frame $\mathrm{V-J}$ vs $\mathrm{U-V}$ color space (UVJ diagram). The UVJ diagram is an empirical relation that is extensively used in literature (e.g., \cite{Wuyts2007,Williams2009}) in order to distinguish quiescent galaxies from star-forming galaxies (SFGs), including those SFGs that are heavily reddened. Since quiescent galaxies exhibit red colours, while SFGs blue colours, the position of our galaxies in this plane tests whether their blue and red intrinsic classification is consistent with literature studies. 

We show in figure \ref{fig:tortorelli_fig20} the UVJ diagram for three different redshift bins: $\mathrm{z < 0.5}$, $\mathrm{0.5 < z < 1.0}$, $\mathrm{1.0 < z < 2.0}$. The black lines in the $\mathrm{V-J}$ vs $\mathrm{U-V}$ panels separates the quiescent galaxies region (light red), from the star-forming one (light blue). The adopted selection criteria is based on \cite{Williams2009}. The $\mathrm{U - V > 1.3}$ and $\mathrm{V - J < 1.6}$ criteria are redshift-independent and prevent the contamination from unobscured and dusty SFGs, respectively. The diagonal selection criteria is instead redshift-dependent. Galaxies are quiescent according to:
\begin{equation}
\begin{split}
\mathrm{(U - V)} >& \ 0.88 \times \mathrm{(V - J)} + 0.69, \quad [0.0 < z < 0.5]\\
\mathrm{(U - V)} >& \ 0.88 \times \mathrm{(V - J)} + 0.59, \quad [0.5 < z < 1.0]\\
\mathrm{(U - V)} >& \ 0.88 \times \mathrm{(V - J)} + 0.49, \quad [1.0 < z < 2.0]\\
\end{split}
\end{equation}

The $\mathrm{U}$, $\mathrm{V}$ and $\mathrm{J}$ magnitudes are computed for each simulated galaxy belonging to the approximate Bayesian posterior distribution. We build the rest-frame spectrum multiplying the $5$ template coefficients by the $5$ Kcorrect templates and applying the Milky-way extinction. Then, we integrate the spectrum over the wavelength range spanned by the $\mathrm{U}$, $\mathrm{V}$ and $\mathrm{J}$ Johnson broad-band filters. The red and blue contours represent the $1\sigma$, $2\sigma$ and $3\sigma$  distributions of our red and blue galaxies in the UVJ diagram. At all redshifts, the distributions of red and blue galaxies are peaked in the quiescent and star-forming galaxies regions, respectively. We also observe the expected effect of the increased star-forming activity at high redshifts, resulting in the distribution of blue galaxies moving towards bluer $\mathrm{V - J}$ colours and in that of red galaxies moving towards bluer $\mathrm{U - V}$ colours. Furthermore, the red galaxies distribution becomes more concentrated towards redder colours with decreasing redshift, as expected from the passive evolution of the stellar populations of red galaxies. Therefore, our red/blue galaxies separation is consistent with empirical studies in literature.


\bibliographystyle{unsrt}
\bibliography{tortorelli_jcap_bibliography}



\end{document}